\documentclass[a4paper,fleqn,usenatbib]{mnras}

\usepackage{newtxtext,newtxmath}

\usepackage[T1]{fontenc}
\usepackage{ae,aecompl}

\usepackage{graphicx}	
\usepackage{amsmath}	
\usepackage{amssymb}	
\usepackage{pdflscape}

\title[The Auriga Stellar Haloes]{The Auriga Stellar Haloes: Connecting stellar population properties with accretion and merging history}

\author[A. Monachesi et al.]{Antonela Monachesi$^{1,2,3}$\thanks{E-mail: amonachesi@userena.cl},
Facundo A. G\'omez$^{1,2,3}$, Robert J. J. Grand$^{4,5}$, \newauthor Christine M. Simpson$^{4}$, Guinevere Kauffmann$^{3}$, Sebasti\'an Bustamante$^{4}$, 
 \newauthor  Federico Marinacci$^{6,7}$, R\"udiger Pakmor$^{4}$, Volker Springel$^{4,5,3}$, Carlos S. Frenk$^{8}$,
\newauthor Simon D. M. White$^{3}$, and Patricia B. Tissera$^{9,10}$
\\
$^{1}$Instituto de Investigaci\'on Multidisciplinar en Ciencia y Tecnolog\'ia, Universidad de La Serena, Ra\'ul Bitr\'an 1305, La Serena, Chile\\
$^{2}$Departamento de F\'isica y Astronom\'ia, Universidad de La Serena, Av.
Juan Cisternas 1200 Norte, La Serena, Chile\\
$^{3}$Max-Planck-Institut f\"ur Astrophysik, Karl-Schwarzschild-Str. 1, 85748 Garching, Germany\\
 $^{4}$Heidelberger Institut f\"ur Theoretische Studien, Schloss-Wolfsbrunnenweg 35, 69118 Heidelberg, Germany\\
   $^{5}$Zentrum f\"ur Astronomie der Universitat Heidelberg, Astronomisches Recheninstitut, Monchhofstr. 12-14, 69120 Heidelberg, Germany\\
$^{6}$Department of Physics, Kavli Institute for Astrophysics and Space Research, MIT, Cambridge, MA 02139, USA\\
$^{7}$Harvard-Smithsonian Center for Astrophysics, 60 Garden Street, Cambridge, MA 02138, USA\\
$^{8}$Institute for Computational Cosmology, Department of Physics, University of Durham, South Road, Durham DH1 3LE, UK\\
$^{9}$Departamento de Ciencias F\'isicas, Universidad Andr\'es Bello, Av. Rep\'ublica 220, Santiago, Chile\\
$^{10}$Millennium Institute of Astrophysics, Av. Rep\'ublica 220, Santiago, Chile}

\date{Accepted XXX. Received YYY; in original form ZZZ}

\pubyear{2017}

\begin{document}
\label{firstpage}
\pagerange{\pageref{firstpage}--\pageref{lastpage}}
\maketitle

\begin{abstract}
We examine the stellar haloes of the Auriga simulations, a suite of thirty cosmological magneto-hydrodynamical high-resolution simulations of Milky Way-mass galaxies performed with the moving-mesh code {\sc arepo}. 
We study halo global properties and radial profiles out to $\sim 150$ kpc for each individual galaxy. The Auriga  haloes are diverse in their masses and density profiles; mean metallicity and metallicity gradients; ages; and shapes, reflecting the stochasticity inherent in their accretion and merger histories. A comparison with observations of nearby late-type galaxies shows very good agreement between most observed and simulated halo properties. However, Auriga haloes are typically too massive. We find a connection between population gradients and mass assembly history: galaxies with few significant progenitors have more massive haloes, possess large negative halo metallicity gradients and steeper density profiles. The number of accreted galaxies, either disrupted or under disruption,  that contribute 90\% of the accreted halo mass ranges from
1 to 14, with a median of 6.5, and their stellar masses span over three orders of magnitude.  The observed halo mass--metallicity relation is well reproduced by Auriga and is set by the stellar mass and metallicity of the dominant satellite contributors. This relationship is found not only for the accreted  component but also for the total (accreted + \emph{in-situ}) stellar halo. Our results highlight the potential of observable halo properties to infer the assembly history of galaxies.
\end{abstract}




\section{Introduction}
\label{sec:intro}

Stellar haloes of large galaxies like our Milky Way (MW) are thought to be formed primarily through the
accretion and merger of smaller satellite galaxies \citep{Searle_Zinn78}.
As a result of this
merger and disruption activity, stellar haloes are expected to possess a large
amount of substructure in the form of extended stellar streams and
small satellite galaxies, extending to large galactocentric radius \citep[e.g.,][]{Johnston96, Helmi_White99, BJ05, Cooper10, Gomez13}, as well as to exhibit large halo-to-halo variations in their properties, due to stochastic variations in halo merger history \citep[e.g.,][]{BJ05, DeLucia_helmi08, Cooper10, Gomez12, Tissera12}. 
Their constituent stars are fossil records of the hierarchical merging process; their ages and metallicities reflect the properties of the interstellar medium in the satellites at the time of their formation. They thus provide a unique window into reconstructing the mass assembly history of galaxies. 

However, stellar haloes are faint (reaching surface brightnesses of $\mu_V \sim 35~\rm{mag/arcsec}^2$), very extended (out to a few hundreds of kiloparsecs from the galactic center), and represent only a few percent of the overall mass and light of a galaxy. Hence, detecting them is an observationally expensive and challenging task.
 Over the past few decades, integrated light studies have detected the faint diffuse component of nearby galaxies uncovering several stellar streams \citep{Malin_hadley97, Shang98, Mihos05, Martinezdelgado10, Mihos13, Watkins15, Merritt16}; more recently significant progress has been made in controlling the scattered light that limits our ability to measure the surface brightness profiles of galactic haloes \citep[see e.g.,][]{Dsouza14, vanDokkum14, Trujillo_fliri16}.
 
Nevertheless, one of the best approaches for
 mapping the structure and properties of stellar haloes is still to resolve their individual stars. This allows us to obtain detailed age and metallicity information
 and enables effective faint surface brightness levels ($\mu_V \sim 34~\rm{mag/arcsec}^2$) to be reached. 
 Due to their proximity, the individual stars of the MW's and M31's haloes have been extensively studied (e.g., \citealt{Newberg_yanny05, Carollo07, Carollo10, Ivezic08, Juric08, Bell08, Bell10, Sesar11, Deason13, Xue15, Carollo16, Slater16, FA17} for MW halo studies; \citealt{Kalirai06, McConnachie09, Gilbert12, Gilbert14, Ibata14} for M31 halo studies). While both stellar haloes are highly structured and present similarities, such as steeply declining power-law like density profiles and large spatial extents, their detailed properties are significantly different. The MW stellar halo appears to have a broken power-law density profile (although see \citealt{Slater16}), has a weak or absent radial metallicity gradient and is rather light, with an estimated mass of $(4-7)\times 10^8~ \rm{M}_{\odot}$. M31, on the other hand, has a rather massive stellar halo ($\sim 10^{10}~ \rm{M}_{\odot}$) that can be best described with a single power-law profile and its metallicity radial profile shows a continuous negative gradient of $-0.01 \rm{dex}~\rm{kpc^{-1}}$ over $\sim 100$ kpc. These differences suggest that the two galaxies have had very different accretion and merger histories (see e.g. \citealt{Deason13,Gilbert14, Harmsen17, Amorisco17, Dsouza_bell18}). 
 
 Clearly, information gained from just two galaxies is insufficient to constrain galaxy formation models. 
Because of this, the study of resolved stellar haloes in other large nearby galaxies outside the Local Group has received increased attention during the last few years  
\citep[e.g.,][see also the review by \citealt{Crnojevic17_rev}]{Harris_harris02, Mouhcine05c, Harris07a, Harris07b, Barker09, Mouhcine10, Bailin11, M13, Rejkuba14, Greggio14, Okamoto15, Peacock15, M16a, Crnojevic16, Harmsen17, Tanaka17}. Detailed stellar population information is obtained mostly with HST observations, such as those from the GHOSTS survey \citep{RS11, M16a}. These are, however, pencil-beam observations and should be complemented by panoramic views of the galaxies  to understand and fully characterise features that may provide information about their assembly histories (see e.g. the panoramic view of M81 by \citealt{Okamoto15} and also the PISCeS survey by \citealt{Crnojevic16}). 

One of the main results from the GHOSTS survey is that 
   stellar halo properties are very diverse among the sample of eight galaxies studied (including the MW and M31) that are otherwise similar in morphology, mass, and luminosity. There is a large range in the median metallicities and radial metallicity profiles of these galaxies' haloes \citep{M16a}. A similar diversity is observed in the slopes of their power-law surface brightness profiles and inferred stellar halo masses \citep{Harmsen17}. Interestingly, the diversity in stellar halo masses was also found in integrated light studies of a different set of nearby galaxies from the Dragonfly survey \citep{Merritt16}. Another important result from the GHOSTS survey is the discovery of a tight correlation between the stellar halo mass and halo metallicity at 30 kpc along the minor axis \citep{Harmsen17}. The more massive the stellar halo, the more metal rich it is; this likely reflects the properties of the dominant satellite contributors to the accreted halo \citep[see e.g.][]{Deason16, Bell17,Dsouza_bell18}.
All these observed properties are important probes of the physics of stellar halo formation and need to be interpreted and contrasted against models in order to improve our understanding of halo formation and help discriminate
between different formation scenarios. 

Early theoretical models that only took into account the accreted component of haloes predicted this observed diversity and attributed it to stochasticity  in the merger history (e.g. \citealt{BJ05,DeLucia_helmi08, Cooper10, Tumlinson10, Gomez12}). However, these studies considered only $\lesssim 10$ different simulations, thus undersampling the range of possible merger histories. More recently, \citet{Amorisco17} used a large number of idealised dark matter only minor merger simulations to link the halo assembly history of MW-mass galaxies to the properties of their stellar haloes. He showed that the stellar halo mass of a galaxy can inform us about its merger and accretion history, but with significant scatter. On average, galaxies with low mass stellar haloes have experienced a phase of ``fast growth" at early redshifts ($z$) and then have had a quiet accretion history until the present day. On the other hand, large stellar haloes have experienced, on average, a phase of ``fast growth" at intermediate redshifts which maximises the accreted stellar mass by $z= 0$. \citet{Deason16}, using 45 cosmological dark-matter only N-body simulations showed that massive stellar haloes are primarily built from a few large satellite galaxies rather than from many low-mass satellites \citep[see also][]{Cooper10, Amorisco17a}. Importantly, \citet{Deason16} found a correlation between the average metallicity of the accreted stellar material and the mass-weighted average of its contributor satellites, which matches remarkably well the relationship subsequently discovered in the GHOSTS survey between the stellar halo mass and halo metallicity at 30 kpc along the minor axis \citep{Harmsen17}. This relationship was further investigated by \citet{Dsouza_bell18} using the accreted component of $\sim 4600$  galaxies from the Illustris cosmological hydrodynamical simulations \citep{Vogelsberger14a, Vogelsberger14b, Genel2014}, demonstrating it to exist over three orders of magnitude in accreted stellar mass.
Despite the insight gained from all these studies, they either i) lack statistical power \citep{BJ05, Cooper10, Gomez12}; ii) are not based on fully hydrodynamical cosmological simulations \citep[][]{BJ05, Cooper10, Gomez12, Deason16, Amorisco17a}; or iii) do not have the very high resolution needed to analyse in detail the properties, especially the gradients, of individual haloes \citep{Dsouza_bell18}.

In addition to the accreted stars, semi-analytical and hydrodynamical simulations of the formation of MW-like galaxies predict that the inner regions of stellar haloes host an {\it in-situ} stellar component
composed both of stars born in the host's galactic disc that are later ejected into
the halo due to interactions with subhaloes or molecular clouds, and of stars formed in streams of gas stripped from infalling satellites 
\citep[e.g.,][]{Benson04, Zolotov09, Purcell10, Font11, McCarthy12, Tissera13, Pillepich14, Tissera14, Cooper15, M16b, Elias18}. This component should be confined close to the disc plane and more metal-rich than the accreted halo
\citep[e.g.][]{Pillepich15, M16b}. However, the prominence of this component in mass and extent can vary by large factors from model to model ranging from being dominant at radii of as large as 30 kpc \citep[e.g.,][]{Font11, M16b, Elias18} to being significant only at $<$ 5 kpc \citep[e.g.,][]{Zolotov09, Pillepich15}. This in-situ halo diversity in simulations is strongly driven by the details of the modelling of sub-grid physical processes, such as star formation and feedback \citep[see discussion on this in][]{Zolotov09, Cooper15}, and is also partly due to the definition of \emph{in-situ} halo, which varies between studies, and to numerical resolution.

In this paper we analyse the stellar haloes of the Auriga simulated galaxies introduced in \citet[][hereafter G17]{Grand17}. These are thirty very high resolution cosmological magneto-hydrodynamical simulations performed with the moving-mesh code {\sc arepo}.
The main advantage of this work over previous numerical studies is the very high resolution obtained for a relatively large number of individual  hydrodynamically simulated haloes; this is the largest dataset of currently available haloes at this mass resolution.  The high resolution allows us to study and analyse in detail the properties of each individual halo, rather than averaging the properties of lower resolution simulations. The relatively large number of haloes allows us to start relating observable properties to the merger and accretion history of each individual galaxy in a statistical manner, quantifying a mean and scatter of stellar halo properties.
 
To provide a meaningful connection between the measured observable properties of stellar haloes and the mass assembly history of galaxies, in this work we consider not only spherically averaged properties, but also other structural properties that can be readily compared 
with observations, such as projected stellar halo shapes, metallicities 
along one axis, projected power law density slopes, etc. A fair and detailed comparison with observations is important in order to ensure that any mismatch is not a consequence of the way the comparison is performed but rather due to the physics implemented in the models. This is also an advantage over many other studies, which give, for example, spherically averaged stellar halo properties that cannot be obtained from observations. 

We test predictions from this suite of simulations against the available results for stellar haloes of nearby galaxies (e.g., GHOSTS, M31, MW, Dragonfly survey). The results and predictions from this work should be useful for comparison with future observations, such as those that are being carried out or planned for current facilities e.g. HSC/Subaru or DESI and future facilities such as LSST, ELT, GMT, and WFIRST. These will greatly increase our knowledge of the stellar properties of galactic haloes in the Local Universe, gained currently from only a handful of galaxies. 

We describe the simulations and nomenclature we use in Section~\ref{sec:meth}.
Section~\ref{sec:general} presents the main properties of 
the Auriga stellar haloes, such as surface brightness, metallicity, age, and axis ratio profiles, as well as accreted mass fractions. All the profiles are presented as a function of spherically averaged radius as well as projected along one direction to
facilitate comparison with observations.
We analyse the mass assembly of the accreted stellar haloes in Section~\ref{sec:satel} and connect observable properties of stellar haloes with the mass accretion history of each galaxy. In Section~\ref{sec:compa} we compare our results with observations of individual galaxies in a detailed and quantitative way, highlighting both the agreements and the mismatches that we find. We discuss our results in Section~\ref{sec:dis} and summarise and conclude in Section~\ref{sec:concl}.

Throughout the paper we use the term ``halo'' to refer to the stellar halo, unless otherwise stated.

\section{Methodology and definitions}
\label{sec:meth}

\begin{figure*}
\centering
	\includegraphics[width=2\columnwidth]{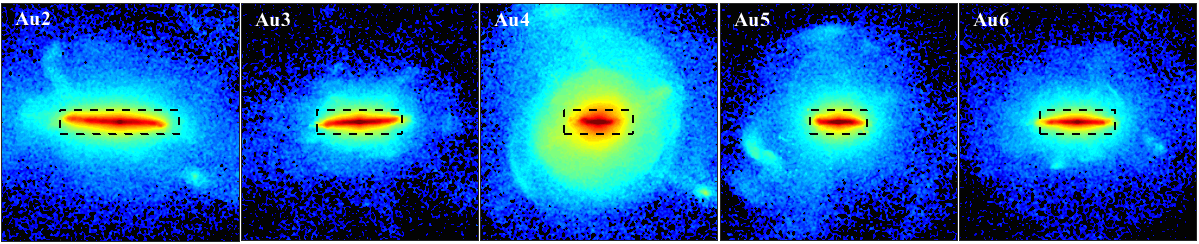}
	\includegraphics[width=2\columnwidth]{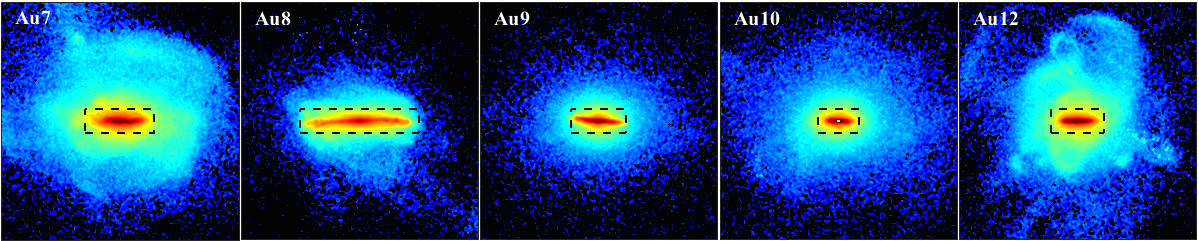}
	\includegraphics[width=2\columnwidth]{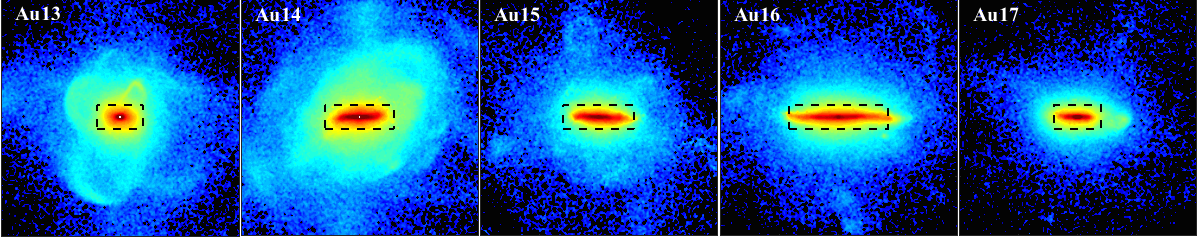}
	\includegraphics[width=2\columnwidth]{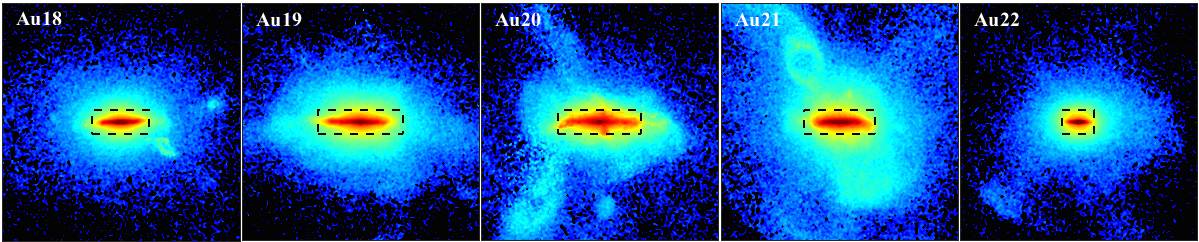}
	\includegraphics[width=2\columnwidth]{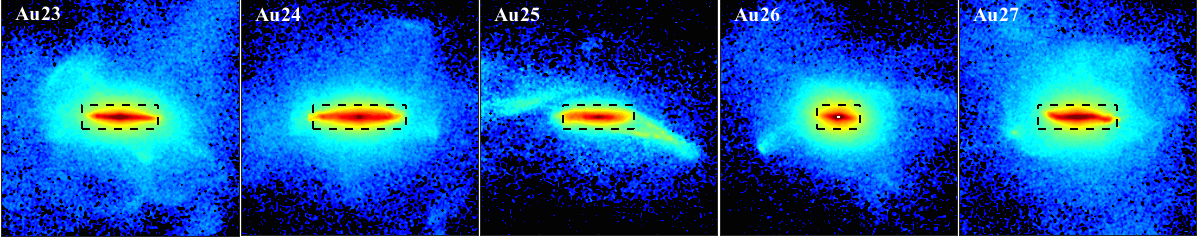}
	\includegraphics[width=2\columnwidth]{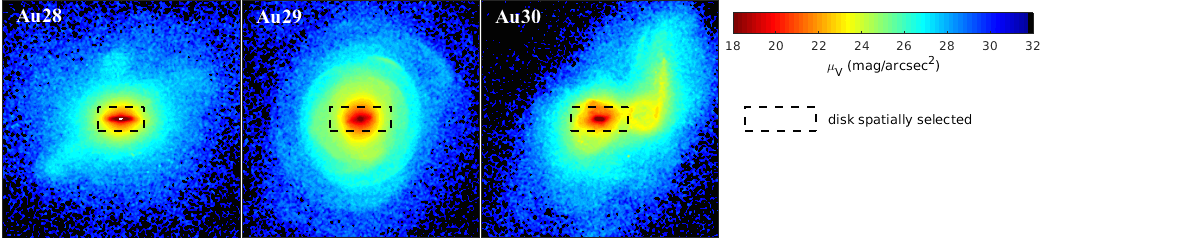}
    \caption{Surface brightness maps in the V-band of the Auriga galaxies,
    seen edge-on at $z=0$ in a $(200\times 200)\, \rm{kpc}^{2}$ square. Only stellar particles that are gravitationally bound to the main galaxy at $z=0$ are plotted, i.e. particles bound to surviving satellite galaxies at $z=0$ are not shown in this figure. The black-dashed box indicates our spatial identification of the disc component, with a 10 kpc distance above and below the disc's plane and a length twice the disc optical radius. Halo stars selected spatially include all stars outside this box. This selection is used when comparing with observations for which it is not possible to select halo stars kinematically.}
    \label{fig:surfacebright}
\end{figure*}

\subsection{The Auriga Simulations}
\label{sec:sims}

The Auriga simulations are a suite of thirty cosmological magneto-hydrodynamical zoom
simulations of MW-sized dark matter haloes. 
A detailed description of these simulations can 
be found in G17. Here we briefly describe their main features. 

Candidate haloes were first selected from a parent dark matter only cosmological simulation of the EAGLE project \citep{Schaye15},
carried out in a periodic cube of side 100$h^{-1}$Mpc. A $\Lambda$CDM cosmology was adopted, with parameters $\Omega_{m}  = 0.307$, 
$\Omega_{b}  = 0.048$,  $\Omega_{\Lambda}= 0.693$,  and  Hubble  constant  $H_{0}$  = 100 $h$ km s$^{-1}$ 
Mpc$^{-1}$, $h=0.6777$ \citep{planck}. Haloes were selected to have a narrow mass range of $1 < M_{200}/10^{12}\rm{M}_{\odot}<2$, comparable to that of 
the MW, and to satisfy an isolation criterion at  $z=0$ (see G17 for details 
on the process of host halo selection).  By  applying a  multi-mass 
`zoom-in' technique, each halo was re-simulated at higher resolution with the state-of-the-art N-body and moving mesh magnetohydrodynamics code {\sc arepo} \citep{Springel10, Pakmor16}. 

Gas was added to the initial conditions and 
its evolution was followed by solving the equations of ideal magnetohydrodynamics on an unstructured Voronoi mesh. The typical mass of a dark matter
particle is $\sim 3 \times 10^{5}$ M$_{\odot}$, and the baryonic mass resolution is 
$\sim 5 \times 10^4$ M$_{\odot}$. The physical gravitational softening length grows with the scale factor up to a
maximum of 369 pc, after which it is kept constant. The softening length of gas cells is scaled by the mean 
radius of the cell, with a maximum physical softening of 1.85 kpc. 

The simulations include a comprehensive model for 
galaxy formation physics which includes important baryonic processes, such as primordial and metal-line cooling \citep{Vogelsberger13}; a sub-grid model for 
the interstellar medium that utilises an equation of state representing a two-phase medium in pressure equilibrium \citep{Springel03}; a model
for the star formation and stellar feedback that includes a phenomenological wind model \citep{Marinacci14, Grand17} and metal enrichment from SNII, SNIa and AGB stars \citep{Vogelsberger13};
black hole formation and active galactic nucleus feedback \citep{Springel05, Marinacci14, Grand17}; a spatially uniform, time-varying UV background after reionization at redshift six \citep{Faucher-giguere09, Vogelsberger13} and magnetic fields \citep{Pakmor13, Pakmor14}.
The model was specifically developed for the {\sc arepo} code and was calibrated to reproduce several observational results such as the stellar
mass to halo mass relation, galaxy luminosity functions and the history of the cosmic star formation rate density. 

In this work, we analyse 28 out of the 30 Auriga galaxies, denoted by `AuN' with N varying from 1 to 30. We exclude from our analysis Au1 and Au11, which are not isolated.
Au1 has a nearly equal mass companion within $R_{200}$, the radius within which the halo's mean density is equal to 
200 times the critical density of the universe, and Au11 is undergoing a major merger at redshift zero. All of the Auriga galaxies are forming stars at $z=0$ and
most of them have a disc; only three out of 30 (Au 13, 29 and 30) do not show extended discs at $z=0$, but rather a spheroidal morphology. Nevertheless they present a small disc component as shown by G17 in their Figs. 2 and 3.

In Figure~\ref{fig:surfacebright} we show
$V-$band maps of stellar surface brightness for all the Auriga galaxies analysed in this work, seen edge-on.
Discs are aligned with the XY plane as in G17.
Only star particles that at $z=0$ are gravitationally bound to the main 
galaxy are used to create the maps and for the analysis presented in this paper,
unless otherwise stated; stars which are bound to a distinct satellite of the main galaxy are excluded. Photometry was obtained using \citet{BC03} stellar population synthesis models. We estimate
the luminosity of each stellar particle, treated here as a single stellar population of a given age, mass and metallicity, in several broad-bands. 
We currently record luminosities for $U, V, B, K, g, r, i,z$ bands without modelling the effects of dust extinction. 
A visual inspection of Figure~\ref{fig:surfacebright} reveals differences between, e.g, disc size (see Table 1 and G17 for a discussion on disc sizes origin)
and the amount of substructure and stellar halo shapes among the Auriga galaxies. 
The diversity in morphological properties of these simulated galaxies
reflects the stochasticity inherent to the process 
of galaxy formation and evolution \citep[e.g.][]{BJ05, Cooper10, Tumlinson10}.

\subsection{Definition of stellar halo}
\label{sec:def}

The definition of a galaxy's stellar halo is not
straightforward. Several definitions have been used in previous work, both
in numerical and in observational studies. We will
use here two definitions: 1) a kinematic decomposition to allow comparison with
previous numerical work and 2) an observationally-motivated definition for a
more consistent comparison with observations, for which kinematic
decomposition of the data is not possible.

{\bf 1) Kinematic definition:} Theoretically, the stellar halo of a disc galaxy is commonly defined as the
kinematic component which is not rotationally supported. This is usually
characterised by the orbital circularity parameter, defined as: 
$\epsilon = J_{z}/{J(E)}$ \citep{Abadi03}, where $J_z$ is the angular momentum
around the disc symmetry axis and $J(E)$ is the maximum specific angular
momentum possible at the same specific binding energy, $E$. We selected the
subset of star particles with $\epsilon < 0.7$ as the spheroidal component. We note that this is done for all galaxies, including the three galaxies mentioned in Section~\ref{sec:sims} that do not have extended discs. Although perhaps in those cases it is more relevant to look at the haloes globally, we decided to select the spheroidal component using the same methodology for all galaxies for consistency in the analysis.
The chosen circularity value ($\epsilon < 0.7$) was used by e.g., \citet{Marinacci14,M16b, Gomez17}
to isolate disc particles from the spheroidal component of the galaxy.
Other weaker (e.g., $\epsilon < 0.8$) and more restrictive (i.e. $\epsilon <
0.65$) constraints for the halo were adopted by \citet{Font11, McCarthy12,Cooper15, M16b} and by
\citet[][]{Tissera13,Tissera14, M16b}, respectively. 
Following \citet{Cooper15}, particles in the spheroidal
component that lie within 5 kpc from the galactic centre are defined as bulge.
We note that the kinematic halo component is selected regardless of the  origin of the
stellar populations, i.e. whether the stellar particles are born \emph{in-situ} or in satellite galaxies.

{\bf 2) Observational definition:} Observers have made several different selection criteria to isolate
halo stars from the galactic disc. Moreover, the criteria are generally
different for the MW and for external galaxy studies.
For the MW, kinematic or photometric selections can be made. For
external galaxies, there is to date no kinematic information available that allows the disc to be isolated from the halo. Thus, the stellar halo
is generally defined as the population located beyond a certain galactocentric distance and outside the disc plane (see e.g. \citealt{Mouhcine05b, M13, Rejkuba14, M16a}). 
The downside of this approach is that it is hard
to define the end of the disc. As a result, it is often questionable whether
what is observed is disc or halo. Studies of edge-on galaxies that confine
their stellar halo studies to the stars above 5 or 10 kpc from the disc plane
are safer. However if galaxies are significantly inclined,
contamination from the disc can be hard to account for. In these cases,
photometric cuts may be made in order to avoid, for instance, the most metal-rich stars 
which are generally attributed to the disc. 
Note that, observationally, it is only possible to obtain projected quantities rather than
spherically averaged quantities, as generally done in numerical work (see~\citealt{M16b}). 

Based on the previous discussion, we use a spatial selection criterion to  define our 
observationally-motivated stellar halo. We define the Z-axis as the direction
perpendicular to the disc plane. All stellar particles that, projected on the X-Z
plane, are located \emph{outside} $|X| = R_{\rm opt}$ 
and $|Z| = 10$ kpc (more than $5-10$ times larger than the typical scale height of all Auriga galaxies; see G17) are considered to be part of the stellar halo. Here, $R_{\rm opt}$ 
is the optical radius, defined
as the radius at which the surface brightness in the $B-$band reaches 
$ \mu_{B} = 25~\rm{mag/arcsec}^2$ when seen face-on\footnote{We note that there is no significant difference in the optical radius when calculating it using the edge-on surface brightness maps. This is because the optical radius mostly coincides with the truncation radius of the disc.}. We note that in some cases the discs are lopsided and thus the radius at which
the galaxy reaches this limiting surface brightness on one side is not the same as on the other
side. We chose in these cases the maximum radius as the end of the disc 
in order to minimise disc contamination as much as possible. 
We note that beyond 10 kpc along the minor axis, all our galaxies
have $ \mu_{B} > 25~\rm{mag/arcsec}^2$ which assures us that we are not looking at the
disc component on the Z-axis either. 
The dashed-black rectangle superimposed on each galaxy in
Figure~\ref{fig:surfacebright} indicates this region. All
stellar particles outside the rectangle are considered in the
observationally-motivated definition of stellar halo.

This stellar halo definition has no dependence on $\epsilon$ or the stellar particle's origin.
Thus it is readily comparable to observations. Note that, except for \citet{Pillepich15} 
and \citet{M16b}, no other
numerical work has used a spatially selected definition of stellar halo, 
which is arguably the most useful definition if the goal is to
compare with observations of external galaxies with no kinematic information.
 
\begin{figure*}
\centering
\includegraphics[width=2\columnwidth]{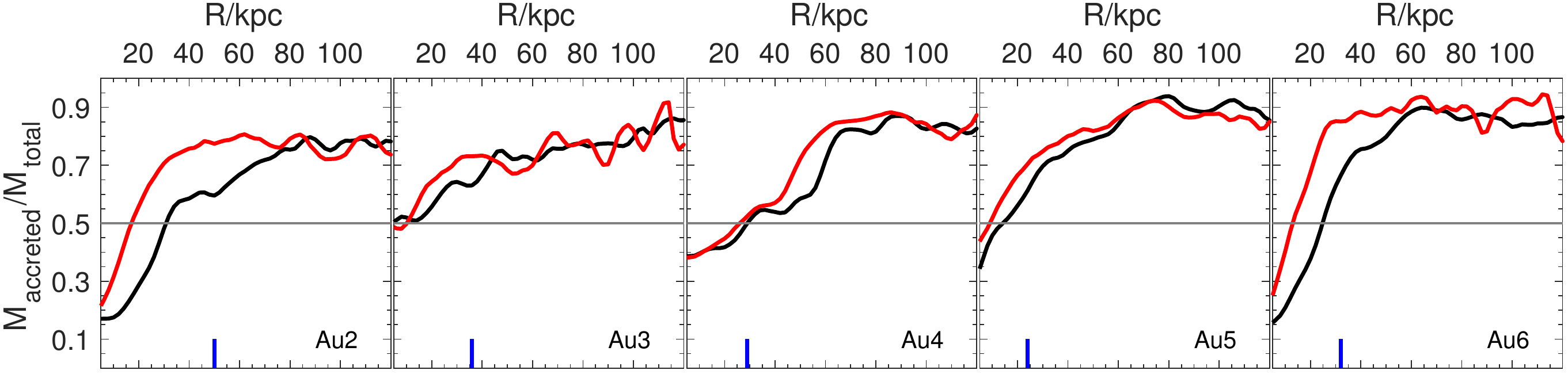}
	\includegraphics[width=2\columnwidth]{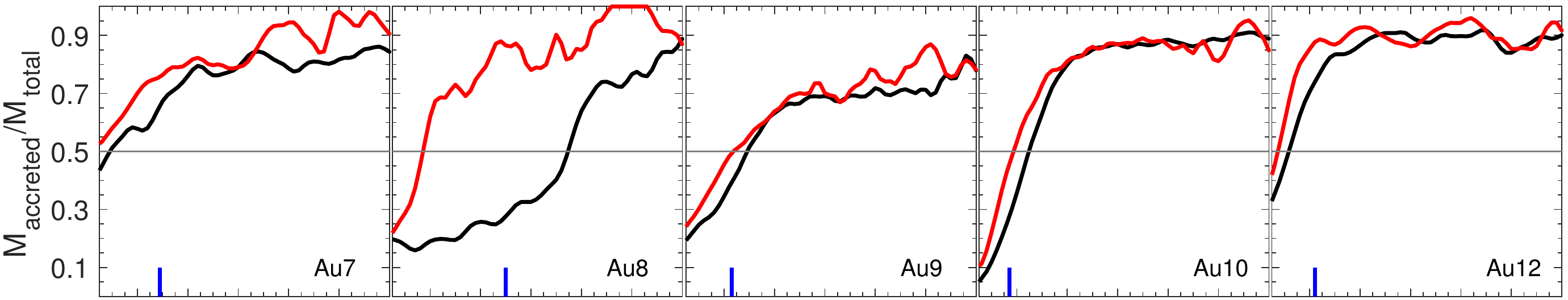}
		\includegraphics[width=2\columnwidth]{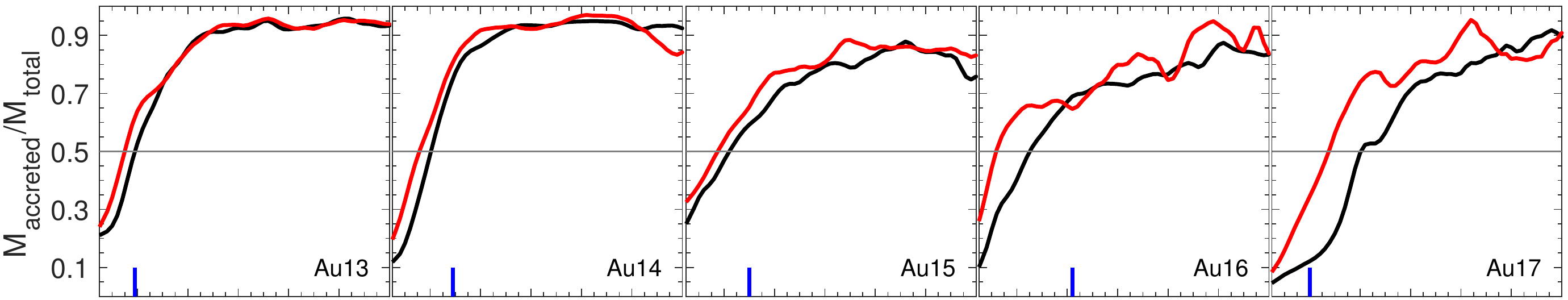}
			\includegraphics[width=2\columnwidth]{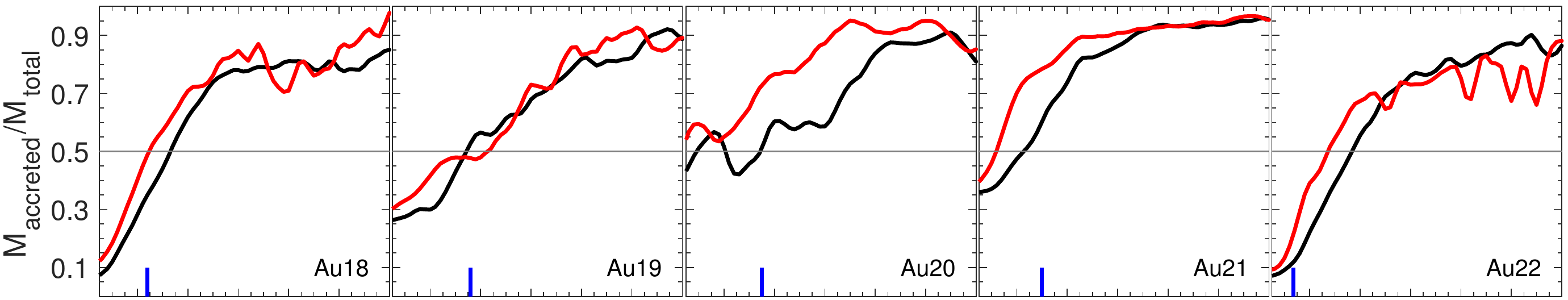}
				\includegraphics[width=2\columnwidth]{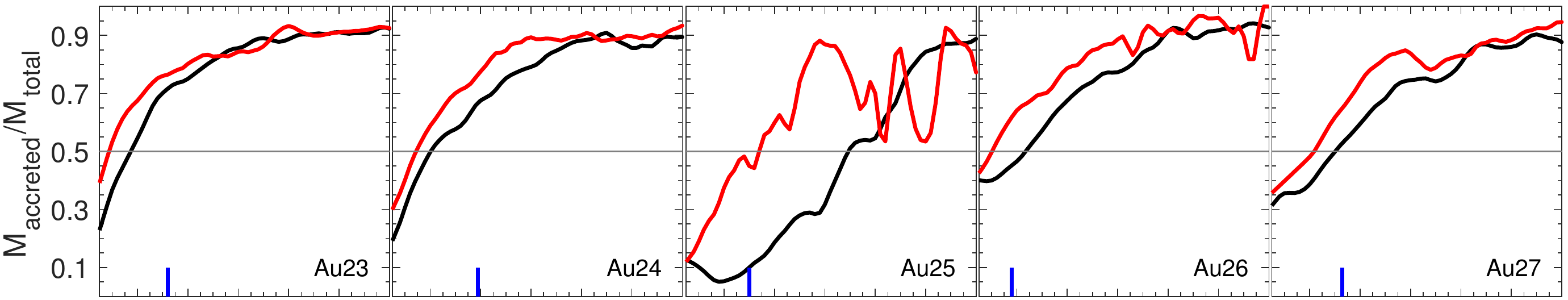}
				\includegraphics[width=2\columnwidth]{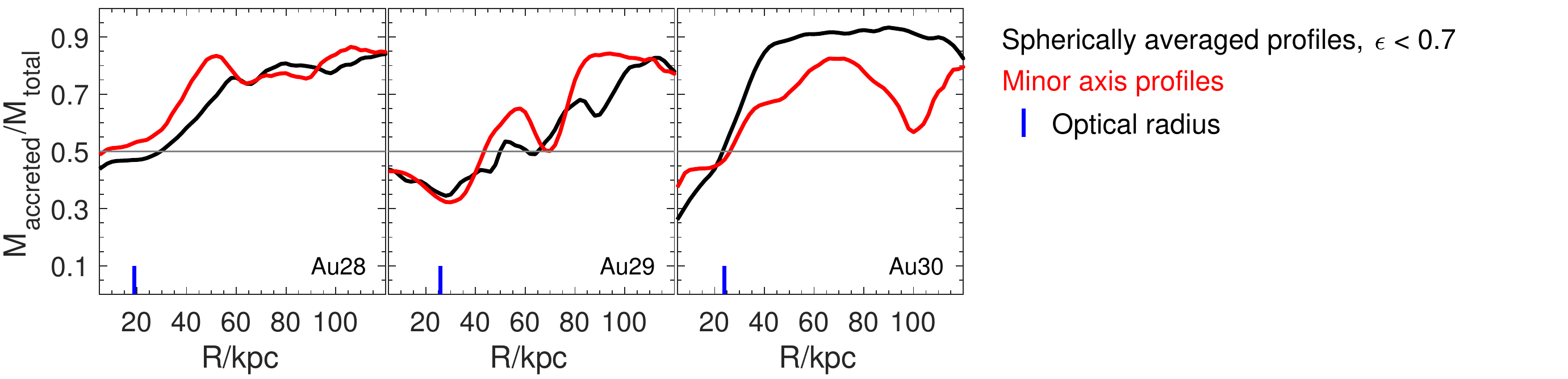}
	\caption{Local accreted mass fraction of the stellar halo as a function of $R$ in spherical concentric shells (black) and on the minor axis (red) for halo particles selected kinematically. The optical radius of each galaxy is indicated as a blue vertical line. The accreted stellar halo dominates typically beyond one optical radius; in some extreme cases this happens only beyond two optical radii. Along the minor axis, the accreted stellar halo component dominates typically beyond 20 kpc; in several cases it dominates beyond 10 kpc. The horizontal line in each panel indicates a local accreted stellar halo fraction of 50\%.}
 \label{fig:exsitu_frac}
\end{figure*}

\subsection{Definition of accreted and \emph{in-situ} component}
\label{sec:accre}

The {\it accreted} stellar component is defined as all stellar particles born
in satellite galaxies, i.e. particles that were bound to a satellite galaxy in the first 
snapshot in which they are identified (`birth time') and that at $z=0$ are gravitationally bound to the main 
galaxy. This definition disregards whether at `birth time' the satellites are outside or
inside the host's virial radius. We note that the stellar particles called `endodebris'
in \citet{Tissera13} and `commuters' in \citet{Snaith16}, defined as stars formed within the host's virial radius from gas brought in by satellites, are classified here in part as accreted component (formed inside the orbiting satellites), and in part as {\it in-situ} component (formed in recently stripped streams of gas).

All the other stellar particles that are bound to the host galaxy at `birth time'
are defined as the {\it in-situ} component. 
As we show in the following sections, the {\it in-situ} stellar halo can reach 
large galactocentric distances, in some extreme cases out to 100 kpc. This is due
to stars that are born from gas stripped from satellite galaxies. These
are included within our definition of the {\it in-situ} component (see a discussion on the
`stripped-gas' \emph{in-situ} component in \citealt{Cooper15}). However, their contribution to the \emph{in-situ} halo
is not very significant. We estimate how much of the \emph{in-situ} halo was born within a spatially
defined disc region, i.e. a cylinder covering a X-Y region of one optical radius and 10 kpc height
on the Z-axis. Most of the \emph{in-situ} halo
($\gtrsim 75\%$) located at distances of  $|Z| \geq 10$ kpc from the disc mid-plane at the present day has its birth radius within
the spatially defined disc. 
We note that \citet{Pillepich15} classified as ``ex-situ'' or accreted those stars formed from gas that was stripped less than 150 Myr earlier than birth time.
  
The two definitions of stellar halo used in this work contain both accreted and \emph{in-situ} stellar populations.
It is nonetheless interesting to highlight that some galaxies show a significant rotationally supported accreted component, i.e.
stellar particles that satisfy $\epsilon > 0.7$. These stellar particles are excluded from the kinematically defined stellar halo and branded 
as  an ``ex-situ" disc component. The properties of these ex-situ discs are analyzed in \citet{Gomez17}.

\section{General properties of the Auriga stellar haloes}
\label{sec:general}

We describe in this section the main properties of the Auriga stellar haloes.  
In what follows we show spherically averaged, azimuthally averaged (2D projected), and projected along the
minor axis properties of each galaxy. The minor axis quantities are 
computed in $15^{\circ}$ projected wedges on the $Z-$coordinate (perpendicular to the disc plane). 
To increase the numerical resolution and smooth out sudden variations due to the presence of substructure, we include 
stellar particles located within diametrically opposed wedges. 

As we discuss below, some results, in particular halo metallicity gradients, can significantly change if they are constructed 
from spherical concentric shells around the galactic center or from projected wedges along the
minor axis \citep[see also][]{M16b}. This difference is important when models
are used to compare with and interpret observations. Typically, azimuthally-averaged, let alone spherically-averaged,
quantities cannot be obtained observationally, and thus these are measured along a given direction. 

We note that only the kinematically selected halo is shown when plotting the spherically averaged profiles. Since the spatial selection avoids the disc region, spherically (and azimuthally) averaged profiles for this selection can only be made beyond the optical radius of each galaxy. Instead, the profiles along the minor axis are presented both for the kinematical and spatial halo selection.

Unless otherwise stated, the line convention for all the figures presented in this section is as follows. Black colours represent spherical (or azimuthal in the case of the surface brightness) profiles whereas red colours represent minor axis profiles. Solid lines are used for the overall (accreted $+$ \emph{in-situ}) halo and dashed lines are accreted-only. Dotted blue lines represent the minor axis profiles for the overall (accreted $+$ \emph{in-situ}) halo spatially defined, i.e. without circularity constraint. 

Table~\ref{t1} lists the main properties of the Auriga stellar haloes that we derive and discuss in this section. 

\subsection{Accreted mass fraction profiles}
\label{sec:massfra}

The total accreted mass fraction of the Auriga galaxies, $f^{\rm tot}_{\rm acc}$,  is $ < 0.2$, as expected for MW-mass galaxies whose stellar mass budget is dominated by their \emph{in-situ} population 
\citep[see e.g.][]{Dsouza14}. Au17 and
Au22 show the lowest values of $f^{\rm tot}_{\rm acc} \sim 0.02$ (see G17 for a list of  $f^{\rm tot}_{\rm acc}$ for each galaxy).

For the kinematically defined stellar halo, the accreted mass fraction $f^{\rm kh}_{\rm acc}$ is of course 
larger than $f^{\rm tot}_{\rm acc}$ and it varies as a function of galactocentric distance. Figure~\ref{fig:exsitu_frac} 
shows the spherically averaged and projected minor axis $f^{\rm kh}_{\rm acc}$ profiles, where the local fraction of accreted mass, i.e. the fraction of stellar halo stars at a given radius which are accreted, is shown at each radius.    
Given the different disc sizes of 
the Auriga galaxies (see Table~\ref{t1} and G17 for a discussion on this), the extent of the \emph{in-situ} halo contribution will vary 
from galaxy to galaxy. The blue vertical line on the bottom of each panel indicates the galaxy optical radius.
This figure shows that the Auriga stellar haloes have an \emph{in-situ} population that typically dominates out to the optical radius for the spherical profiles,
beyond which the accreted component begins to dominate. 
Interestingly, some galaxies such as Au8, Au9, Au10, Au17, Au18, Au22 and Au25 have a dominant \emph{in-situ} halo component 
($f^{\rm kh}_{\rm acc} < 0.5$)   even beyond $R_{\rm opt}$. The reason for this is two-fold. 
Au8 and Au25 have experienced recent violent satellite interactions
that ejected \emph{in-situ} material to large galactocentric distances. On the other hand, as we show later 
in Section~\ref{sec:satel}, Au10, Au17, Au18, and Au22 have had 
extremely quiet late merger histories and thus have very low mass accreted stellar haloes. In those cases, as we will show in Section~\ref{sec:density}, the \emph{in-situ} halo material in the outer regions originates from disc-satellite interactions at early times ($\gtrsim 8$ Gyr ago) that ejected disc material to large galactocentric distances. 
 
Along the minor axis, the accreted component  becomes dominant ($f^{\rm kh}_{\rm acc} > 0.5$)  at shorter distances, typically at 20 kpc and in several cases it dominates beyond 10 kpc. This is not surprising since most of the stellar halo  material
beyond $R_{\rm opt}$ along this direction is expected to be accreted. Note, however, that on those galaxies that
have experienced either strong interactions or very quiet merger histories, the \emph{in-situ} halo dominates beyond 
$R_{\rm opt}$, even along the minor axis. 
The spatially defined stellar halo accreted mass fraction is not shown in Figure~\ref{fig:exsitu_frac}. However, we note that, as we will show in the next figures, the properties of the kinematically and spatially defined stellar haloes along the minor axis overlap with each other.

 We note that Fig.~\ref{fig:exsitu_frac} shows that the accreted mass fraction does not reach 100\%. The fraction of \emph{in-situ} material is generally lower than 10-20\% beyond 100 kpc, which is nevertheless a non-negligible fraction. The presence of \emph{in-situ} material at those radii is mainly due to two effects: 1) stars formed at large distances in streams of gas stripped from infalling gas-rich satellites; 2) \emph{in-situ} stars that were born in the disc of the main galaxy and are then scattered out to large radii because of major mergers. Effect 1)
has been studied in detail by \citet[see  their  discussion on the `stripped-gas' \emph{in-situ} halo component]{Cooper15}. In addition, Cooper et al. also find stars formed outside the disc from gas that has been smoothly accreted onto the halo. These `smooth gas' \emph{in-situ} stars tend to form at the same time and place as the stripped-gas population, suggesting that their formation is associated with the same gas-rich accretion event. Effect 2) occurs in the event of a major merger. This event significantly perturbs and even destroys the disc, generating tidal arms that populate the outer galactic regions with \emph{in-situ} material, thus scattering \emph{in-situ} stars out to very large radii. We see evidence of this in a few Auriga galaxies, like Au4, Au25 and Au29, in which the fraction of \emph{in-situ} stars reaches $\sim 20$ \% at distances of $\sim 100$ kpc.

Table~\ref{t1} lists the total \emph{in-situ}  and accreted masses of the kinematically defined stellar haloes. In general, we find that the Auriga stellar haloes have a massive ($\sim 1\times10^{10}~\rm{M}_{\odot}$) \emph{in-situ} population which varies at most by a factor of six among the models, i.e. the \emph{in-situ} haloes are all similar in mass. Nevertheless, it is a more centrally concentrated component than the accreted one, with a typical half-mass radius for the \emph{in-situ} halo of $\sim 10$ kpc (with the inner 5 kpc region excluded, as this is considered to be part of the bulge), compared to the $\sim 25$ kpc for a typical half-mass radius of the accreted haloes. The accreted haloes vary by an order of magnitude in their masses (from $\sim 0.1~\rm{to}~2 \times10^{10}~\rm{M}_{\odot}$). A detailed analysis of the different channels of \emph{in-situ} halo formation is deferred to future work.

\begin{figure*}
	\includegraphics[width=2\columnwidth]{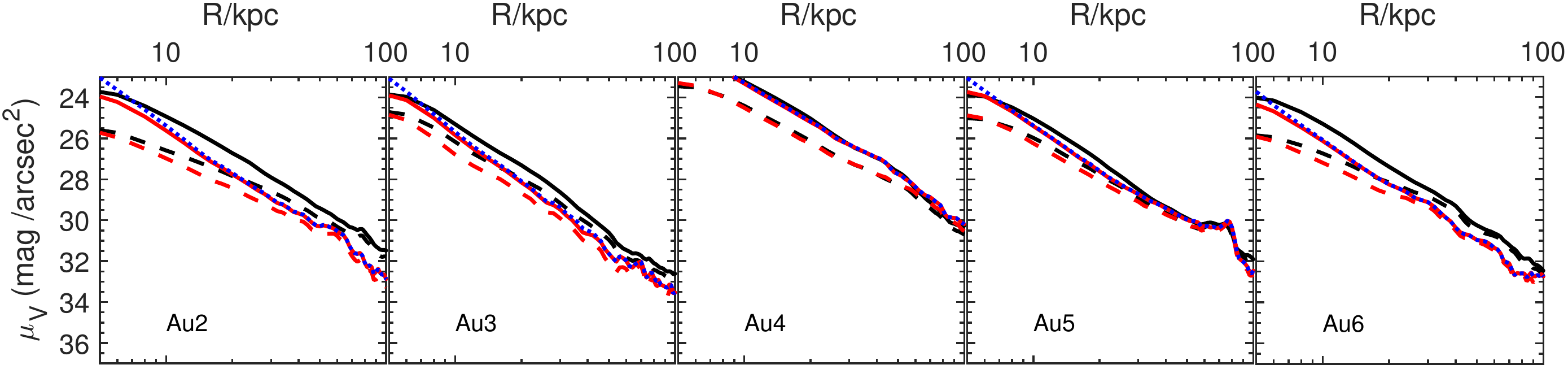}
	\includegraphics[width=2\columnwidth]{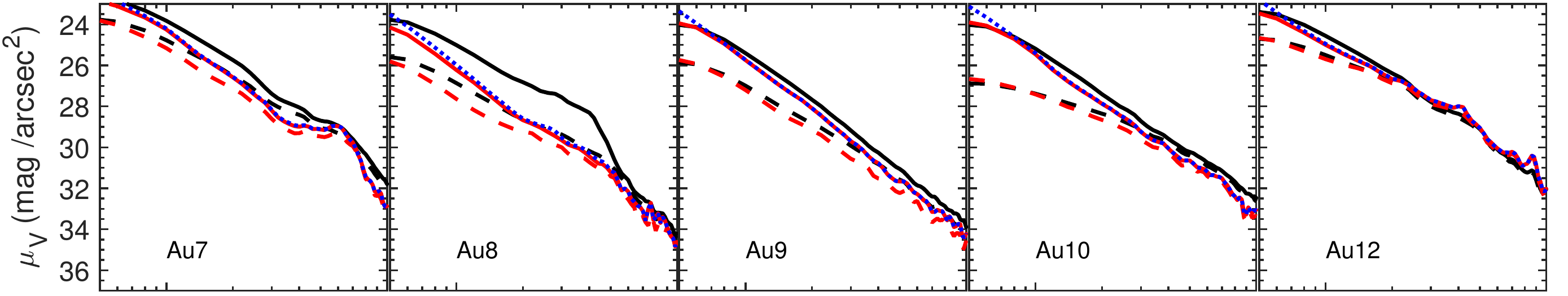}
	\includegraphics[width=2\columnwidth]{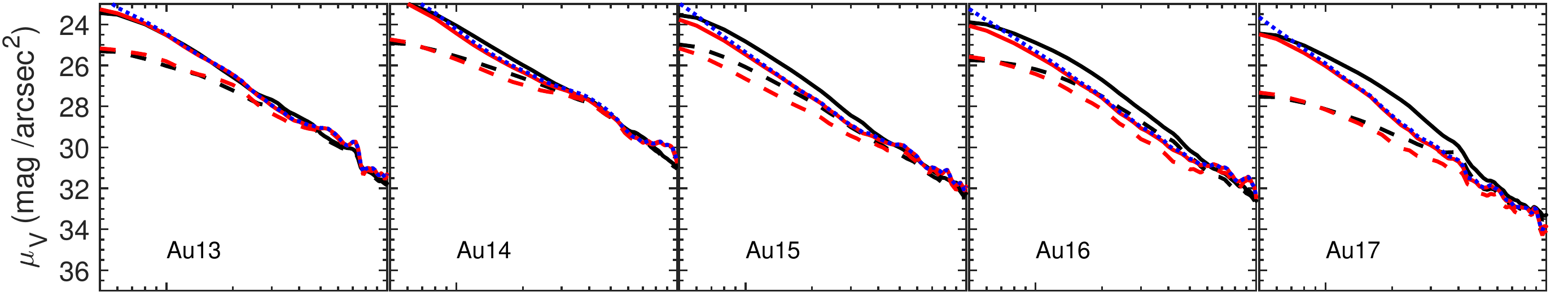}
	\includegraphics[width=2\columnwidth]{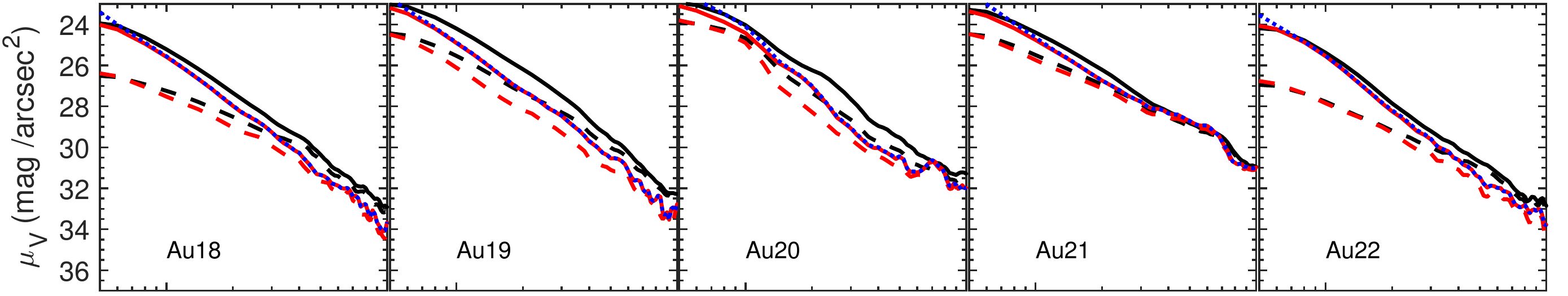}
	\includegraphics[width=2\columnwidth]{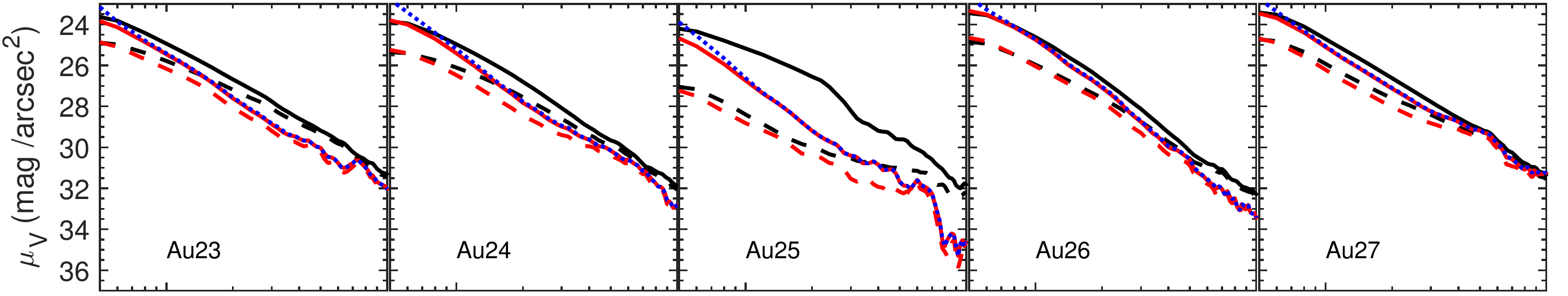}
	\includegraphics[width=1.98\columnwidth]{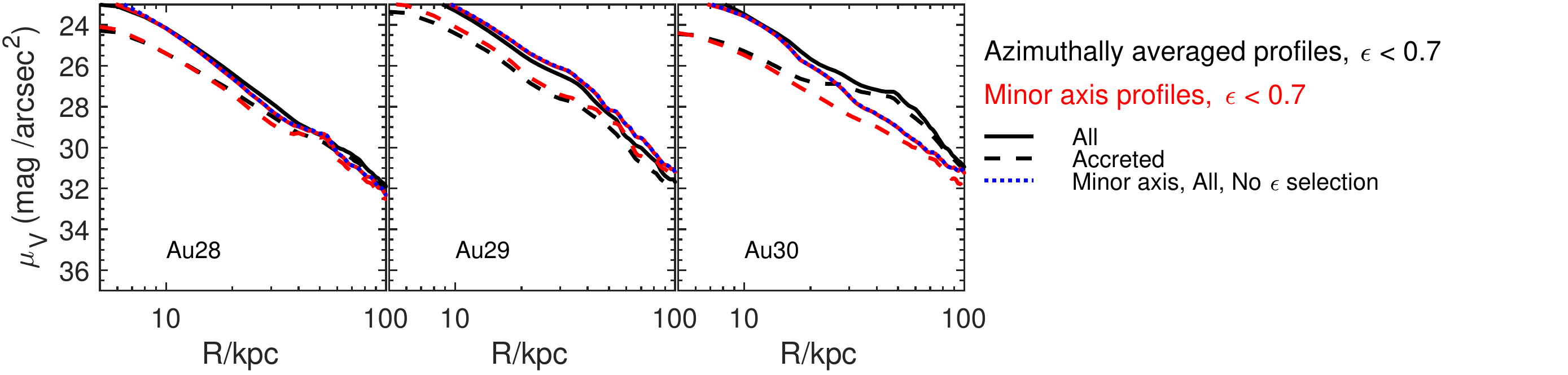}
	\caption{Surface brightness profiles, azimuthally averaged on a projected edge-on view (black) and on the minor axis (red) of each galaxy, for halo particles selected kinematically. Solid (dashed) lines are the overall (only accreted) stellar halo profiles. The blue dotted lines are the overall minor axis profiles for the spatially selected halo, i.e. without circularity constraint. Most total surface brightness profiles can be approximately fitted with single power law functions, with slopes between $-2.5$ and $-4$. The majority of these profiles do not present strong breaks; only in 20\% of cases these profiles require fitting with a broken power law. Note that the total minor axis profiles from a kinematic and spatial selection are indistinguishable.}
 \label{fig:densprof_spher}
\end{figure*}

\subsection{Surface brightness profiles}
\label{sec:density}

We show in Figure~\ref{fig:densprof_spher} the azimuthally-averaged surface brightness (SB) profiles of the kinematically defined Auriga stellar haloes. 
The profiles are computed after orienting each galaxy to an edge-on view. 

In all cases, the SB (2D) profiles can be approximately fitted by a power-law, with a slope varying between $-2.5$ and $-4$ 
(values listed in Table~\ref{t1}); these are typical values found observationally (see Section~\ref{sec:corre}).
Note that 
very similar slopes have been obtained for the stellar density profiles in other numerical studies \citep{BJ05, Deason13}. 
For completeness, we note that the spherically-averaged (3D) stellar density profiles have slope values between $-3.5$ and $-5$. 
In general, for both azimuthally and spherically averaged, we find that these profiles do not present strong breaks. Noticeable breaks can be seen in only 20\% of the Auriga galaxies (namely Au7, Au8, Au17, Au20, Au25 and Au30). These particular cases could be better fitted with a broken-power law.

We find significant variation in the slopes as well as in the normalisation values of the profiles (see Table~\ref{t1}). This translates into large variations of SB of the stellar haloes 
at both small and large galactocentric distances, with approximately three orders of magnitude range in SB at any given radius. 

The SB profiles of the kinematically defined haloes obtained using only accreted stellar particles are generally flatter, with slopes ranging from approximately
$-1.8$ to $-3$. While the variation in the slopes is similar to the one obtained with the overall stellar haloes, their normalisations
show a significantly larger scatter (see Table~\ref{t1}). Even at 10 kpc, there is a four magnitude range in SB for the accreted component, with values of $\mu_{V}$ ranging from $25$ to $29\,~ \rm{mag/arcsec^2}$. This difference is, at most, one order of magnitude for 
the overall haloes at 10 kpc, reflecting how dominant the \emph{in-situ} halo population is in the inner galactic regions.

From Figure~\ref{fig:densprof_spher} it is thus possible to appreciate that the prominence 
 of the \emph{in-situ} stellar component, i.e. its total mass and extent, varies significantly from halo to halo. The \emph{in-situ} stellar halo
 component can dominate the light, and therefore
  mass, out to galactocentric distances ranging between 5 and 30 kpc. As already highlighted and discussed in section \ref{sec:massfra}, we note that even at 100 kpc the contribution from the \emph{in-situ} component to the halo is small, but non-negligible. Au25 presents an extreme case, in which the \emph{in-situ} contribution
  dominates even at 50 kpc. As discussed in \citet{Gomez17}, this galaxy undergoes a violent interaction with a
$10^{11.5}\rm{M_{\odot}}$ companion $0.9$ Gyr ago. This close interaction significantly perturbs the host disc, generating two strong and
radially extended tidal arms that populate the outer galactic regions with \emph{in-situ} material.

In Figure~\ref{fig:densprof_spher} we also present  projected SB profiles obtained along the minor axis (on the projected wedges). 
We show both the kinematically and spatially selected stellar haloes for all particles. As expected, very good
agreement is found between these two halo selection criteria. Note however that this is only valid for profiles obtained along the 
disc minor axis since the material we see outside the disc plane is dominated by particles with non circular orbits. 

In comparison with the azimuthally averaged profiles, the \emph{in-situ} component extends to smaller distances along the minor axis.
However, we still find 
that it typically dominates out to $\lesssim 20$ kpc. As we already discussed in Section~\ref{sec:accre}, most of these \emph{in-situ} star particles formed within the spatially-defined disc and thus are not associated with star formation from gas recently stripped from satellite 
galaxies.  In some extreme cases, i.e. Au17, Au18, and Au22  the \emph{in-situ} component dominates the light (and mass) along the minor axis out to 
larger distances. As already discussed in Sections~\ref{sec:massfra} and later in~\ref{sec:satel}, this is due to early major mergers ($\gtrsim 8$ Gyr ago) and a subsequent quiet merger history in these galaxies.

Both in the azimuthally averaged and in the projected minor axis SB profiles, there are significant over- and under-dense regions at large distances. These wiggles are due to coherent substructure in the stellar haloes produced by accretion events. Note that these wiggles are less evident in the azimuthally averaged profiles than along a given line of sight.

\subsection{Metallicity profiles}
\label{sec:metal}

\begin{figure*}
\centering
	\includegraphics[width=2\columnwidth]{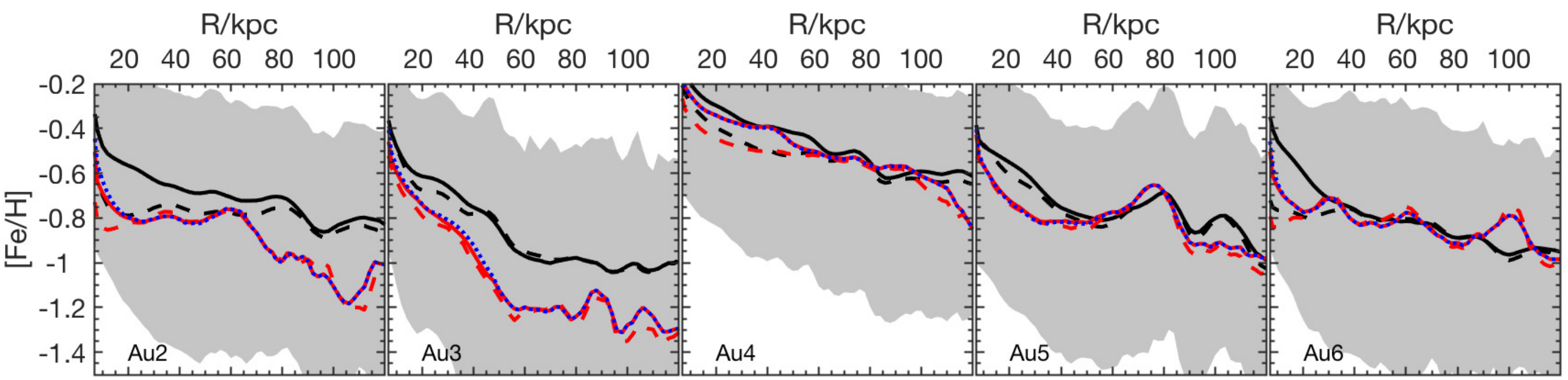}
	\includegraphics[width=2\columnwidth]{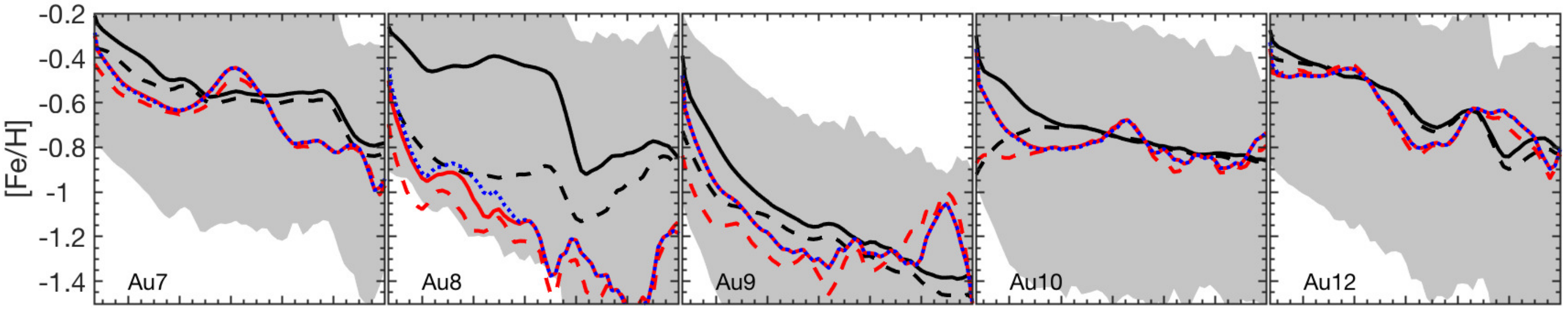}
	\includegraphics[width=2\columnwidth]{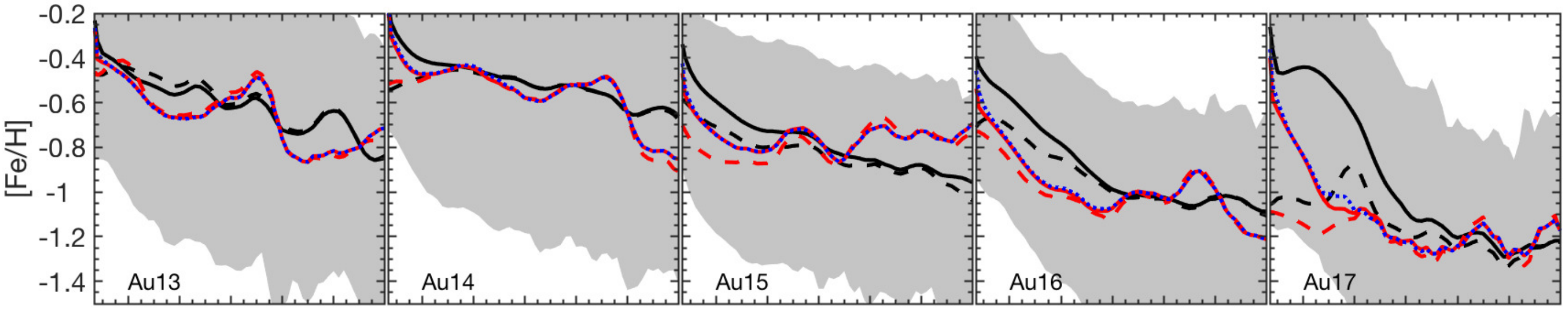}
	\includegraphics[width=2\columnwidth]{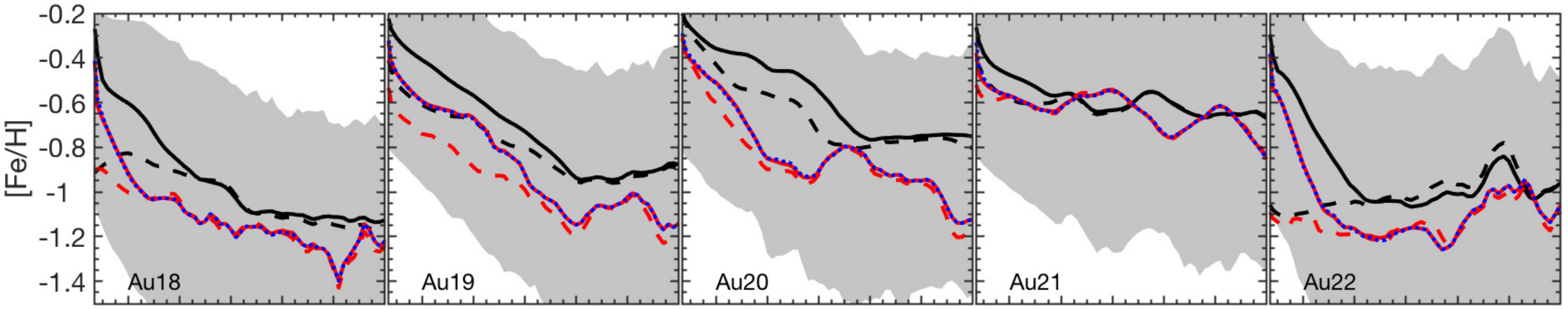}
	\includegraphics[width=2\columnwidth]{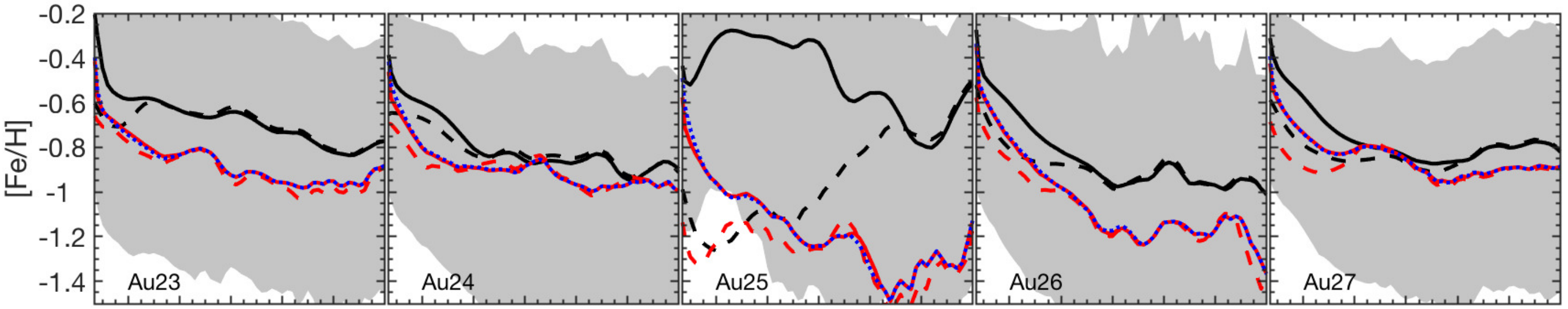}
	\includegraphics[width=2\columnwidth]{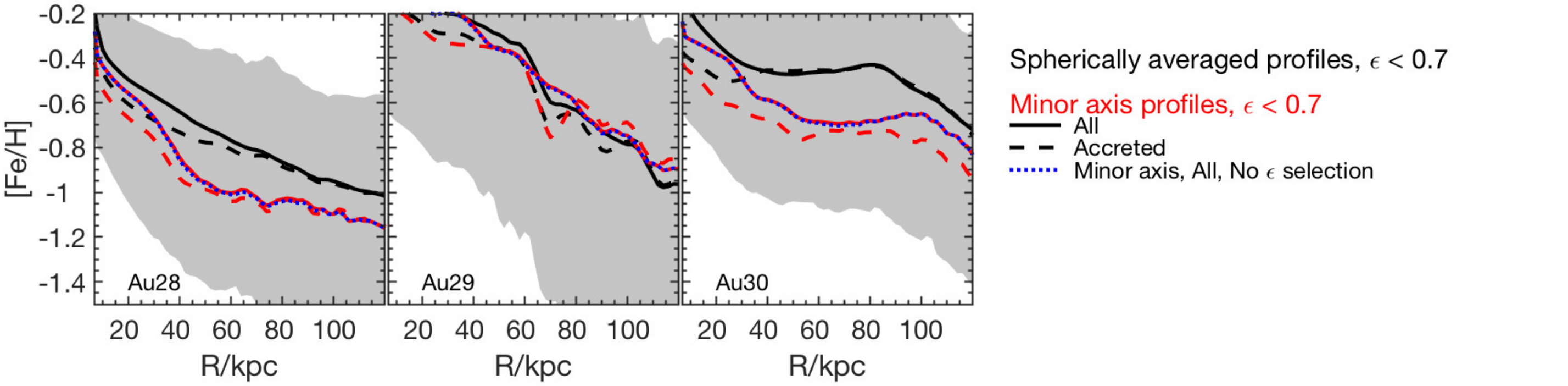}
	\caption{Median [Fe/H] spherically averaged (black) and minor axis (red) profiles for halo stars. Line conventions are as in Fig.~\ref{fig:densprof_spher}. Shaded areas indicate the 10 and 90 percentiles of the [Fe/H] values at each spherical radius. There is great diversity in the profiles and the median [Fe/H] at a given radius among the Auriga haloes. There are also significant differences between the spherically averaged and minor axis profiles, due to the larger contribution of \emph{in-situ} halo stars on the disc plane when computing the spherically averaged profiles. As in Fig.~\ref{fig:densprof_spher}, the minor axis profiles for the overall spatially defined halo overlap with those of the kinematic selection.}
 \label{fig:metprof_spher}
\end{figure*}

Figure~\ref{fig:metprof_spher} shows the median metallicity [Fe/H] profiles for the kinematically selected halo stars
computed in spherical shells around the galactic centre. 
The profiles are shown between 10 and 120 kpc from the galactic centre. These are the regions 
generally targeted by observations of external stellar haloes, given the difficulty of isolating halo stars in the very inner regions 
of a disc galaxy \citep[e.g.][]{Mouhcine05b, Barker09, Bailin11, M13, Greggio14, M16a, Gilbert12, Harmsen17}. 

As in the SB profiles, there is a great diversity in the [Fe/H] profiles of the Auriga haloes. 
The median [Fe/H]  values at a given radius vary significantly, with differences of up to 0.8 dex at e.g. $\sim 60$ kpc. Shaded areas represent the values between the 10 and 90 percentiles. We can see that there is a large distribution of [Fe/H] values at each radius.
In general, the median halo [Fe/H]  values decrease as a function of galactocentric distance, thus the spherically averaged 
profiles show negative gradients, in agreement with previous results \citep{Font11, Tissera14, M16b}.
However, when the accreted star particles are considered separately (dashed lines), the profiles generally show flatter behaviour 
and lower [Fe/H] values, especially in the inner galactic regions. 
The contribution of the more metal rich \emph{in-situ} component is 
what causes the overall profiles to rise in the central regions. We find that, as expected, the \emph{in-situ} halo component is always more metal rich than the accreted component, in some extreme cases showing differences of up to 0.6 dex in the median [Fe/H] at a given radius (e.g., Au17).
 
The projected median [Fe/H] profiles along the minor axis of each galaxy are also presented in Figure~\ref{fig:metprof_spher} as red lines. These profiles 
are more useful to compare with and to interpret observations since observed stellar halo [Fe/H] profiles
are typically obtained along galaxy minor axes \citep[e.g.][]{Sesar11, Gilbert14, Rejkuba14, M16a, Peacock15}.
Very different behaviours are found in the overall [Fe/H] profiles along the minor axis, ranging from very negative gradients (e.g. Au19)
to very flat profiles (e.g. Au15). We also find large differences in the values of the median metallicity at any given radius with values ranging from -0.4 dex (Au12) to -1.2 dex (Au22) at 40 kpc. These profiles also show significant wiggles, mostly beyond 40 kpc, which are due to the presence of substructure found along this line of sight. As we show in Section~\ref{sec:compa}, the diversity in the 
[Fe/H]  profiles along the minor axis is reminiscent of the variety of profiles seen in the observational data, and reflects the
different accretion histories of these galaxies (see Sections~\ref{sec:assembly}, \ref{sec:corre}, and \ref{sec:massmetal}).

As already highlighted in \citet{M16b}, we find significant differences between the spherically averaged and minor axis
[Fe/H] profiles. Not only are the median [Fe/H] values generally larger in the spherical profiles (up to $\sim 0.4$ dex), 
at least within the inner 50 kpc, but also some profiles show significantly different gradients. This is due to the larger contribution 
of \emph{in-situ} material along the disc galactic plane that is taken into account when computing the spherically averaged profiles. 
Contrary to the spherically averaged profiles, the accreted [Fe/H] profiles on the minor axis follow closely the profiles obtained from the overall halo for galactocentric distances $\gtrsim 20$ kpc, which highlights the lesser contribution from \emph{in-situ} halo stars along the disc minor axis.

Figure~\ref{fig:metprof_spher} also shows the overall profiles (accreted + \emph{in-situ}) for the spatially defined stellar halo. \emph{The profiles obtained from the kinematically and spatially defined stellar haloes are indistinguishable along the minor axis}. Furthermore,
as shown in \citet{M16b}, the circularity threshold used to kinematically define the stellar halo does not affect the [Fe/H] profiles along the minor axis.  This indicates that the [Fe/H] profile along a galaxy's minor axis is robust against different kinematic 
halo definitions. 

\subsection{Age profiles}
\label{sec:age}

\begin{figure*}
\centering
	\includegraphics[width=2\columnwidth]{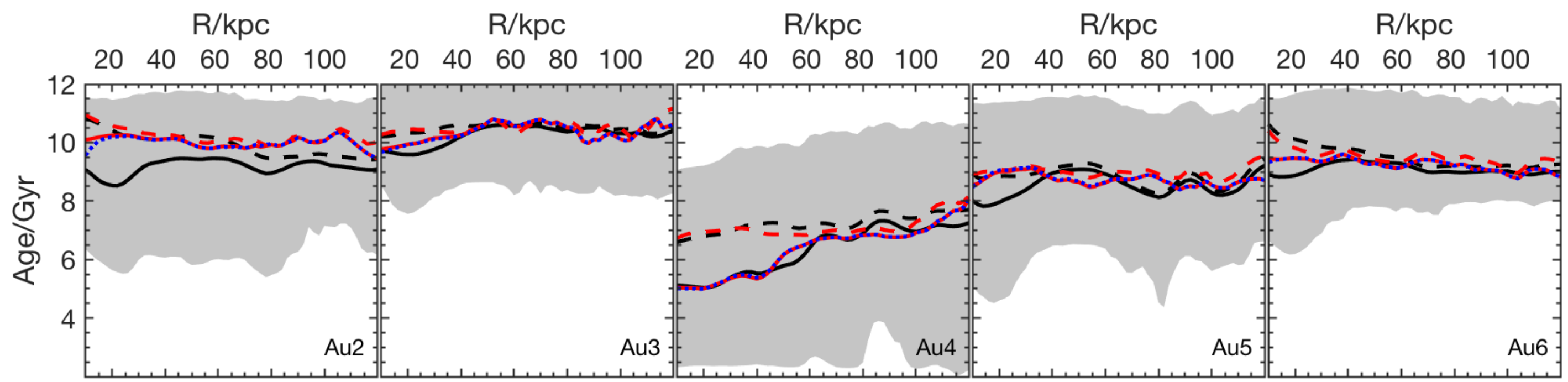}
	\includegraphics[width=2\columnwidth]{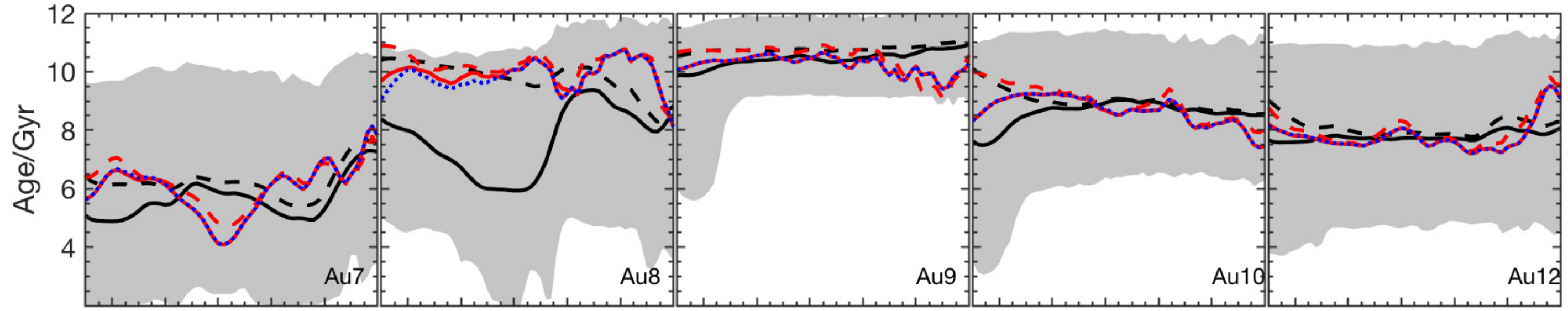}
	\includegraphics[width=2\columnwidth]{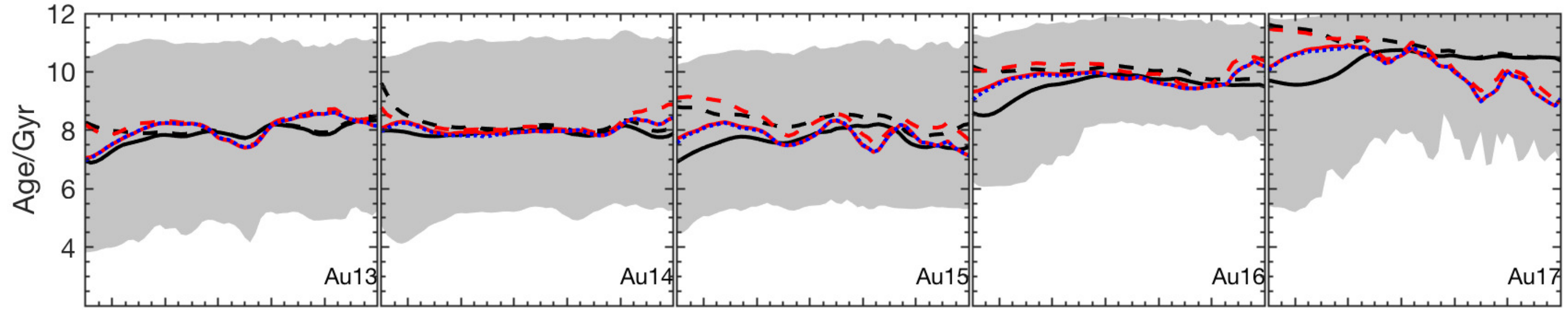}
	\includegraphics[width=2\columnwidth]{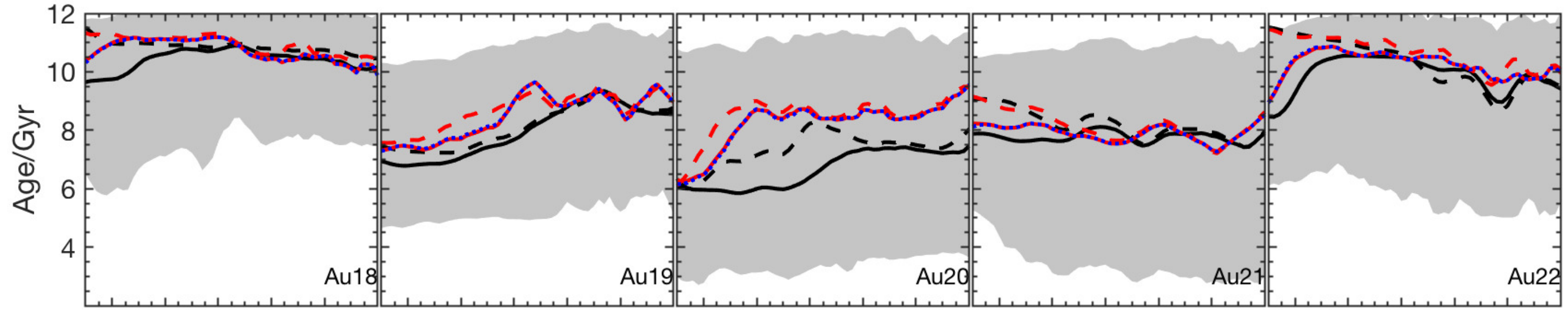}
	\includegraphics[width=2\columnwidth]{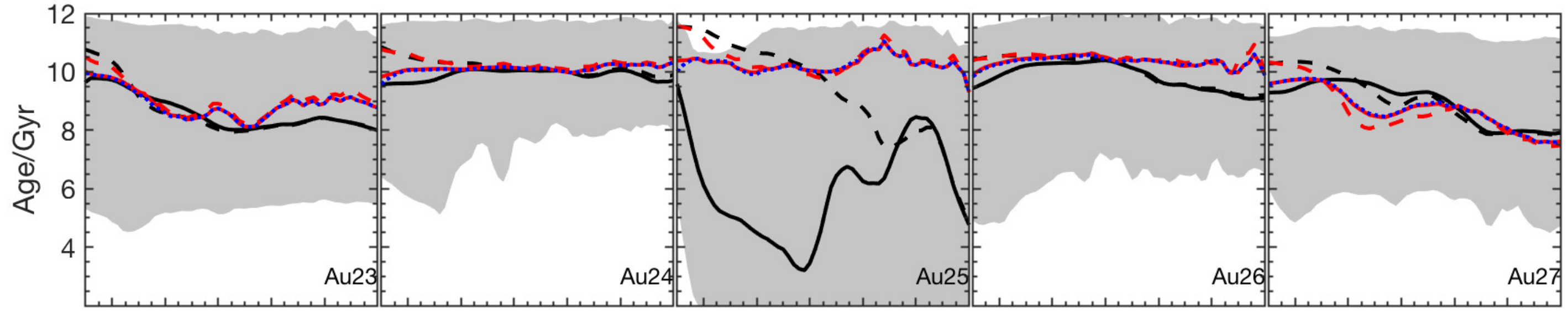}
	\includegraphics[width=2\columnwidth]{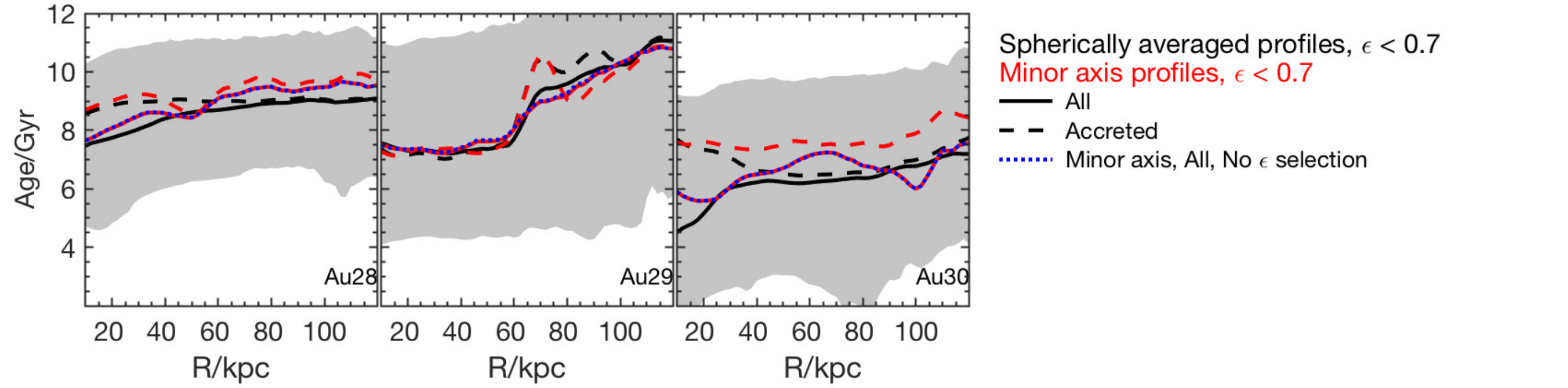}
	\caption{Median age profiles for halo particles. Line conventions and shaded areas are as in Fig.~\ref{fig:metprof_spher}. The profiles are typically flat, however, some galaxies show gradients both positive (e.g. Au19) and negative (e.g. Au25).
The overall halo age is old (mostly $>6$ Gyr) with a large scatter in the median ages among all galaxies, ranging from 6 to 10 Gyr as well as a large spread in the age distribution per galaxy at each radius. The accreted halo is typically older in the inner $\sim 40$ kpc, by $\gtrsim 2$ Gyr.}
 \label{fig:ageprof_spher}
\end{figure*}

In Figure~\ref{fig:ageprof_spher} we show the spherically
averaged and minor axis median age profiles of the kinematically selected halo for each galaxy.
The spherically averaged age profiles are typically flat but some galaxies show gradients both positive (e.g. Au19) and negative (e.g. Au23).
In general, the overall halo age is old, with ages  $>6$ Gyr. There is a large scatter in their median ages, typically ranging from 6 to 11 Gyr, with a median age of all stellar haloes at $\sim 30$ kpc of 7.7 Gyr. 
A few galaxies, however, show lower median ages, with values between 4 and 6 Gyr. These galaxies, namely Au4, Au7, Au8, Au25, and Au30 have experienced either
mergers or close-by interactions with massive satellites during the last $\sim 5$ Gyr. In the cases of Au4 and Au7, we find that these have 
merged with a satellite of $5.3 \times 10^{11}~\rm{M_{\odot}}$ and $2 \times 10^{11}~\rm{M_{\odot}}$, 2.5 and 4 Gyr ago, respectively. Such satellites not only 
bring relatively young material into the halo, but also expel a large amount of \emph{in-situ} disc stars to large galactocentric distance. Indeed, 
there is a clear distinction between the ages of a galaxy's total and accreted halo which indicates that the \emph{in-situ} halo is always younger. 
The difference in age between the two components ranges from 1 to 4 Gyr. Au25 is an extreme case, in which this difference can be as large as 8 Gyr. 
As discussed in \ref{sec:density}, this is due to \emph{in-situ} 
disc stars populating large galactocentric distances as a consequence of a very recent (< 1 Gyr ago) close interaction with a massive satellite that violently disturbed the host disc. Similarly, Au8 fully merged with a $10^{11} \rm{M_{\odot}}$ mass satellite 4 Gyr ago. This satellite was accreted with a very low infalling angle 
and thus significantly perturbed the disc along its major axis. In fact, a very strong ex-situ disc is formed as a result of this interaction (see \citealt{Gomez17}). In addition to the variation in the median ages among the different haloes, individual galaxies have a large spread in age at each radius. This is indicated by the shaded area in Figure~\ref{fig:ageprof_spher}, which represents the region between the 10 and 90 percentile age values.

The median ages of the halo population projected along the minor axis are typically older than 8 
Gyr, with a median age of all stellar haloes at $\sim 30$ kpc of  9.2 Gyr; thus older than the
spherically averaged ages at least in the inner 50 kpc. This, again, is because stellar populations along the minor axis are 
mostly composed  of accreted stars, which are older than their \emph{in-situ} counterparts. The halo age profiles along the minor axis are 
rather flat, with only few galaxies showing some weak gradient
(e.g. Au27 presents a decreasing median age with radius whereas Au4, Au19, and Au29 have an increasing median age with radius). 

Interestingly, there is a significant variation in the accreted halo ages among the Auriga galaxies, with values ranging from $\sim 6-11$ Gyr, 
and a median age of 9.4 Gyr at $\sim 30$ kpc. This is in broad agreement with \citet{Tissera12, Carollo18}, who find 
that their accreted stellar haloes have median ages $> 9$ Gyr when both the `debris' and `endo-debris' (stars formed in satellites 
after falling into the host) are taken into account. On the other hand, \citet{Font06c} used the eleven \citet{BJ05} models 
and found that all the accreted haloes have typical median ages $> 10$ Gyr. In their models, star formation activity 
in satellites is suppressed as soon as the satellites cross the galaxy's virial radius for the first time. Thus, an age difference 
may be expected with our work 
where gas rich satellites can continue forming stars while they interact and disrupt within the main host.
The stacked age distribution of $\sim 400$ low resolution models analysed in \citet{McCarthy12} also 
shows that the accreted age of the spheroidal component is older than $> 10$ Gyr, with a median of $\approx 11.1$ Gyr (no scatter is reported in their work). The age profiles of the Aquarius project were presented by \citet{Carollo18} who found a variety of age gradients for their five simulated haloes.

The wider range of accreted stellar halo ages in this work with respect to other numerical work is partly due to the wider range of accretion histories explored, thanks to our larger
sample of individual galaxies studied (at least three
times more than that in previous studies of individual simulated MW-like haloes). In particular, differences in the accretion times of the most massive contributing satellites provide the wider range of ages for the accreted stellar haloes. 

In addition to the wider range of accretion histories, and the contribution of accreted stars born inside the virial radius of the host galaxy, the younger ages of the accreted haloes may be due to the
stellar feedback model implemented in Auriga. \citet{Simpson18}, who studied the surviving dwarf galaxies in the Auriga simulations, found that subhaloes in their sample are a bit under-quenched at high masses and over-quenched at low masses, in comparison with observations. They argued that
this trend is likely caused by the stellar feedback model employed in the
Auriga simulations, which consists in the combination of a stiff equation of state in
dense gas with a phenomenological wind model
that removes mass from star forming gas and deposits momentum
in lower-density gas. 
The issues discussed in \citet{Simpson18} are likely to result in a population of over-luminous stellar halo building blocks.

\begin{figure*}
	\includegraphics[width=2\columnwidth]{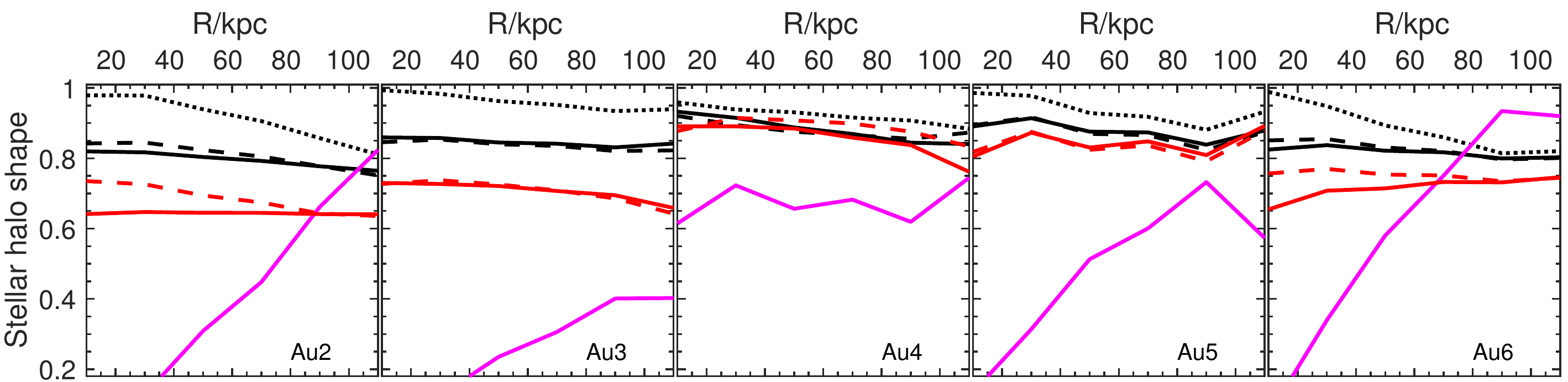}
	\includegraphics[width=2\columnwidth]{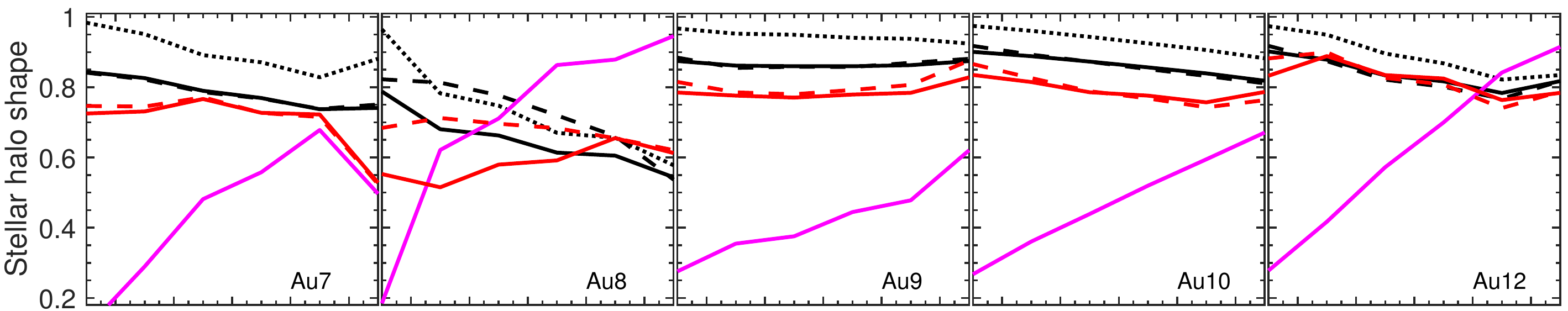}
	\includegraphics[width=2\columnwidth]{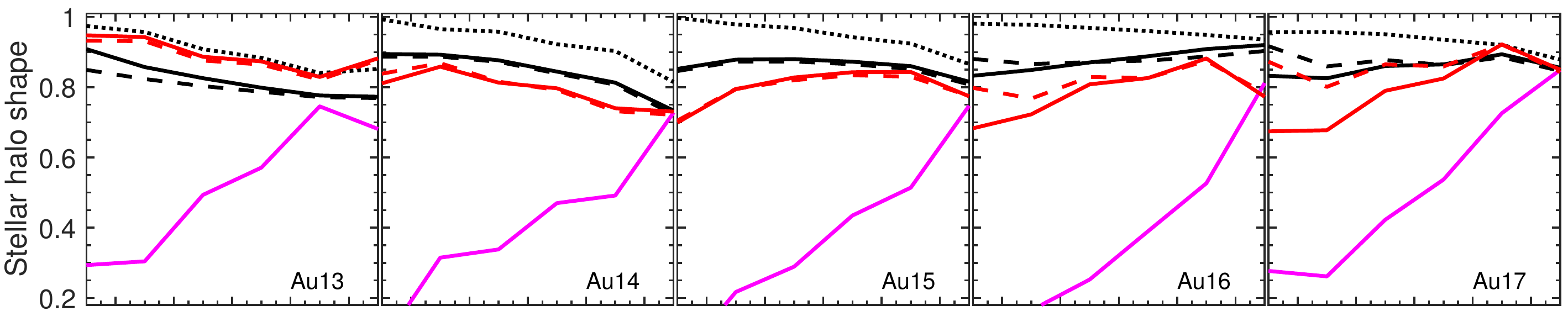}
	\includegraphics[width=2\columnwidth]{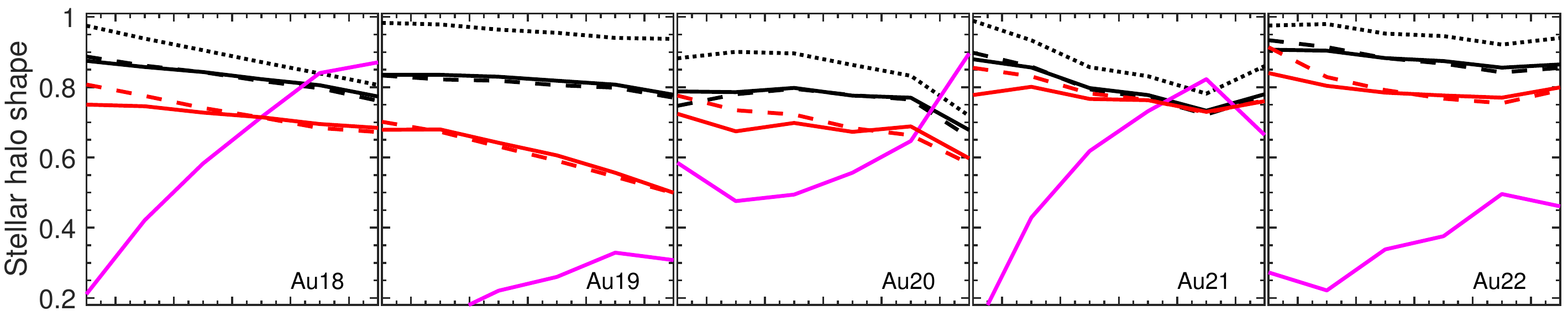}
	\includegraphics[width=2\columnwidth]{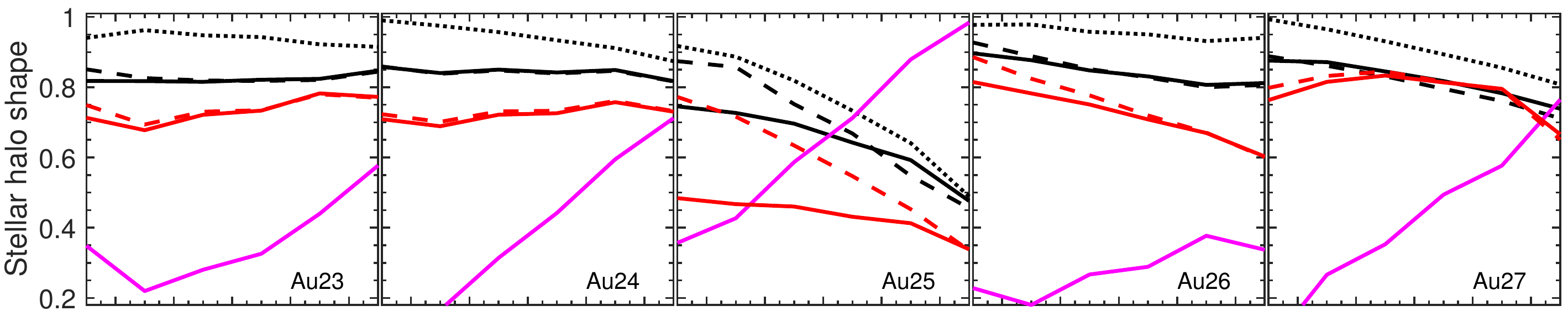}
	\includegraphics[width=2\columnwidth]{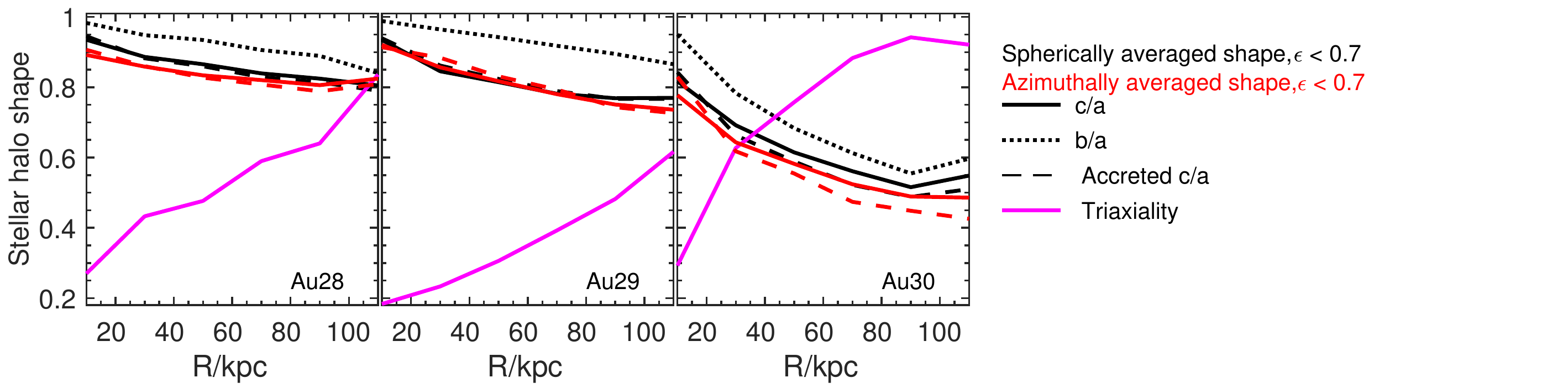}
	\caption{Stellar halo axis ratio profiles for halo particles selected kinematically, obtained from calculating the inertia tensor of the 3D (black lines) and 2D (red lines) mass distribution.
	Solid lines indicate $c/a$ axis ratios, dotted lines show $b/a$. Dashed lines are the $c/a$ profiles when only the
	accreted component of the stellar halo is considered. Magenta lines show the triaxiality parameter as a function of radius for each galaxy.}
 \label{fig:shape}
\end{figure*}

\begin{figure*}
	\includegraphics[width=2\columnwidth]{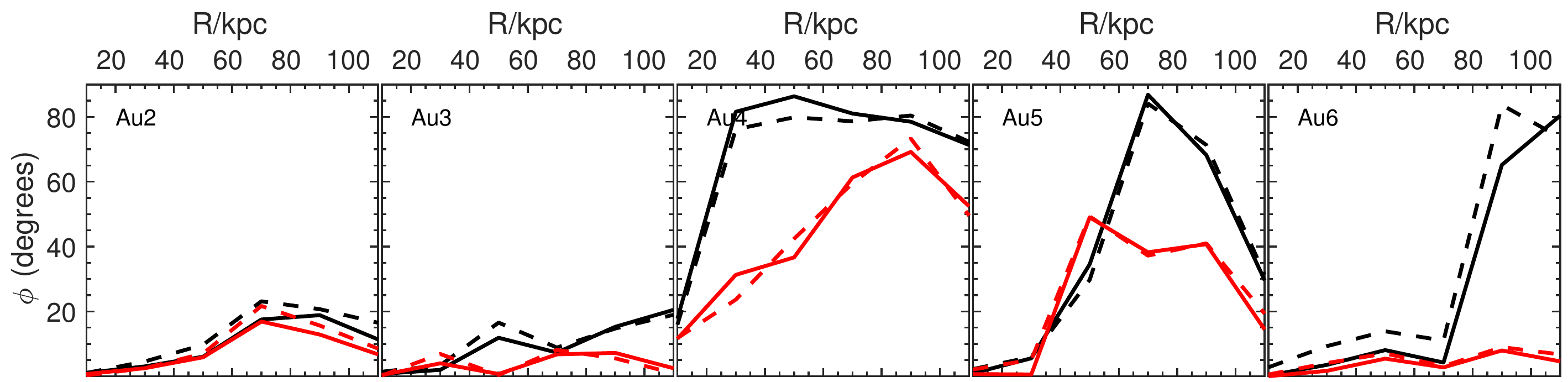}
	\includegraphics[width=2\columnwidth]{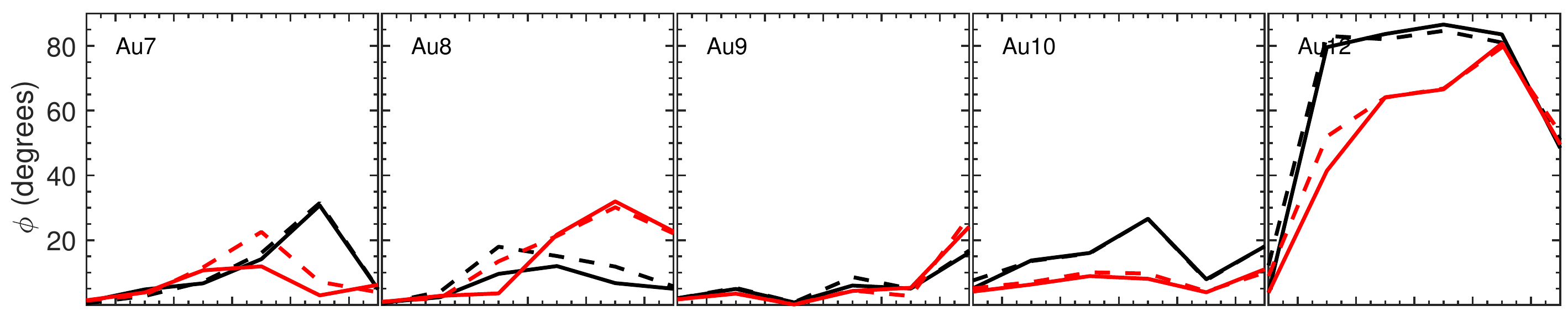}
	\includegraphics[width=2\columnwidth]{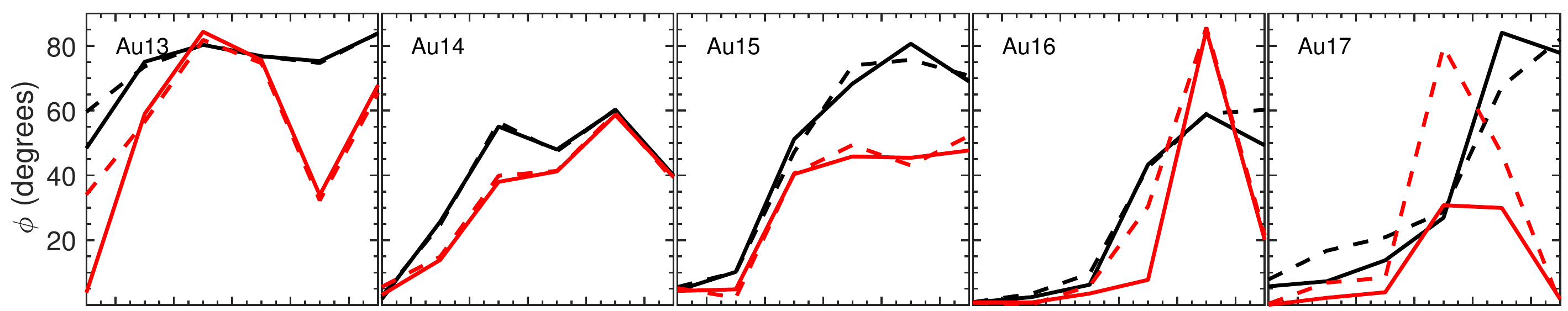}
	\includegraphics[width=2\columnwidth]{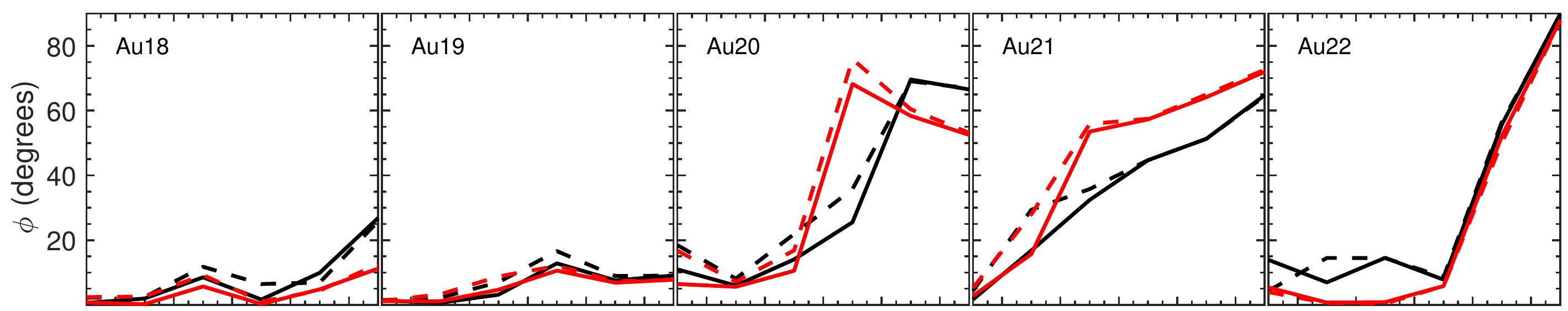}
	\includegraphics[width=2\columnwidth]{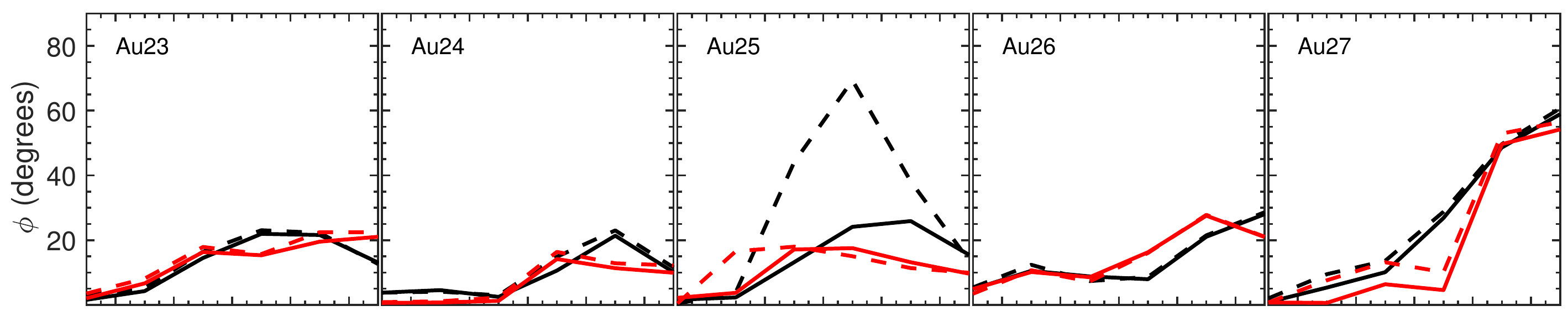}
	\includegraphics[width=2\columnwidth]{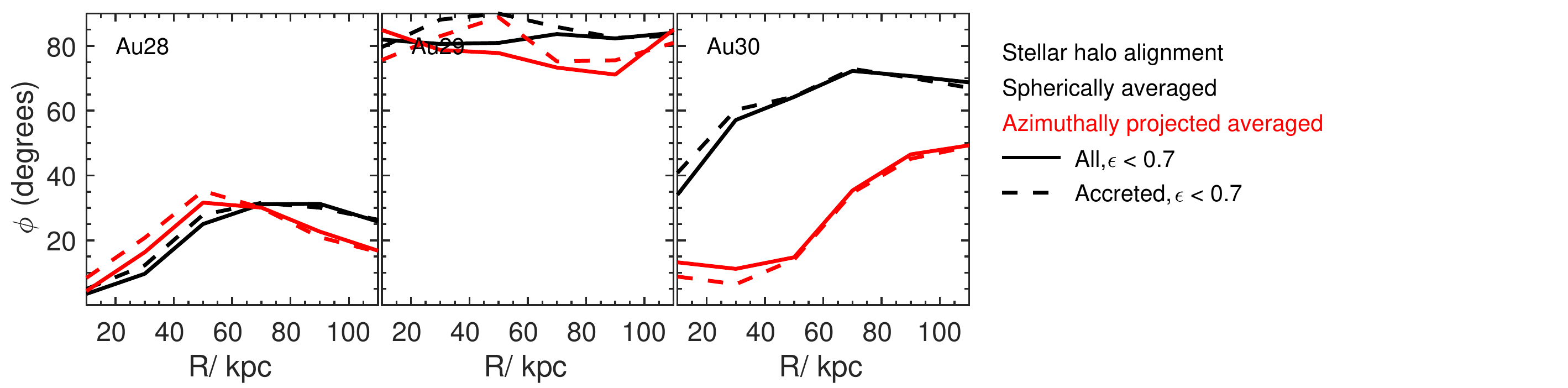}
	\caption{Spherically averaged (black) and projected (red) alignment of the stellar halo ($\Phi$) as a function of radius for all (solid) and only-accreted (dotted) halo particles selected kinematically. The alignment $\Phi$ is calculated as the angle between the minor axis of the disc and that of the halo.}
 \label{fig:rot}
\end{figure*}

\subsection{Shapes: axis ratio profiles}
\label{sec:shape}

Stellar halo shapes provide information about how haloes were assembled.
The time and way in which satellites fall into the halo,  i.e. their orbital distribution, will be imprinted in the present-day halo shape.

We measure the 3D shape of the kinematically defined haloes by calculating the
eigenvalues of the 3D mass distribution inertia tensor following the procedure in \citet{Gomez17a}. Haloes are sliced in 10 kpc width concentric spherical shells around
the galactic center. At each shell, the principal axes of the mass distribution are calculated as the square roots of the corresponding eigenvalues.
Figure~\ref{fig:shape} shows the minor-to-major $c/a$ (solid lines) and intermediate-to-major $b/a$ (dotted lines) axis 
ratios as a function of galactocentric distances for each galaxy. 
We use the triaxiality parameter $T = (1 - b^{2} / a^{2}) / (1 - c^{2} / a^{2})$ to determine the stellar haloes shapes, where $T = 0 ~(1)$ for a 
perfectly oblate (prolate) spheroid. This parameter is shown in each panel of Figure~\ref{fig:shape} as a magenta line.

We find that most Auriga haloes have oblate shapes ($T < 0.5$) within $\sim 50$ kpc; only in four cases the haloes 
are prolate in the inner regions (Au4, Au8, Au20, Au30). Many remain oblate at galactocentric distances larger than 50 kpc. However,
most haloes  become prolate ($T > 0.5$) at distances of $\sim 100$ kpc. Exceptions are haloes Au3, Au19, Au22, and Au26, which keep their oblate shapes out to 100 kpc. 

The degree of flattening varies from halo to halo with typical $c/a$ values between $\sim 0.6 - 0.9$ and a median value of $\approx 0.8$.
In three particular cases (Au8, Au25, Au30) both $c/a$ and $b/a$ values
decrease gradually outwards. These haloes become more flattened at larger radii, with a difference between the inner and  outer axis ratios of up to $\sim 0.3$.
As discussed before, Au8 and Au25 have experienced interactions with massive satellites 4 and 0.9 Gyr ago, 
respectively. Their outer haloes are dominated by a flattened  and extended structure generated during the tidal interaction which contains material from both the satellite and the host disc. 
Similarly, Au30 is currently
merging with a satellite, as evidenced in Figure~\ref{fig:surfacebright} by the large stream of tidally disrupted 
material. Interestingly, all these three galaxies present noticeable breaks in their surface brightness 
profiles (see Figure~\ref{fig:densprof_spher}). 

The accreted stellar halo $c/a$ axis ratios (dashed lines in Fig.~\ref{fig:shape}) are in rather good agreement with
those of the overall stellar halo. As expected, the largest differences are found in Au8 and Au25, presenting less flattened distributions in the inner $\sim 50$ kpc, with $c/a$ differences of up to 0.2. The lack of the strong \emph{in-situ} material tidal arms in the accreted halo is the reason for this difference. The triaxiality parameter for the accreted haloes (not shown in the figure) typically follows the same behaviour as that for the overall haloes, i.e. a tendency to shift from oblateness to prolateness with galactocentric distance.

Figure~\ref{fig:shape} also shows the 2D projected stellar halo shapes.
We compute the inertia tensor eigenvalues of the 2D mass distribution, which is obtained
from an edge-on view of the galaxies. The $c/a_{\rm{projected}}$ axis ratios are shown as a function of azimuthally averaged 2D radius and vary typically between $c/a_{\rm{projected}} = 0.6 - 0.8$ with a median value of $\approx 0.7$. Only two
extreme cases (Au25, Au30) have a value of 0.4 at large galactocentric distances. 
At each radius, most of the Auriga haloes have more flattened shapes in projection than in 3D concentric shells (red vs. black lines in Fig.~\ref{fig:shape}), even beyond 50 kpc. This is the result of the marginalisation of the density 
profile along the intermediate axis, which  enhances the more flattened halo shape. 

The accreted projected axis ratios agree reasonable well with the total $c/a_{\rm{projected}}$, except again for Au8 and Au25.
Our results are consistent with \citet{McCarthy12}, who find that, on average, haloes in the GIMIC simulations within 
the inner 40 kpc are oblate in projection with a median axial ratio of $\sim 0.6$, and show significant
scatter.

Lastly, Figure~\ref{fig:rot} shows the alignment of the stellar halo with the galactic disc as a function of galactocentric distance for the 3D and the 2D projected configuration. 
At each radius, we compute  $\Phi$, the angle between the minor-axis of the disc and that of the halo. All haloes, with the exception of Au4, Au13, Au29 and Au30, which do not have very well-defined discs, are almost perfectly aligned with the disc within 20 kpc. This is expected due to the dominating \emph{in-situ} component in the inner galactic regions. Between 20 and 100 kpc, half of the stellar haloes remain aligned with the disc ($\Phi < 20^{\circ}$). The other half show a halo-disc misalignment with values of $\Phi$ values between $20^{\circ}$ and $90^{\circ}$. 
These results are unaffected when only the accreted halo is considered, except for Au25 whose
accreted halo, as opposed to its overall halo, deviates from alignment with the disc beyond 40 kpc.
The projected alignments are in good agreement with the spherically averaged alignments, apart from a few degrees of difference (typically less than $20^{\circ}$) between their $\Phi$
values in Figure~\ref{fig:rot}. 

\section{Mass assembly of the accreted stellar halo}
\label{sec:satel}

We now analyse the accreted component of the Auriga stellar haloes, i.e. stars that were born in satellite galaxies but belong 
to the host 
galaxy at redshift zero. This component provides insights into the assembly history of galaxies such as when and how many
satellites contributed to the build up of the stellar halo.
In what follows we quantify and characterise the accretion history of each model and establish connections
between their present-day properties and main contributing satellites.

\subsection{Accreted stellar mass and significant progenitors}
\label{sec:sign}

\begin{figure*}
\centering
\hspace{-0.2cm}
	\includegraphics[width=2\columnwidth]{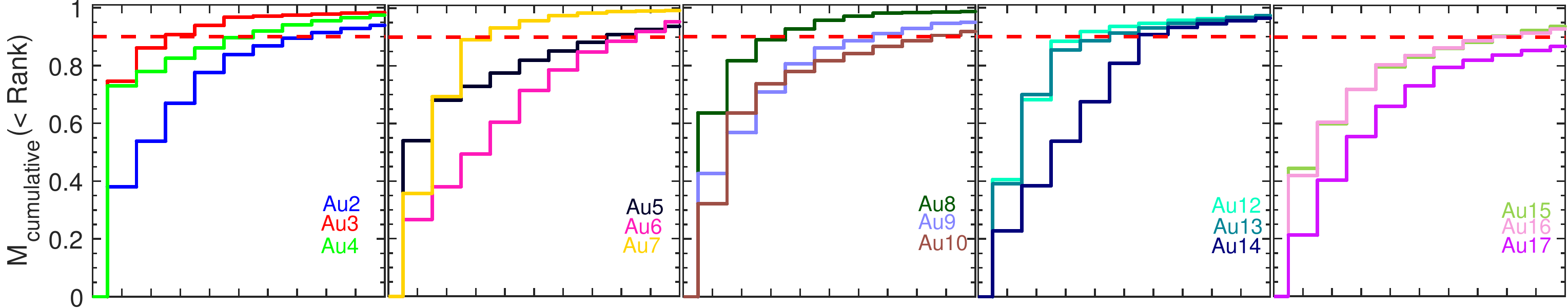}
	\includegraphics[width=2\columnwidth]{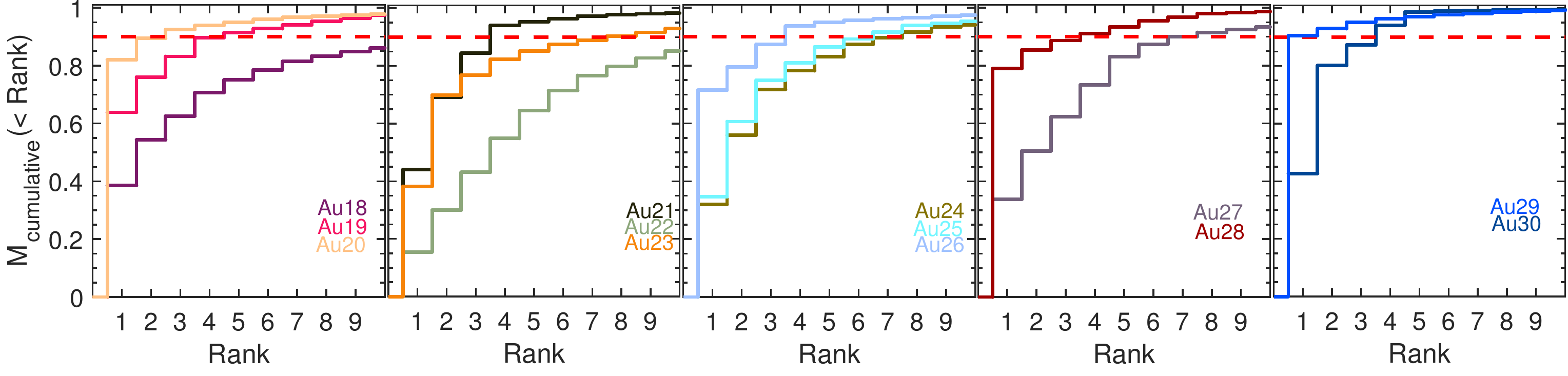}
	\caption{Mass assembly of the accreted stellar haloes. The cumulative mass fraction of the accreted stellar mass is plotted as a function of the rank up to 10 progenitor satellites. Rank equal to one indicates the most significant contributor/massive satellite of the accreted stellar halo. The red dashed line shows 90\% of the accreted stellar mass. The build up of the accreted stellar halo varies significantly from galaxy to galaxy. The number of satellites that contribute 90\% of the accreted stellar mass, i.e. number of significant progenitors, varies from 1 to 14 among the Auriga galaxies, with a median of 6.5.}
 \label{fig:massasse}
\end{figure*}

We show in Figure~\ref{fig:massasse} the cumulative mass fraction of the accreted stellar halo as a function of the number of satellite progenitors. Contributing satellites are ranked from 1 to 10, with number 1 being the most significant contributor. This figure shows that the build up of the Auriga accreted halo varies quite significantly
from galaxy to galaxy.
The number of satellites that contribute 90\% of the accreted stellar mass (hereafter significant progenitors, $N_{\rm sp}$)
varies from 1 to 14 with a median of 6.5.  
On the other hand, the median total number of contributing satellites is 84. Thus, only a
small fraction of all accreted satellites contributes significantly to the accreted halo mass budget (see also Section~\ref{sec:dis}). 
In comparison with other work, the $N_{\rm sp}$ for the Auriga haloes is smaller than that found in the \citet{BJ05} models and larger than those from
the Aquarius stellar haloes by \citet{Cooper10}. In \citet{BJ05}, the 15 largest satellites contribute approximately 80\% of the accreted mass.
In \citet{Cooper10}, on the other hand, the number of satellites that contribute 95\% (not 90\% as we calculated) of the stellar halo mass for the six Aquarius haloes
varies between 1 and 8, with a median of 6\footnote{Note that the number of satellites that contribute 95\% of the stellar halo mass 
is presented by \citet{Cooper10} in Section 4.2, their Fig. 10, and differs from the $N_{\rm{prog}}=M^2_{\rm{halo}}/\sum_i  m^2_{\rm{prog,i}}$ defined 
in their Section 4 and listed in their Table 2.}. 
We calculated this number for the Auriga stellar haloes, i.e. the satellites that contribute 95\% of the halo mass,
and find that this ranges from 5 to 27, with a median of 9. 

Now we characterise the relation between $N_{\rm sp}$ and the stellar halo properties at redshift zero, such as total mass, metallicity and density profile slopes. If correlations exist, they could provide useful diagnostics of galactic assembly 
histories.  As previously shown, the Auriga haloes show a 
great diversity in their present-day properties, despite their similarity in terms of total mass and luminosity. 

\begin{figure*}
	\includegraphics[width=2\columnwidth]{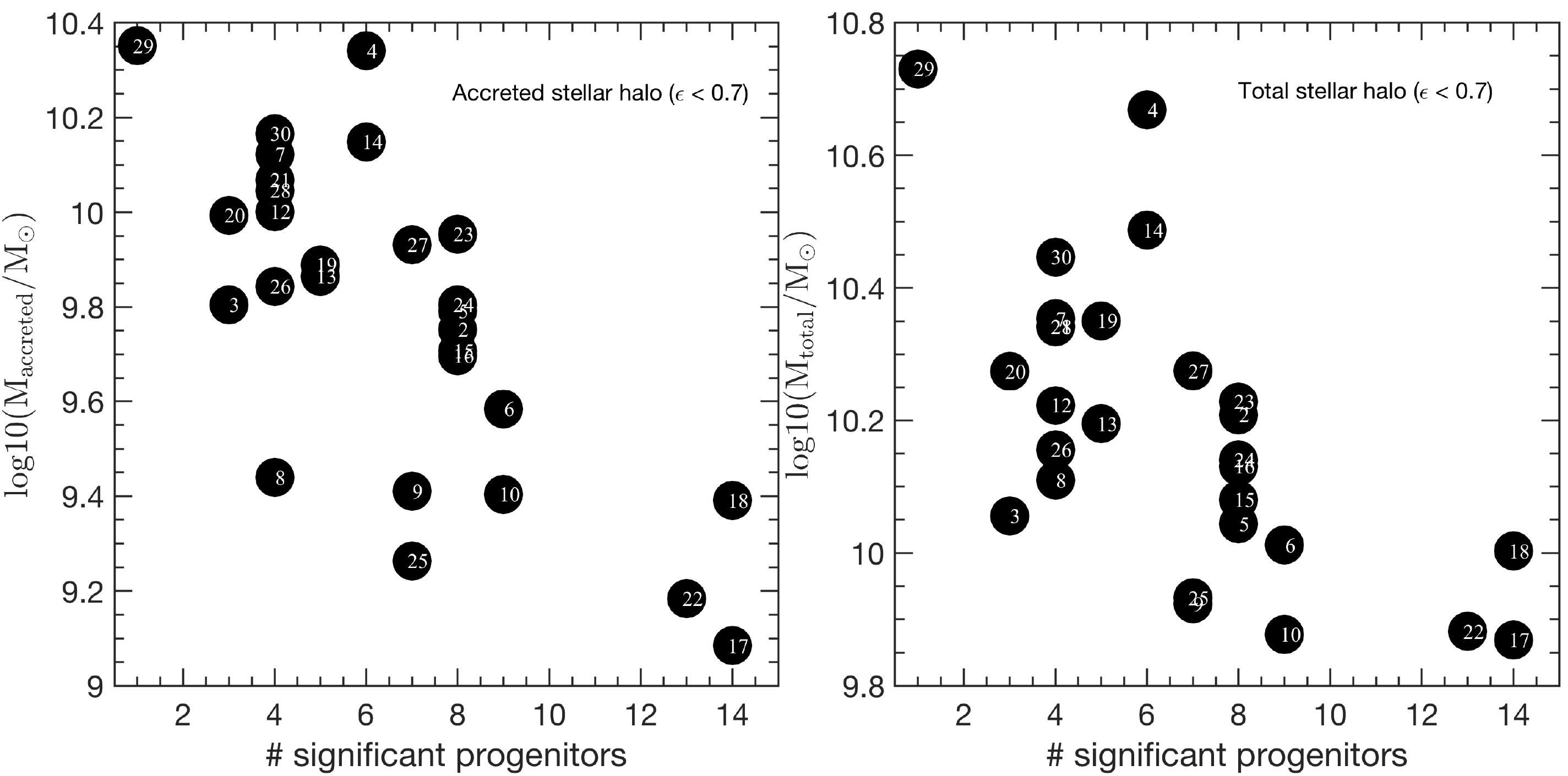}
	\caption{Left: Stellar mass of the accreted stellar haloes, selected kinematically, as a function of the number of significant progenitors, $N_{\rm sp}$. There is a clear trend, though with significant scatter, such that more massive accreted haloes are formed from debris of fewer satellite progenitors. Right: Total stellar mass of the halo as a function of $N_{\rm sp}$. Here the correlation shows a larger scatter, which likely indicates that the \emph{in-situ} component of the stellar halo does not strongly correlate with $N_{\rm sp}$. Numbers inside the circles indicate each galaxy's label.}
 \label{fig:mass_vs_prog}
\end{figure*}

 The left panel of Figure~\ref{fig:mass_vs_prog} shows the kinematically defined stellar halo accreted mass as a function of $N_{\rm sp}$. There is a clear trend: more massive stellar haloes are typically built from 
 fewer significant progenitors \footnote{We note that the trend is robust to variations of the definition of significant progenitors; in particular it still holds if we define significant progenitors as contributing 95\%, 85\% or 80\% of the accreted stellar mass.}. 
 The right panel of Figure~\ref{fig:mass_vs_prog} 
 shows the relation between $N_{\rm sp}$ and the overall mass of the kinematically defined halo, i.e. including the contribution from \emph{in-situ}
 stellar populations. We find that the correlation is weaker when the total mass is considered, indicating that the mass contribution of the \emph{in-situ} component to 
 the stellar halo does not strongly correlate with $N_{\rm sp}$ or, in other words, with the accretion history as already suggested in Section~\ref{sec:massfra}.
 
 \subsection{The connection between stellar population gradients and mass accretion}
\label{sec:assembly}
 
\begin{figure*}
	\includegraphics[width=2\columnwidth]{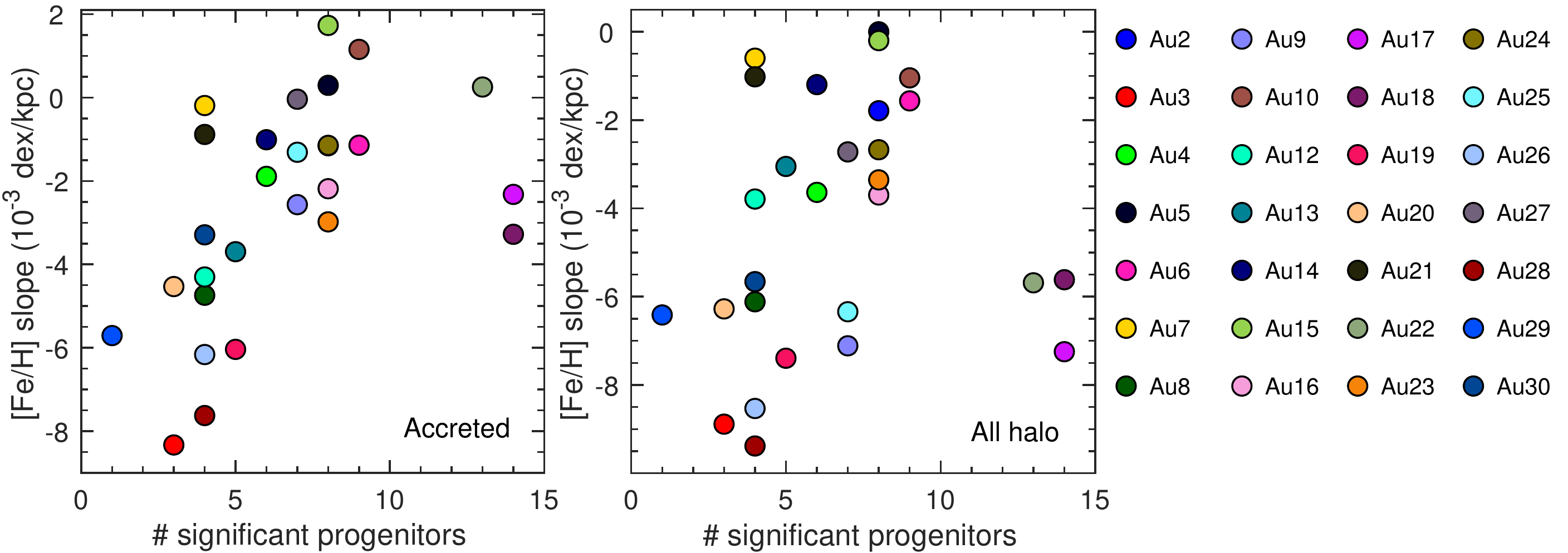}
	\caption{Minor axis [Fe/H] profile slopes for the accreted-only component (left panel) and overall (middle panel) spatially selected halo as a function of the number of significant progenitor satellites, $N_{\rm sp}$. Each color indicates a galaxy, as labelled in the right panel. We find that large negative [Fe/H] gradients are found in systems with small $N_{\rm sp}$. This correlation, although with large scatter, is strong when only the accreted component is considered, but it is also noticeable when the overall halo is considered. The three galaxies which have $N_{\rm sp}> 12$ are outliers in the middle panel. They better follow the relationship when only the accreted component is considered; however the trend is stronger if these three galaxies are not considered.}
 \label{fig:grad_vs_prog}
\end{figure*}

\begin{figure*}
		\includegraphics[width=2\columnwidth]{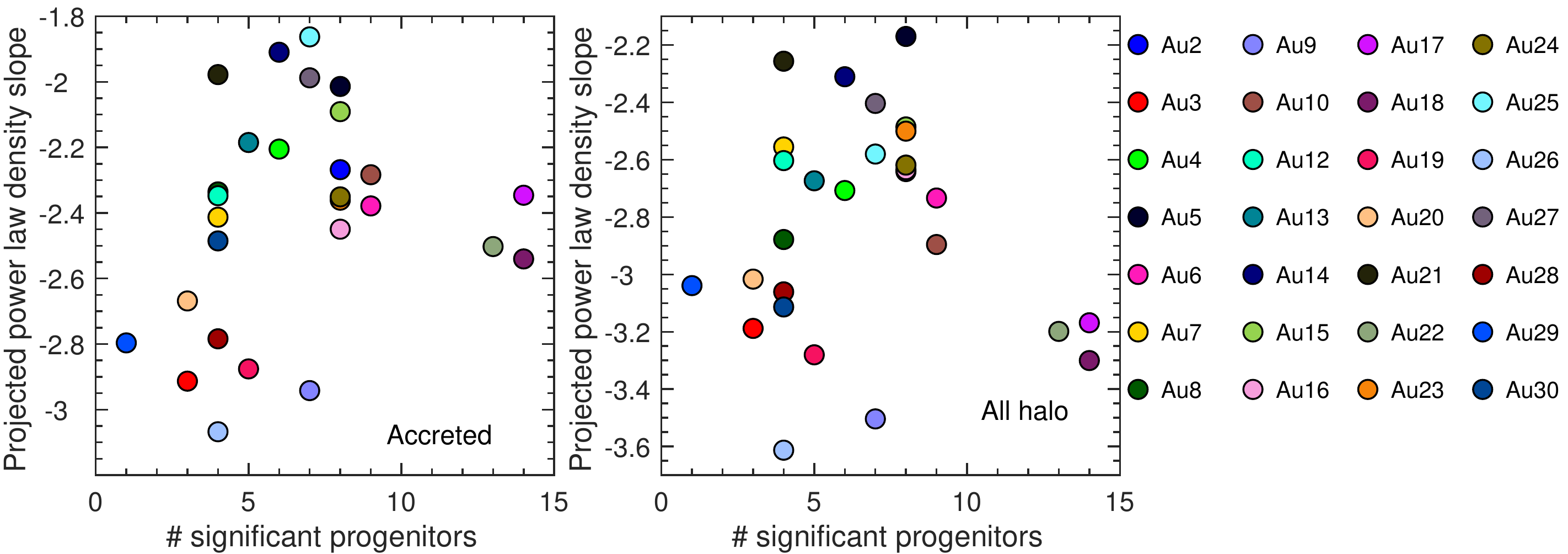}
	\caption{Slopes of the projected minor axis stellar halo density power-law profiles for the accreted-only component (left panel) and the overall (middle panel) spatially defined halo as a function of $N_{\rm sp}$.  Each color indicates a galaxy, as labelled in the right panel. Steeper density profiles are typically found in systems with smaller $N_{\rm sp}$, whereas shallower density profiles are obtained when many satellites contributed 90\% of the accreted stellar mass. This correlation, although with large scatter, is stronger when only the accreted component is considered but it is also noticeable when both \emph{in-situ} and accreted component are considered. The three galaxies which have $N_{\rm sp}> 12$ are again outliers in these correlations.}
 \label{fig:slopedens_vs_prog}
\end{figure*}

 The left and middle panels of Figure~\ref{fig:grad_vs_prog} show the relation between $N_{\rm sp}$ and the slope of the stellar halo
 metallicity profile along the minor axis for the accreted and the overall spatially defined halo, respectively.
 The slope has been calculated by applying a linear fit to the spatially selected halo minor axis [Fe/H] profile between 15 and 100 kpc. The correlation between these two quantities is strong when only the accreted material is considered. 
 Galaxies whose haloes are assembled from fewer 
 $N_{\rm sp}$ (typically less than four) have larger gradients in their minor axis halo [Fe/H] profile. A similar trend is obtained with the projected
 power law density slope, shown
 in Figure~\ref{fig:slopedens_vs_prog}, although this is less noticeable. Stellar haloes with steeper density profiles tend to be assembled from fewer significant progenitors. 
 These correlations are stronger when the properties of only the accreted stellar halo are considered (left panels in Figures~\ref{fig:mass_vs_prog}, ~\ref{fig:grad_vs_prog}, 
 and~\ref{fig:slopedens_vs_prog}) but the trends are also noticeable when the properties derived from the overall stellar halo, having both \emph{in-situ} and accreted components, 
 are considered (right panels). We note that the trends break down for galaxies with many (and low-mass) significant progenitors, i.e.  $N_{\rm sp} > 10$. Note that, as shown in Fig. ~\ref{fig:mass_vs_prog}, these galaxies have the lowest mass accreted haloes.

\begin{figure}
\centering
	\includegraphics[width=0.8\columnwidth]{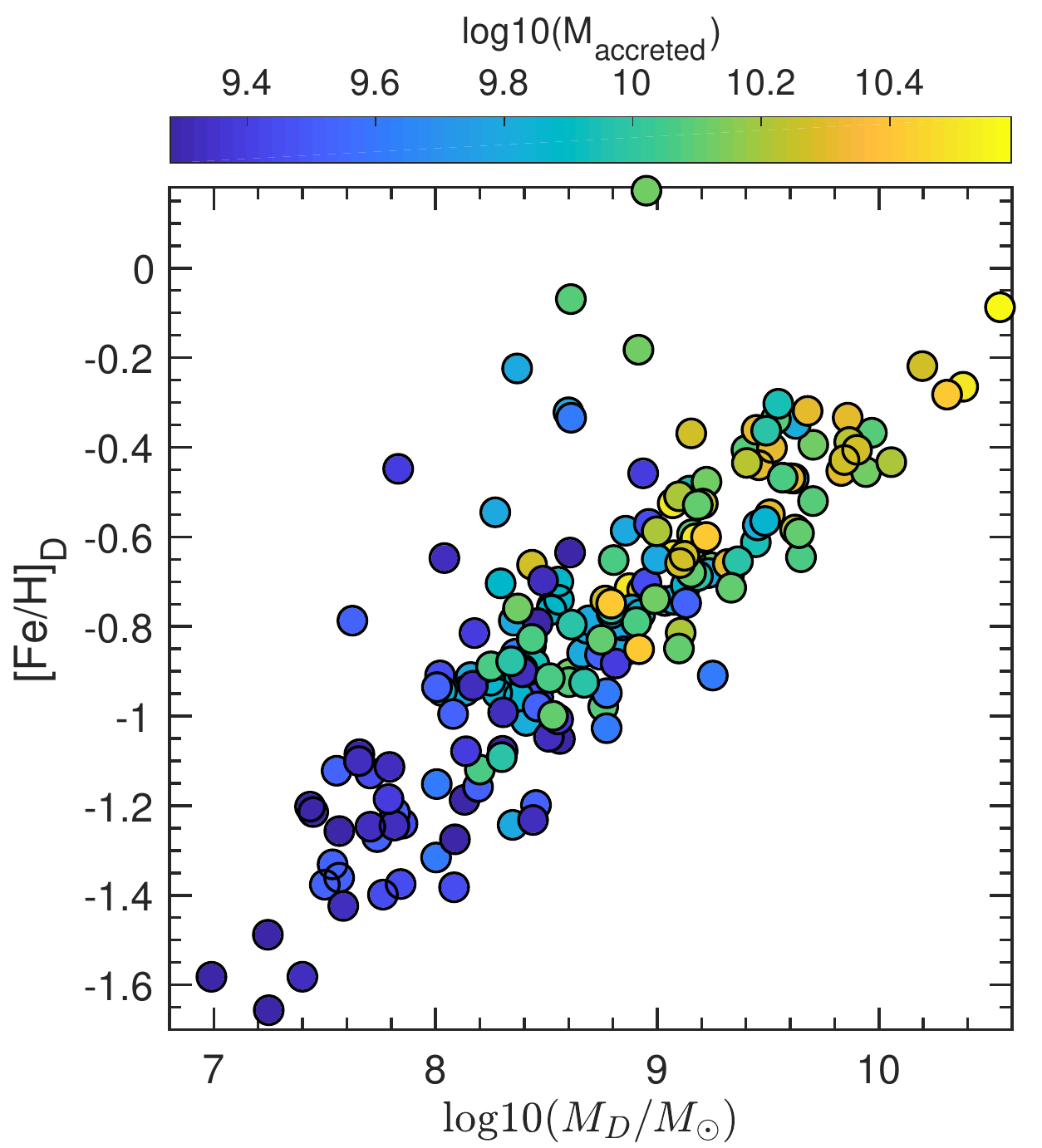}
	\includegraphics[width=0.8\columnwidth]{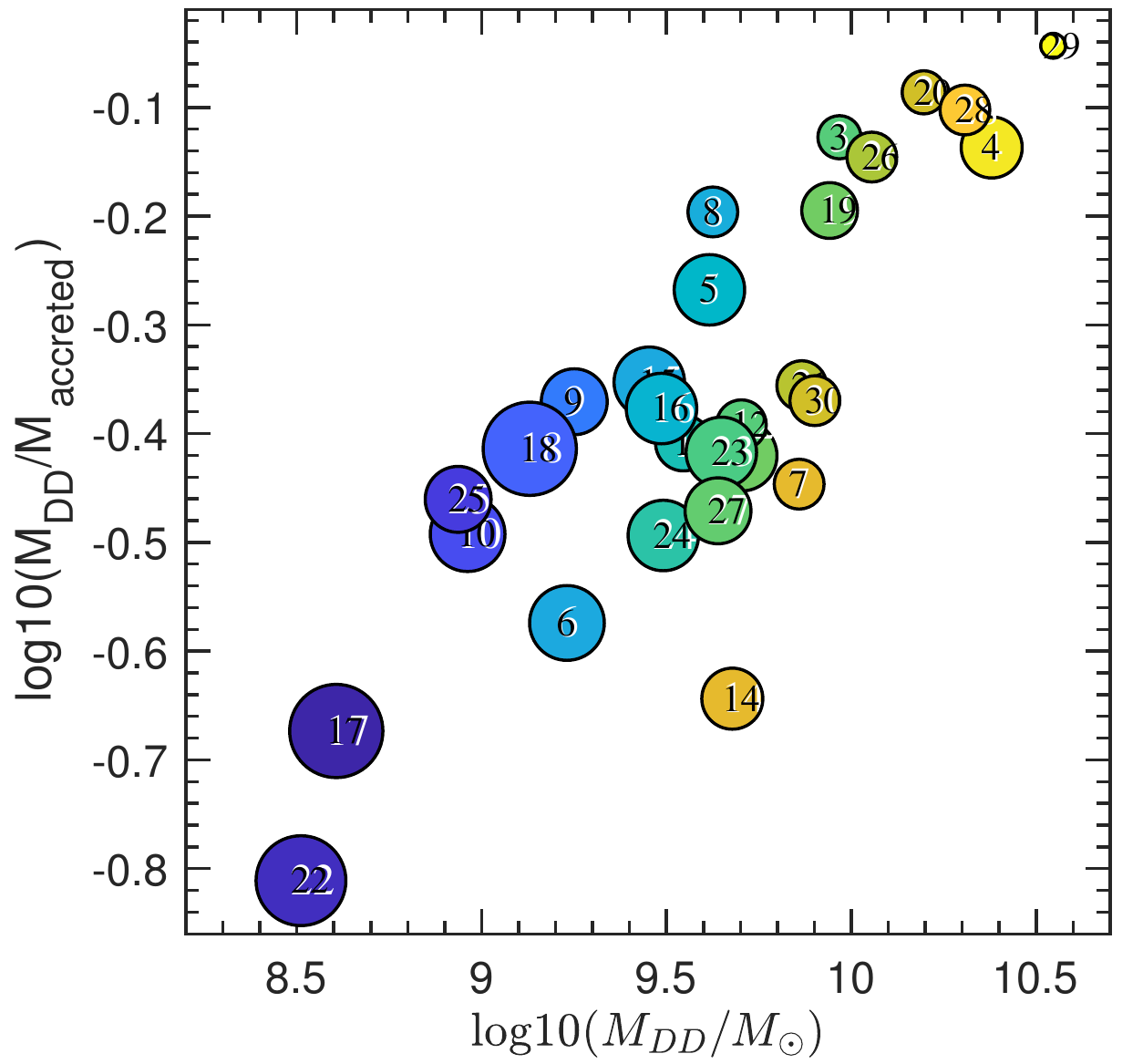}
	\caption{Top: Stellar mass- [Fe/H] relation of satellite debris from the most significant progenitors of each galaxy, which vary in number from 1 to 14. Colors indicate the accreted stellar mass of the main galaxy. The satellites follow a mass-[Fe/H] relation such that more massive satellites are more metal rich. Bottom: Fraction of stellar halo mass as a function of stellar mass for the satellite debris of the single most significant (i.e. massive) progenitor for each galaxy. Circle sizes represent the number of significant progenitors that each galaxy has, which vary from 1 to 14. Numbers inside the circles indicate halo label.}
 \label{fig:mass_debris}
\end{figure}

 The correlations of halo metallicity gradients and density slopes with mass assembly are a direct consequence of the effects of dynamical friction
 in concert with the satellites' mass-metallicity
 relation, shown in the top panel of Figure~\ref{fig:mass_debris}. Large [Fe/H] gradients and steeper slopes in their density profiles are found in haloes where most of 
 their mass comes from one significant contributor (e.g Au3, Au20, and Au29), as can be seen from Figure~\ref{fig:massasse}. 
 In those cases, the largest progenitor is significantly more metal rich than the other contributors due to
 the satellites' mass-metallicity relation. Moreover, large satellites already possess a metallicity gradient before disruption, therefore their outermost metal-poor material is stripped first and farthest from the halo centre.
 Furthermore, dynamical friction is more efficient for massive satellites, which sink to the central regions before disruption, thus preferentially depositing
 their more metal rich material in the inner galactic regions. The less massive satellites contributing to the halo are more metal poor and deposit their stars at all distances. Likewise, the SB (i.e. stellar density) in the inner regions of these haloes will reflect that of the largest progenitor which will be rather high compared with the surface brightness of the lower mass satellites contributing to the outer regions. The difference in [Fe/H] and SB between the inner and outer regions generates the measured steep [Fe/H] and SB gradients. 
 
Conversely, the mass fraction contributed by the most massive halo progenitor in galaxies with 
 several significant progenitors is always smaller than $40\%$, as can be seen from Figure~\ref{fig:massasse} and the bottom panel in Figure~\ref{fig:mass_debris}. Moreover, its contribution is similar to the contribution from the subsequent significant progenitors
 (e.g. Au6, Au17 and Au22). As already mentioned, less massive satellites are less affected by dynamical friction, and thus
 their debris is not deposited preferentially at the center, as is the case for the massive satellites. Moreover, due to the satellites' mass-metallicity
 relation, these less massive satellite progenitors will not have a significant metal rich component.  As a consequence, the radial density
 profiles of haloes with large $N_{\rm sp}$ are not as steep as when few satellites contribute significantly to the inner radii. Furthermore, these tend to have
 flatter [Fe/H] profiles, as the very metal rich population is missing and any pre-existing metallicity gradient that satellites may have is washed out due to the 
 debris distribution. 
 
\begin{figure}
\centering
	\includegraphics[width=\columnwidth]{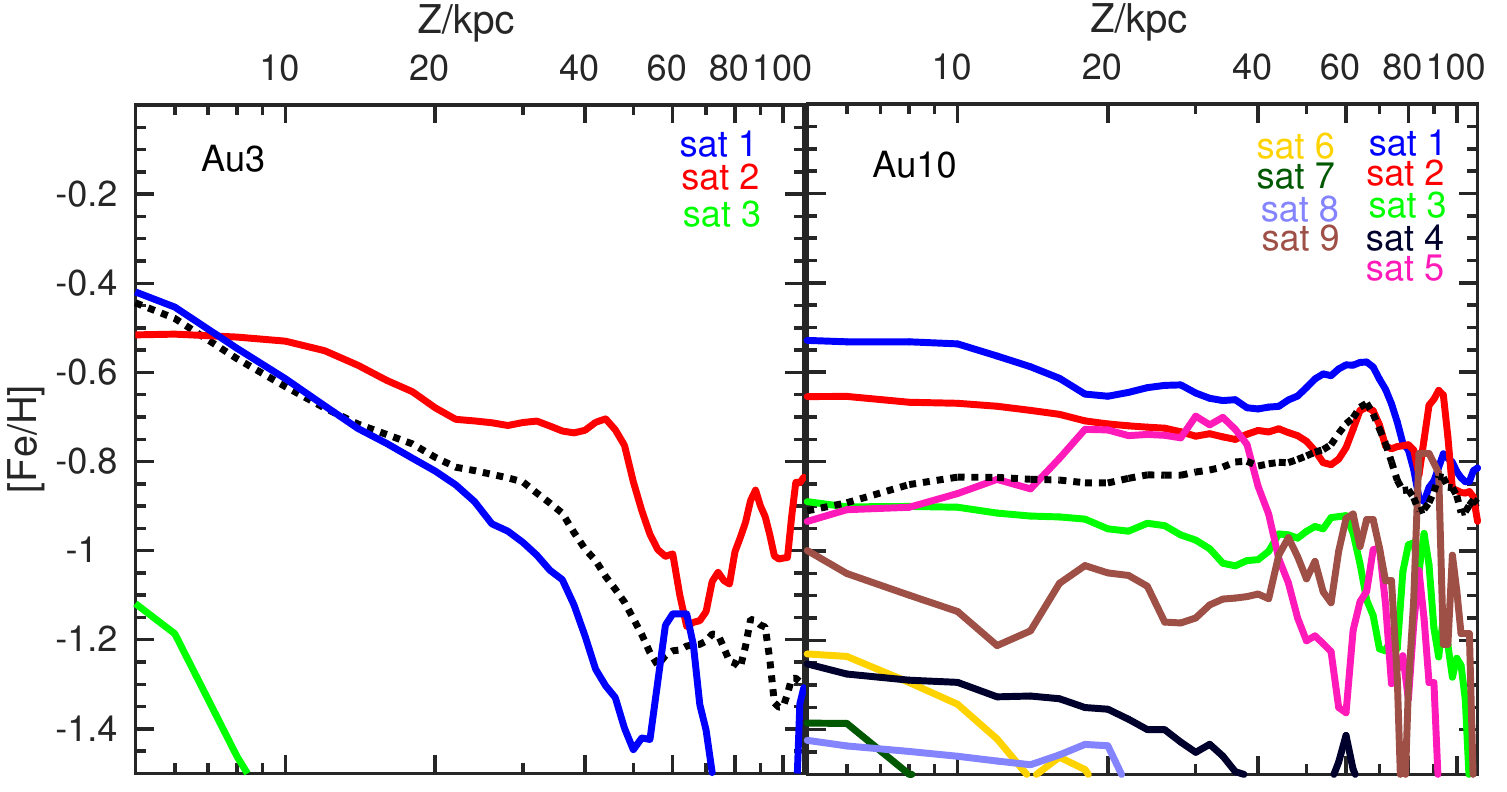}
		\includegraphics[width=\columnwidth]{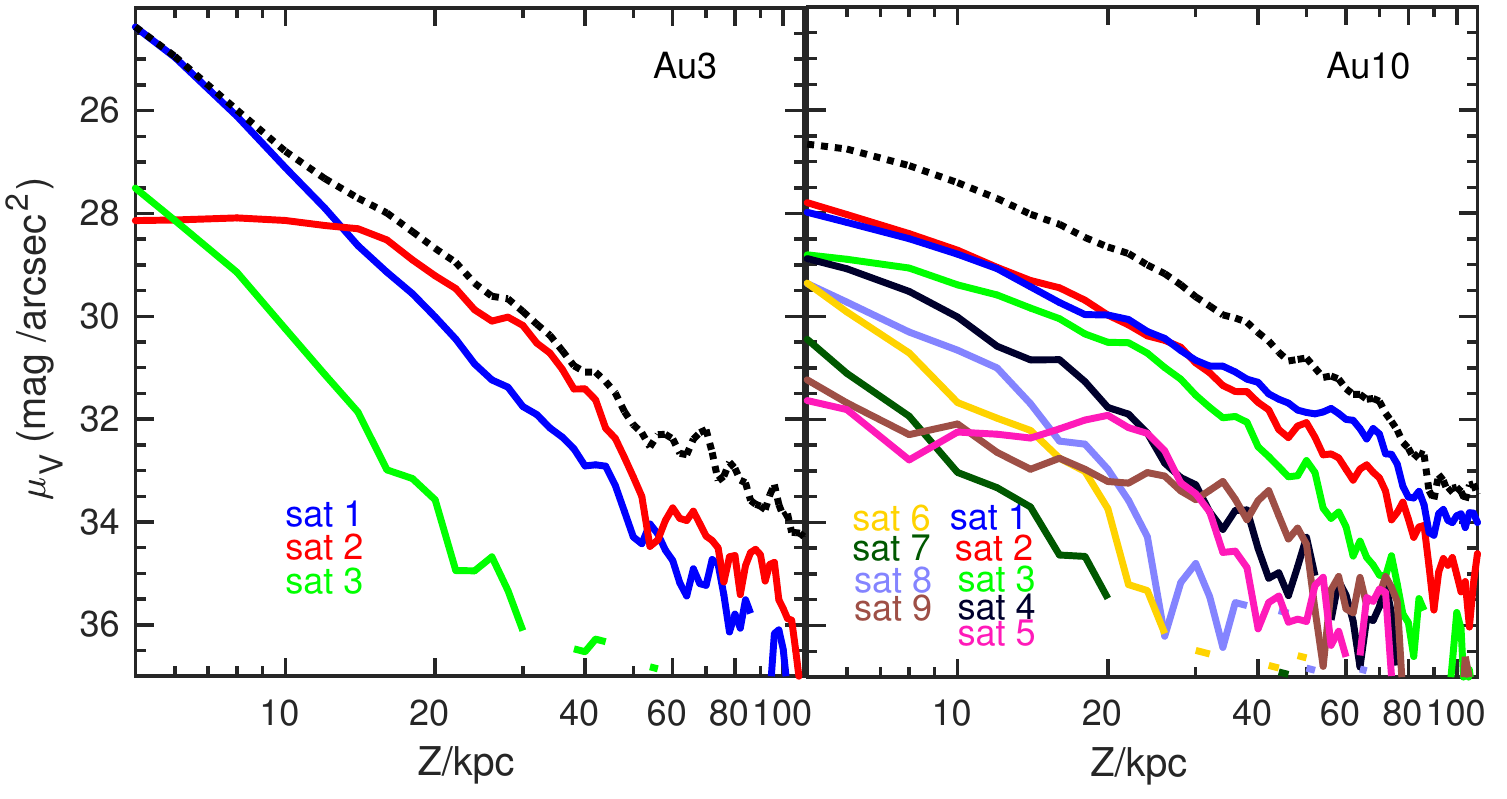}
	\caption{Contribution of the most significant progenitors to the overall acccreted minor axis [Fe/H] (top) and SB profiles (bottom), shown as dotted lines, for two representative haloes. Left: Results from Au3 which has three significant progenitors. Right: Results from Au10 which has nine significant progenitors. The profiles of each individual significant satellite progenitor are shown with different colours. For Au3, the two most massive satellites dominate the inner $\sim 60$ kpc of the [Fe/H] and SB profiles whereas for Au10, the overall [Fe/H] and SB profiles reflect the contribution from several satellite progenitors.}
 \label{fig:sat_profiles}
\end{figure}

We illustrate this in Figure~\ref{fig:sat_profiles}, where we show the surface brightness 
$\mu_V$ (bottom panels) and [Fe/H] (top panels) profiles for each of the significant progenitors
of Au3 ($N_{\rm sp}=3$, left panels)  and Au10 ($N_{\rm sp}=9$, right panels) in
different colours. The total $\mu_V$ and [Fe/H] profiles of the accreted haloes are shown for
comparison as dotted lines.
In Au3, the inner regions of both profiles are clearly dominated by the two most massive satellites.
Beyond $\sim 40$ kpc, however, the properties of the Au3 accreted halo
reflects the contributions from not only the 3 most massive satellites but also from lower mass low
metallicity contributors.  
In Au10, both the inner and outer regions of the $\mu_V$ and [Fe/H] profiles reflect the
contribution from several satellites' debris, which do not have very strong [Fe/H] gradients.

Our results are in agreement with previous numerical work. \citet{Cooper10} have suggested similar correlations
between metallicity gradient and accretion history. However, that study suffers from the small number of 
objects analysed. In addition, it is based on dark matter only simulations, thus 
neglecting important effects that baryonic components, such as the disc and the \emph{in-situ} halo, have on the present-day main properties and morphology of the stellar halo as well as on the orbits and the number of surviving satellites \citep{Zhu16, Zhu17}. 
A relation between the stellar halo mass and number of significant progenitors was also obtained by
 both \citet{Deason16} and \citet{Amorisco17a}, who have shown that the more massive stellar haloes formed from a few bigger satellites. Both of these studies
were done using idealised dissipationless simulations.

\subsection{Connection between surface brightness profile breaks and accretion history}
\label{sec:breaks}

The Auriga galaxies that show noticeable breaks in the surface brightness profiles of their haloes are Au 7, 8, 17, 20, 25, and 30 (see Figure~\ref{fig:densprof_spher}). In this section, we explore the merger histories of these galaxies in particular. We find that they all have an accretion/merger event with a large satellite sometime during the last 4 Gyrs. The most prominent and late accretion events were produced in Au 8 and Au 25 at 4 and 1 Gyr ago, 
respectively. These are the galaxies that show the strongest breaks.

This large satellite dominates the accreted component in the inner regions (within a radius of $\sim 40$ kpc depending on the galaxy size) of the stellar halo and the density profile of the halo reflects the stellar density of such a disrupted satellite. Beyond that radius, the profile reflects the underlying surface brightness of the earlier-accreted satellites. 
Thus, we find that a late big accretion event is the likely cause for the break in the profiles. This is consistent with the results presented in \citet{Deason13}, using the \citet{BJ05} models of accreted-only haloes, where they find that the break in the Milky Way stellar density profile is likely associated with a massive accretion event, although they mention this to be relatively early ($\sim 6-9$ Gyr ago) whereas we find that the accretion should be relatively late for the break to happen. \citet{Deason13} found this result by measuring the orbital properties of the simulated halo stars. Interestingly they confirmed this result with observations of halo stars using information provided by Gaia and the Sloan Digital Sky Survey \citep{Deason18}. 

In terms of number of significant progenitors, we find that there is not a clear correlation. Some of the galaxies with breaks in their profiles, e.g. Au20, have one big significant progenitor (which generates the break since it is a late massive accretion) whereas Au17 has several significant progenitors, and yet the profile has a break because one of these significant progenitors was large and accreted at late times.

One of the goals of this paper is to compare in a consistent way the numerical results with those from observations. For most observations, (particularly the GHOSTS galaxies), a single power law was fit to the density profiles for simplicity. Thus, for consistency in the comparison with the observations, we decided to fit a single power law to the density profiles of the simulations and see what can we learn from observations given a certain density profile slope.

\section{Comparison with observations}
\label{sec:compa}

In this section we compare our results with observational studies of galactic stellar haloes. In particular
we will focus on  the results presented by \citet[][hereafter M16a]{M16a} and \citet[][hereafter H17]{Harmsen17}, who measured stellar halo properties
of individual nearby MW-like galaxies observed as part of the GHOSTS survey. The morphology (spirals) and stellar masses of the GHOSTS galaxies are all within the ranges covered by the Auriga simulations. We also 
compare against studies of the stellar halo properties of our own MW 
(taken from various sources but mainly \citealt{Sesar11, Xue15, BHG16, FA17}) and M31 \citep[from][]{Gilbert12, Gilbert14, Ibata14}.
 
 \subsection{Surface brightness and color/metallicity profiles}
\label{sec:prof}
 
\begin{figure}
\centering
  \includegraphics[width=1\columnwidth]{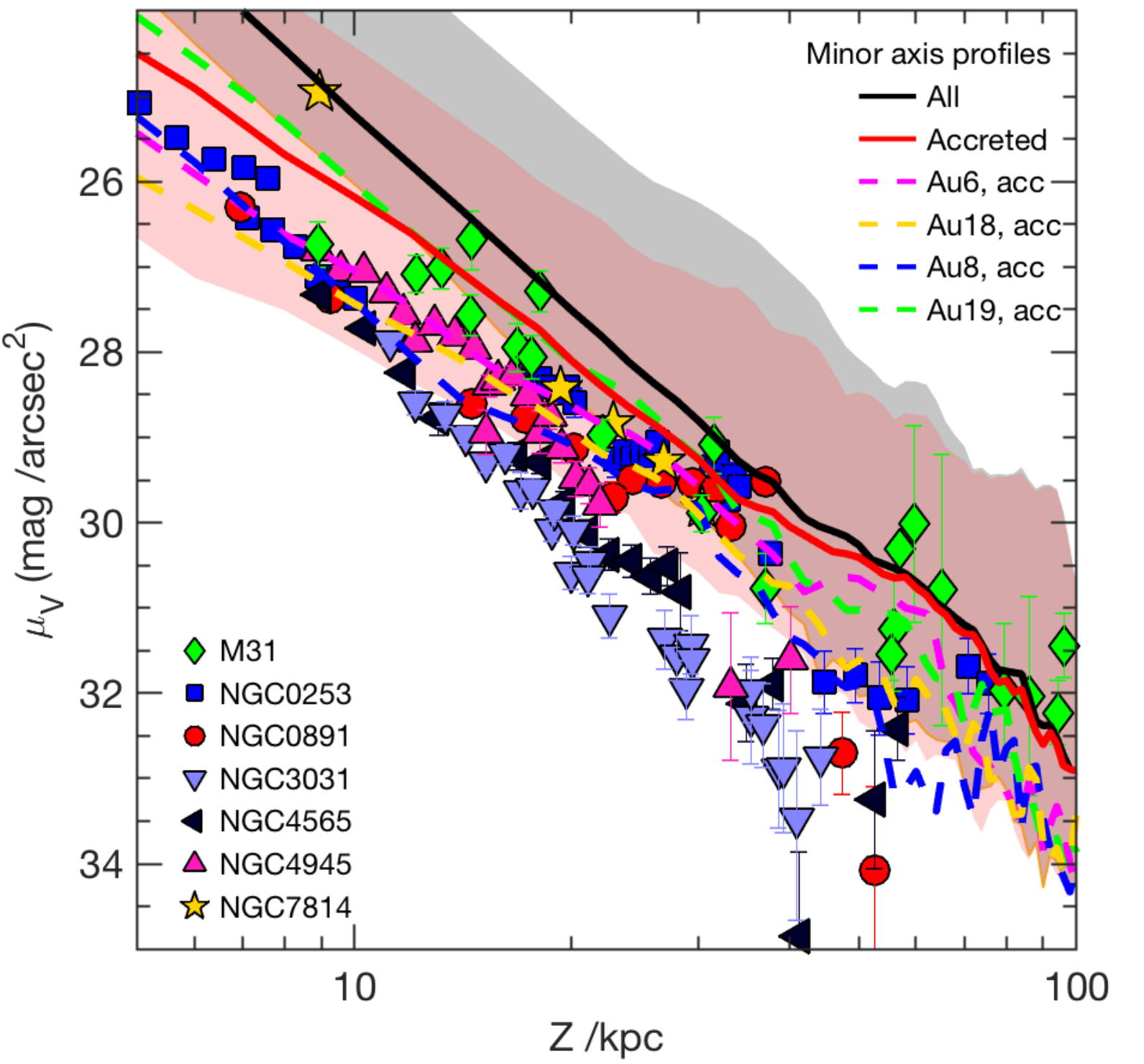}
  \caption{Minor axis projected SB profile for the spatially selected halo, i.e. without kinematic selection. Solid black (red) lines indicate the median value of all Auriga simulations for the overall (only accreted) halo. Shaded areas correspond to 5 and 95 percentiles. Coloured dashed lines are the profiles of the accreted stars of few individual Auriga galaxies. Different symbols represent the observed minor axis data for the stellar haloes of GHOSTS galaxies \citep{Harmsen17} and M31 \citep{Gilbert12}. Note that there is observational data from $\sim$5 kpc, thus we show the simulation profiles in this figure down to that radius, 5 kpc below our definition of spatially selected halo.}
  \label{fig:surf_data}
\end{figure}

 The black solid line in Figure~\ref{fig:surf_data} shows the median projected minor-axis SB profile of the Auriga spatially-defined stellar haloes, i.e. including  all stellar particles independently of their circularity parameter\footnote{We recall that halo properties along the minor axis obtained from particles kinematically and spatially selected are indistinguishable (see Figure~\ref{fig:densprof_spher}). Thus, the trend observed in Figure~\ref{fig:surf_data} is independent of halo selection.}. 
 The red solid line shows the results obtained when only 
 accreted stellar particles are considered. Shaded areas represent the 5 and 95 percentiles. Results from individual observed galaxies (H17) are shown with different coloured symbols.
 Most observational data lie below the median of the Auriga simulations, indicating that the Auriga stellar haloes are typically brighter than these observed galaxies. The Auriga profiles obtained only from the accreted component agree better with the observational data, especially in the inner $\sim 25$ kpc. 
 This suggests that either {\it i)} the Auriga galaxies have a more extended \emph{in-situ} stellar halo than observed or 
 {\it ii)} the Auriga galaxies are sitting on typically more massive dark matter haloes than those of the GHOSTS galaxies 
 (see \citealt{Pillepich14} and \citealt{Cooper13} for a relation between halo and stellar halo mass). Note that, even when only the accreted component is considered, the Auriga haloes are typically brighter than most of the GHOSTS galaxies at large galactocentric distances.
 M31, on the other hand, follows quite well the median profile of the Auriga simulations.
  
\begin{figure}
  \includegraphics[width=1\columnwidth]{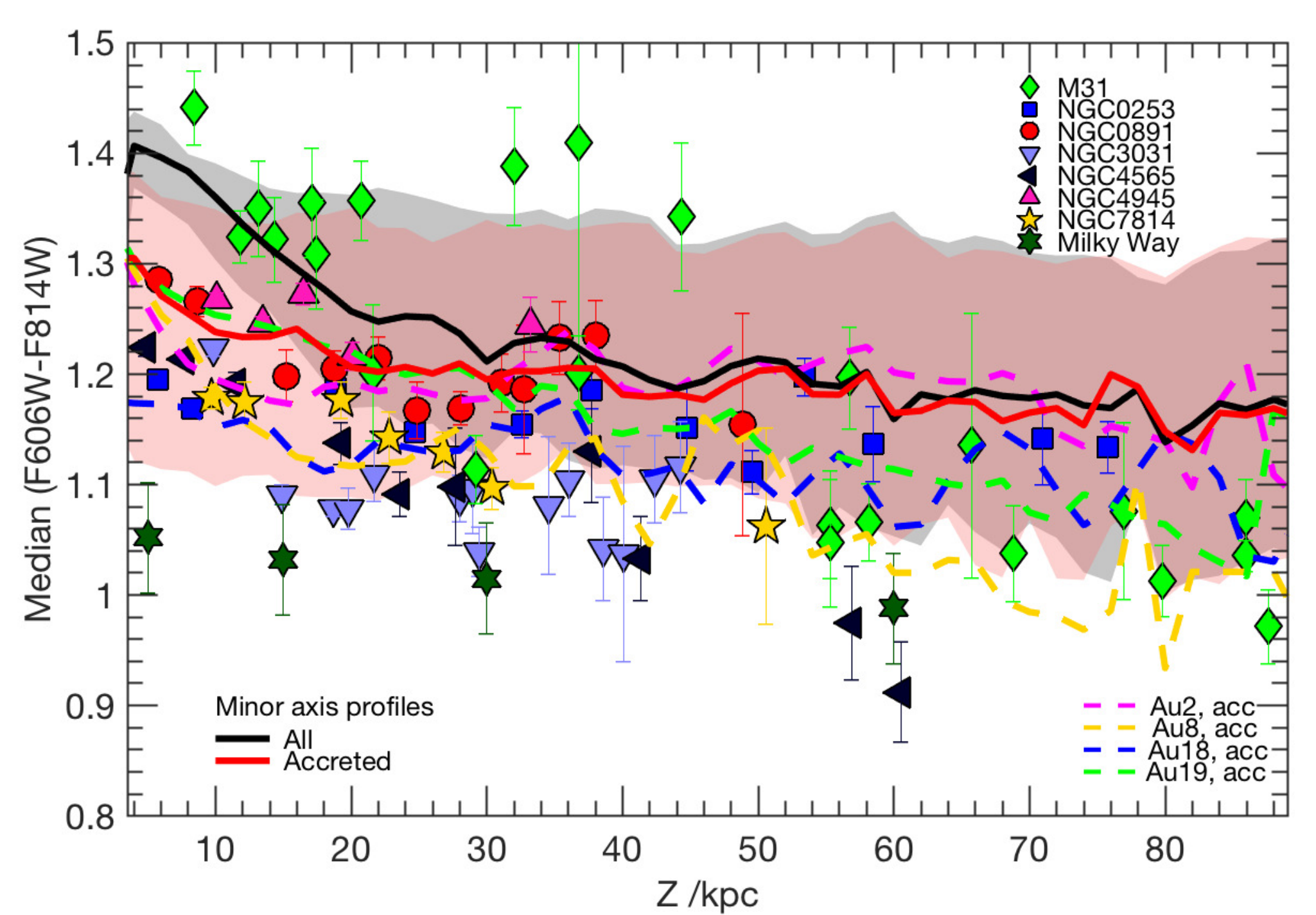}
  \caption{Minor axis projected RGB color profile for the spatially selected halo, i.e. without kinematic selection. Lines and symbols are as in Fig.~\ref{fig:surf_data}. The data for the stellar haloes of GHOSTS galaxies are from \citet{M16a}, those for the MW are from \citet{Sesar11, Xue15, FA17} and for M31 from \citet{Gilbert14}. As in Fig.~\ref{fig:surf_data}, note that there is observational data from $\sim$5 kpc, thus we show the simulation profiles in this figure down to that radius, 5 kpc below our definition of spatially selected halo.}
  \label{fig:color_data}
\end{figure}

\begin{figure}
\centering
  \includegraphics[width=0.85\columnwidth]{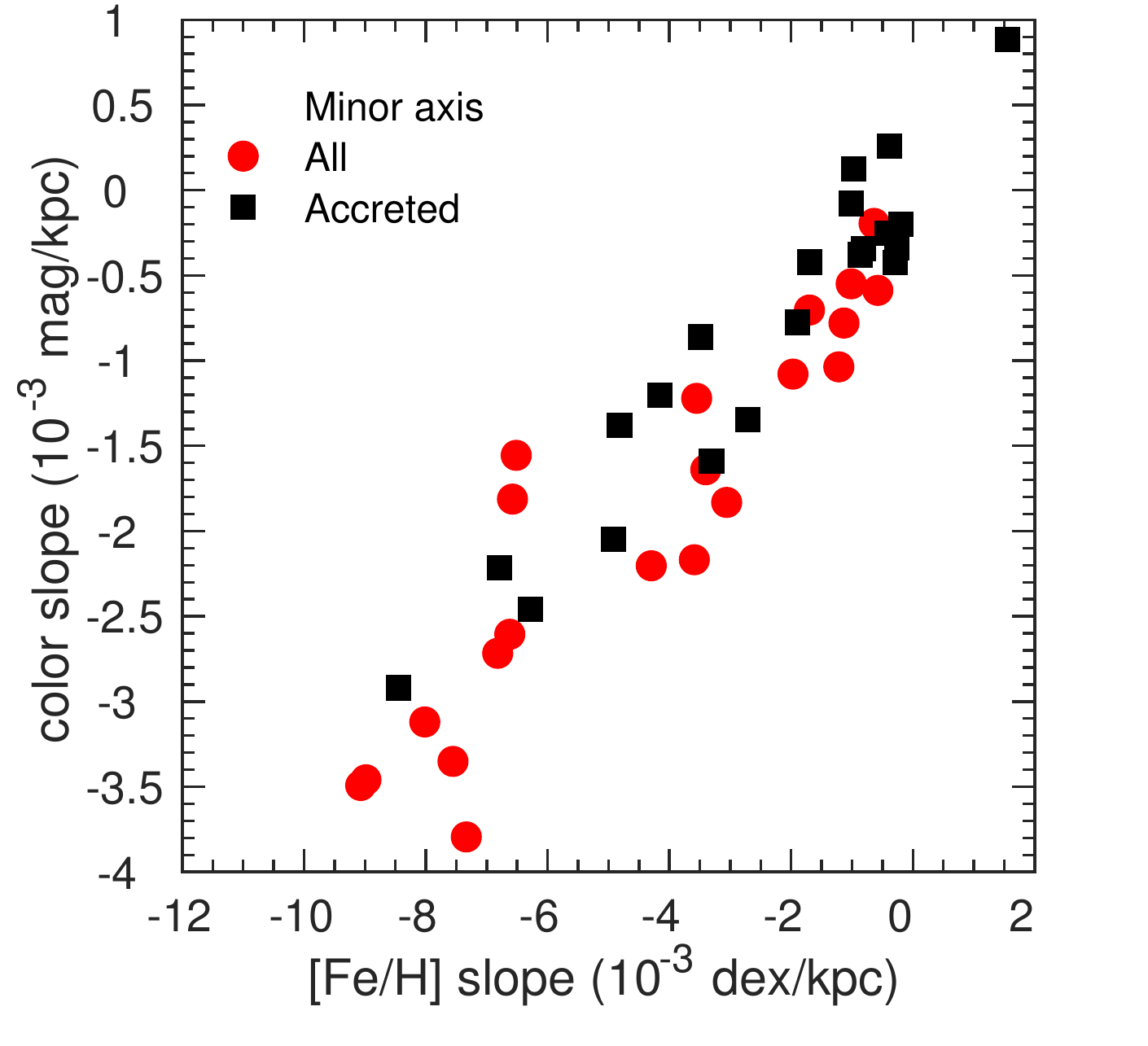}
  \caption{Minor axis projected RGB color profile slopes as a function of [Fe/H] profile slopes for the overall (only accreted) halo of each Auriga galaxy in red (black). The clear correlation demonstrates that RGB color profile gradients reflect actual [Fe/H] profile gradients. }
  \label{fig:color_fe_grad}
\end{figure}

 In Figure~\ref{fig:color_data} we compare against observations the median minor-axis projected colour profiles of the Auriga galaxies.
 The colour profiles of the GHOSTS galaxies were presented in M16a. 
 They were obtained by computing the median colour of the RGB stars that are located within the tip of the RGB (TRGB) and  $\sim$ one 
 magnitude below it. M16a also presented median metallicty values, obtained using the median color-metallicty relationship 
 found by \citet{Streich14} based on globular clusters. 
 The uncertainties in these estimated metallicities are however rather significant. 
 Thus, in this work, we will focus on the median color profiles for a fair comparison with the data. 
To compute the median color profiles of the Auriga galaxies we follow the method described in \citet{M13}. 
This involves converting stellar particles into synthetic populations of RGB stars. Briefly,  the 
 age, metallicity and mass of each stellar particle is used to generate a synthetic color
magnitude diagram using the IAC-star code \citep{Aparicio_gallart04}. We adopt the BaSTI stellar
library \citep{Pietrinferni04} and bolometric corrections by \citet{Bedin05} to transform
the theoretical tracks into the HST photometric system. A Kroupa initial mass
function \citep{Kroupa02} is assumed. We then applied the same selection culls as applied to the observational data (see M16a) and used the remaining synthetic RGB stars to calculate the median color (F606W$-$F814W)
profile along the minor axis of each simulated galaxy. For more details we refer the reader to M13. To check that RGB color profiles fairly represent metallicity profiles, we show in Figure~\ref{fig:color_fe_grad} the relation between the slopes in the RGB color and [Fe/H] profiles for the Auriga stellar haloes. The observed strong correlation demonstrates
that the color gradient of RGB stars reflects the metallicity gradient of a system, and vice-versa.

Red and black lines in Figure~\ref{fig:color_data} show the median total and accreted-only
RGB color profiles along the minor axis. The accreted-only profiles seem to agree better with the observed color profiles, which again suggests
that in general the \emph{in-situ} component of the Auriga galaxies extends farther along the minor axis and is much more prominent than in the
observations.
Interestingly, the M31's color profile matches better the color profile of the Auriga galaxies using all stars whereas none of the Auriga profiles seem to be as blue as the MW halo's median color profile. We note that the median RGB (F606W-F814W) colors of the MW and M31 are estimated from their [Fe/H] values at different radii, reported in \citet{Sesar11, Xue15, FA17} for the MW and in \citet{Gilbert14} for M31,
using the Streich et al. (2014) relationship between metallicity and
RGB (F606W-F814W) colour, and assuming [$\alpha$/Fe]$ = 0.3$.

To further compare with observations, we also show individual SB and RGB color profiles of accreted stars of few Auriga haloes in Figs~\ref{fig:surf_data} and~\ref{fig:color_data}, respectively (coloured dashed lines). Note the similarities between the observed and simulated profiles.

\subsection{Axis ratio and age profiles}
\label{sec:compashape}

H17 presented the projected axis ratios $c/a_{\rm{projected}}$ for the haloes of GHOSTS galaxies. 
They find values between $0.4-0.75$ at a projected radius of $\sim 25$ kpc, assuming alignment between the halo and the disc. The shape of M31's halo in the inner $\sim40$ kpc was estimated by \citet{Ibata05}
who found evidence for an extended disc-like structure which is aligned with the
stellar disc and has $c/a \approx 0.6$. In the outer regions, M31's halo shape
has been estimated in \citet{Gilbert12} and
\citet{Ibata14} using pencil beam kinematical data and panoramic imaging,
respectively. Both studies find that M31 has a prolate halo at large radii, having
its major axis aligned with the minor axis of M31's disc. They find a mass
flattening $q \sim 1.01-1.06$ which corresponds to a $b/a \approx 0.96$, thus almost spherical at $R\sim 90 $ kpc. 
For the MW, several studies show that the halo is more flattened in the inner $\sim 15$ kpc with $q \sim 0.6$ than farther out $\sim 30-60$ kpc, 
where $q \sim 0.8$ (see the review by \citealt{BHG16} and references therein).

As shown in Section~\ref{sec:shape} we find $c/a_{\rm{projected}}$ values between $\approx 0.4-0.8$ at $\sim 30$ kpc, in good agreement with the GHOSTS observations. In addition, most of the Auriga haloes are prolate
in the outer regions, with a median $c/a \approx 0.8$ which matches well the results for M31's halo as well as the reported flattening for MW's halo in the outer regions.
Moreover, four haloes are anti-aligned with the disc ($\Phi= 90^{\circ}$) at all radii and eight in the outer 80-100 kpc, which also agrees with the result found in M31. We caution the reader that different studies have used different methods to estimate the values of $c/a_{\rm{projected}}$. Bearing this in mind, we note that the shape of most Auriga galaxies appear in a good agreement with the observational results.

Observational constraints on stellar halo ages are, unfortunately, rather limited. 
Only the ages of the MW and M31 stellar haloes have been estimated to date. The MW's inner halo  ($r < 5$ kpc) is estimated to be $\sim 10-12$ Gyr from a sample of field 
stars from the Sloan Digital Sky Survey \citep{Jofre11}. More recently, \citet{Carollo16} estimated a mild age gradient in the MW's halo using a 
large sample of blue horizontal branch (BHB) stars. They found that their mean ages are greater than 12 Gyr out to galactocentric distances of 15 kpc and a decrease in the mean ages
with distance by $\sim 1.5$ Gyr out to 45 kpc. These results for the MW are not really comparable with what we show in Figure~\ref{fig:ageprof_spher} for two reasons. 
First, we discard the inner 5 kpc of the halo, as we assume it to be mostly bulge. Second, the BHB are already a subsample of the halo stars with ages older than 10 Gyr.
Thus, BHBs trace the old stellar population while the existence of a younger one would be not represented. The analysis of the age profiles from the Aquarius project reported recently by \citet{Carollo18} yield a variety of age gradients albeit weaker than current observations of the MW. The authors show that the accreted stars determine a steeper negative age gradients. Only in those cases when the stellar haloes are assembled by small mass satellites the age profile is more comparable to that of the MW. 
For M31's halo, \citet{Brown08} derived the SFH  of a
small region on the sky at $\sim 35$ kpc along the
disc's minor axis using deep HST observations.
They find that the mean age of M31's halo at that location is $\sim 10.5$ Gyr. This is consistent with half of the Auriga haloes, which have median ages of 10 Gyr at $\sim 35$ kpc as shown in
Figure~\ref{fig:ageprof_spher}. 
The study of resolved populations in other nearby galaxy stellar haloes rely mostly on RGB stars, which are indicative of ages older than $> 2$ Gyr but unfortunately do not 
provide a more accurate age measurement than that. 

\subsection{Correlations between stellar halo properties}
\label{sec:corre}

\begin{figure*}
	\includegraphics[width=2\columnwidth]{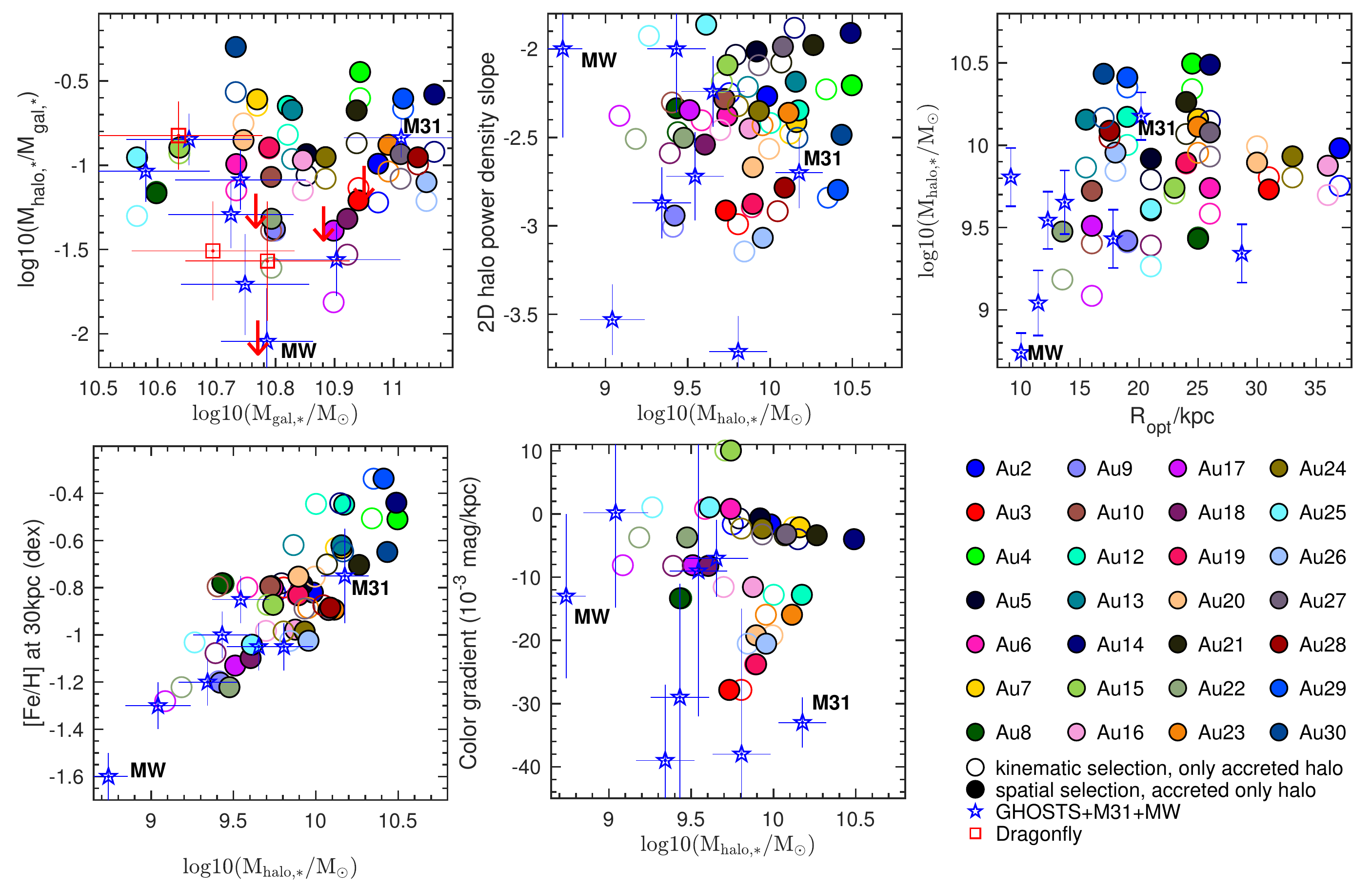}
    \caption{Correlations among stellar halo properties for accreted stellar particles only. Each color indicates a galaxy, as labelled on the right panel. Empty and filled symbols represent quantities obtained for the kinematically-defined and spatially-defined stellar halo, respectively, as explained in Section~\ref{sec:def}. Data from GHOSTS galaxies, M31, and MW are shown as blue stars. Red squares (arrows) indicate detections (upper limits) for Dragonfly galaxies. Auriga reproduces quantitatively the diversity in stellar halo properties found in the observations.}
    \label{fig:scatterplot_acc_obs}
\end{figure*}

\begin{figure*}
	\includegraphics[width=2\columnwidth]{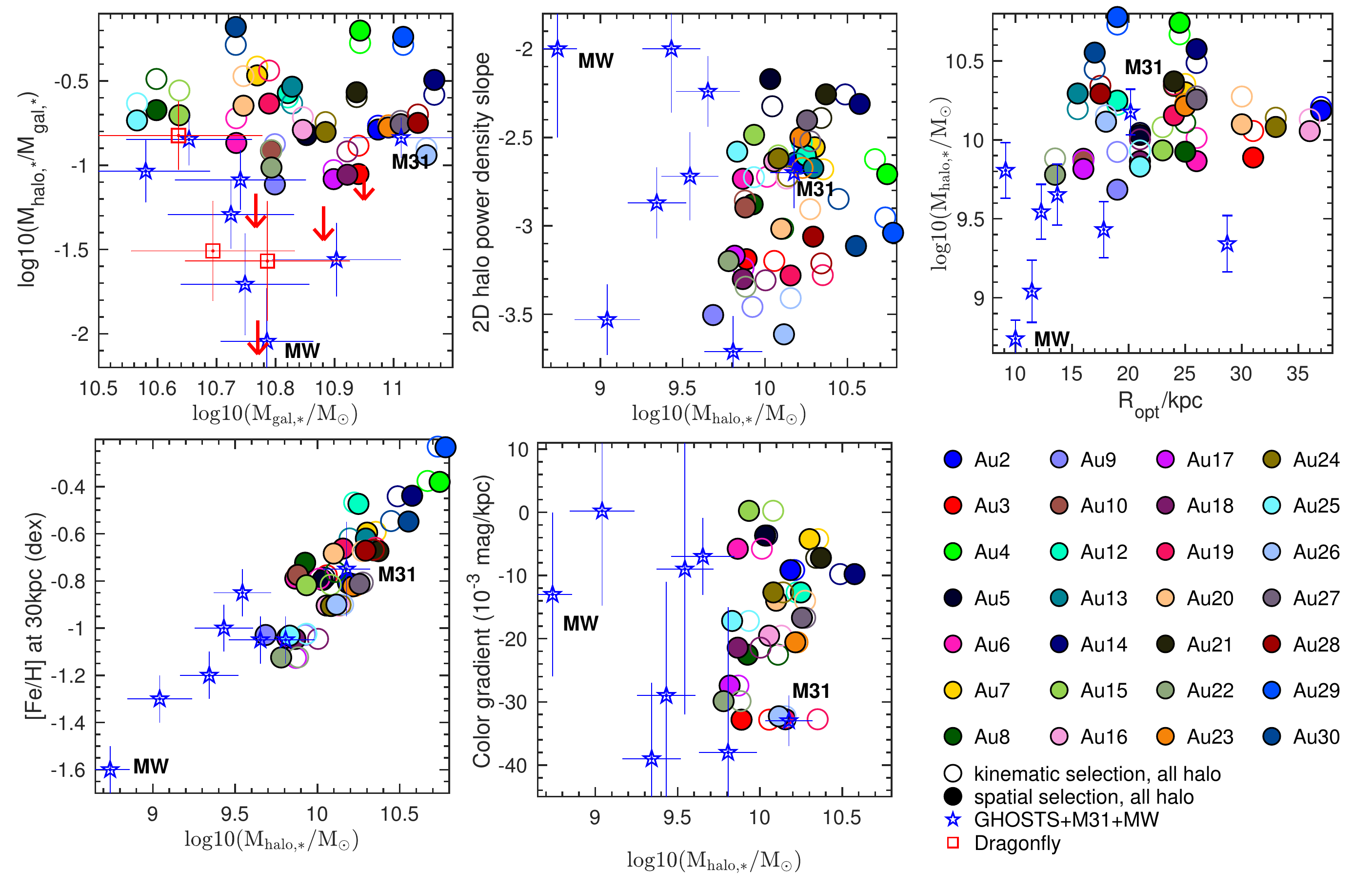}
    \caption{Correlations among stellar halo properties when both \emph{in-situ} and accreted halo particles are considered. Each color indicates a galaxy, labelled on the right panel. Empty and filled symbols represent quantities obtained for the kinematically-defined and spatially-defined stellar halo, respectively, as explained in Section~\ref{sec:def}. Observed data from GHOSTS galaxies, M31, and MW are shown as blue stars. Red squares (arrows) indicate detections (upper limits) for Dragonfly galaxies. Auriga reproduces the diversity in stellar halo properties found in the observations. However, the \emph{in-situ} stellar halo mass of the Auriga galaxies is about an order of magnitude larger than for observed galaxies.}
    \label{fig:scatterplot_all_obs}
\end{figure*}

We now correlate the stellar halo properties of the Auriga galaxies and compare them with observations. Figure~\ref{fig:scatterplot_acc_obs} shows these correlations for the accreted component
 of the stellar halo and Figure~\ref{fig:scatterplot_all_obs}
shows the results when all particles of the stellar halo, both \emph{in-situ} and accreted, are considered. 
The empty and filled symbols in all panels represent the quantities obtained for the kinematically-defined and spatially-defined stellar halo, as explained in Section~\ref{sec:def}, respectively. 

The panels in both figures show the fraction of stellar halo mass to total
stellar mass as a function of total stellar mass\footnote{Here the total stellar mass is considered as the stellar mass contained within a virial radius.} (top-left panels); the projected stellar halo density slope as a function of stellar halo mass (top-middle panels);
 the stellar halo mass as a function of optical radius (top-right panels); and
 the [Fe/H] at 
 30 kpc along the minor axis, and color gradient slope along the minor axis as a function of stellar halo mass (bottom-left, -middle panels, respectively).  

All the observed quantities are from individual galaxies presented in H17, shown in these figures as blue stars,
and we refer the reader to that work for details on how they were obtained. In the top-left panels, we also show the Dragonfly galaxies \citep{Merritt16} using the values presented in H17, as red squares for the detections and red arrows for upper limit values. 
The quantities presented for the Auriga galaxies are calculated
in the same way, or as closely as possible, as done for the observations, as explained below.

Briefly, H17 estimated the total accreted stellar halo mass of the GHOSTS galaxies as follows. First,  
the mass enclosed within two ellipses of equal axis ratios at minor axis radii of 10 and 40 kpc is estimated. The size of 
this region is determined by the region covered by the observations and along the major axis it ranges from 1.5 to 5.5 disc optical radii.
The resulting mass is multiplied by a factor of three to account for the mass outside the elliptical shells; in particular for the 
mass in the inner regions since there is little halo mass beyond 40 kpc along the minor axis.
The factor of 3 was estimated by comparing the total stellar mass of the accreted halo 
models of \citet{BJ05} to their stellar mass within the same elliptical region.
For this comparison, we estimate the total spatially-defined stellar mass of the Auriga haloes in a similar way. 
We calculate the stellar mass beyond 1.5 optical radii along the major axis and 10 kpc above the disc plane and 
multiply this by a factor of three, to mimic what was done for the observations. 
We tested here the accuracy of this approach and found that there is a 40\% uncertainty in the total accreted stellar mass when
estimating it using the factor of three as in H17, which is within the errorbars of the observational results 
(see also \citealt{Dsouza_bell18}). 
  
The [Fe/H] and (F606W-F814W) color gradients of each Auriga stellar halo (panels bottom-left and -middle) 
are calculated along the projected disc's minor axis. The median metallicity is computed at 30 kpc along the minor axis
to match the metallicity presented in H17 for the GHOSTS galaxies. The color gradient slope
 is calculated from linear fits to the color profiles of the Auriga galaxies along the minor axis, again as done for GHOSTS galaxies. 
 The 2D density slopes (top-middle panels) are obtained from power-law fits to the SB profiles along the minor axis.
 
A first important result to note from Figures~\ref{fig:scatterplot_acc_obs} and \ref{fig:scatterplot_all_obs} is that \emph{the diversity observed in 
 stellar halo properties for MW-mass galaxies, such as total stellar halo mass, slopes in the density and color profiles,
  and median metallicity values, is reproduced by the Auriga galaxies.}
 As with the observed galaxies presented in H17, none of the panels in these figures shows strong
 correlations for the Auriga galaxies, except for the stellar halo mass -- [Fe/H] relation (bottom-left panel) which we discuss below. 
 
 The range of projected stellar halo density profile slopes agrees rather well with those obtained from observations. However, 
 the accreted only profiles have slopes always flatter than $\sim -3$ whereas some observations have steeper slopes than that. This discrepancy may indicate the presence of \emph{in-situ} haloes in at least some observations.
  Steeper density profiles for the Auriga haloes are obtained when both accreted and \emph{in-situ} stellar halo components are considered. 
  
 The diversity in stellar halo [Fe/H] is also in agreement with the observations, when only the accreted component is considered
 (bottom-left panel in Fig.~\ref{fig:scatterplot_acc_obs}).  More importantly, the Auriga galaxies follow the stellar halo mass -- [Fe/H] correlation
  found in the observed galaxies very nicely, notwithstanding the larger stellar halo masses which consequently imply larger metallicity values (see below).
 Interestingly, the Auriga stellar haloes present a variation in their minor axis RGB color profile gradients in good agreement with the observations, ranging from flat color profiles to very steep negative gradients.
 
The disc sizes (identified here as optical radius) are not strongly correlated with stellar halo mass (top-right panels), although there is
a slight tendency for larger discs to have more massive stellar haloes.  We note that the discs of the Auriga galaxies are typically larger
than those of the eight observed galaxies that we use to compare the models with. See G17 for a discussion about disc sizes in the Auriga simulations.
 
  The most noticeable disagreement between the observations and the Auriga simulations is the stellar halo masses, which are
 substantially higher in Auriga, by as much as an order of magnitude when the accreted plus \emph{in-situ} components are considered (e.g., top-middle 
 panel in Fig.~\ref{fig:scatterplot_all_obs}). Given  that the simulations match the stellar mass - metallicity relation, it follows that a disagreement exists also between the absolute values of the halo metallicites and those of the observations, when the \emph{in-situ} component is considered, as seen in Fig. \ref{fig:scatterplot_all_obs}. The accreted stellar halo masses agree better with the observations (e.g., top-middle panel 
 in Fig.~\ref{fig:scatterplot_acc_obs}), which suggests
 that the Auriga haloes have an overly large \emph{in-situ} component. Moreover, some Auriga galaxies have fractions of stellar halo mass to  total  stellar mass 
  larger than found in the observations, especially when both accreted and \emph{in-situ} components are considered (top-left panels).
  
 Summarising, in general the Auriga stellar haloes are able to reproduce \emph{quantitatively} most observational results, when only the accreted component is considered. When both the accreted and \emph{in-situ} components are considered, the most notable disagreement between the simulations and observations is the stellar halo masses, which are larger (up to an order of magnitude) than the observed halo masses. This results also in a disagreement between the absolute halo metallicity values of the simulations and observations.

\subsection{The stellar halo mass -- [Fe/H] relation}
\label{sec:massmetal}

H17 discovered a relationship between the total stellar halo mass of nearby galaxies and the [Fe/H] of the stellar halo at 30 kpc along the minor axis. This is likely to be connected to the tight observed luminosity-metallicity relation in dwarf galaxies \citep[e.g.,][]{Kirby13}.
 This is the strongest correlation between observed stellar halo properties of nearby galaxies.
Figs.~\ref{fig:scatterplot_acc_obs} and~\ref{fig:scatterplot_all_obs}  (bottom-left panel) show that the Auriga galaxies reproduce this relationship quite nicely, such that more massive stellar haloes
are more metal rich. Here we investigate the origin of this relationship and demonstrate that it provides insights on the halo mass assembly and accretion history (see also \citealt{Dsouza_bell18}, a parallel study using the Illustris simulations reaching similar conclusions).

\begin{figure*}
	\includegraphics[width=2\columnwidth]{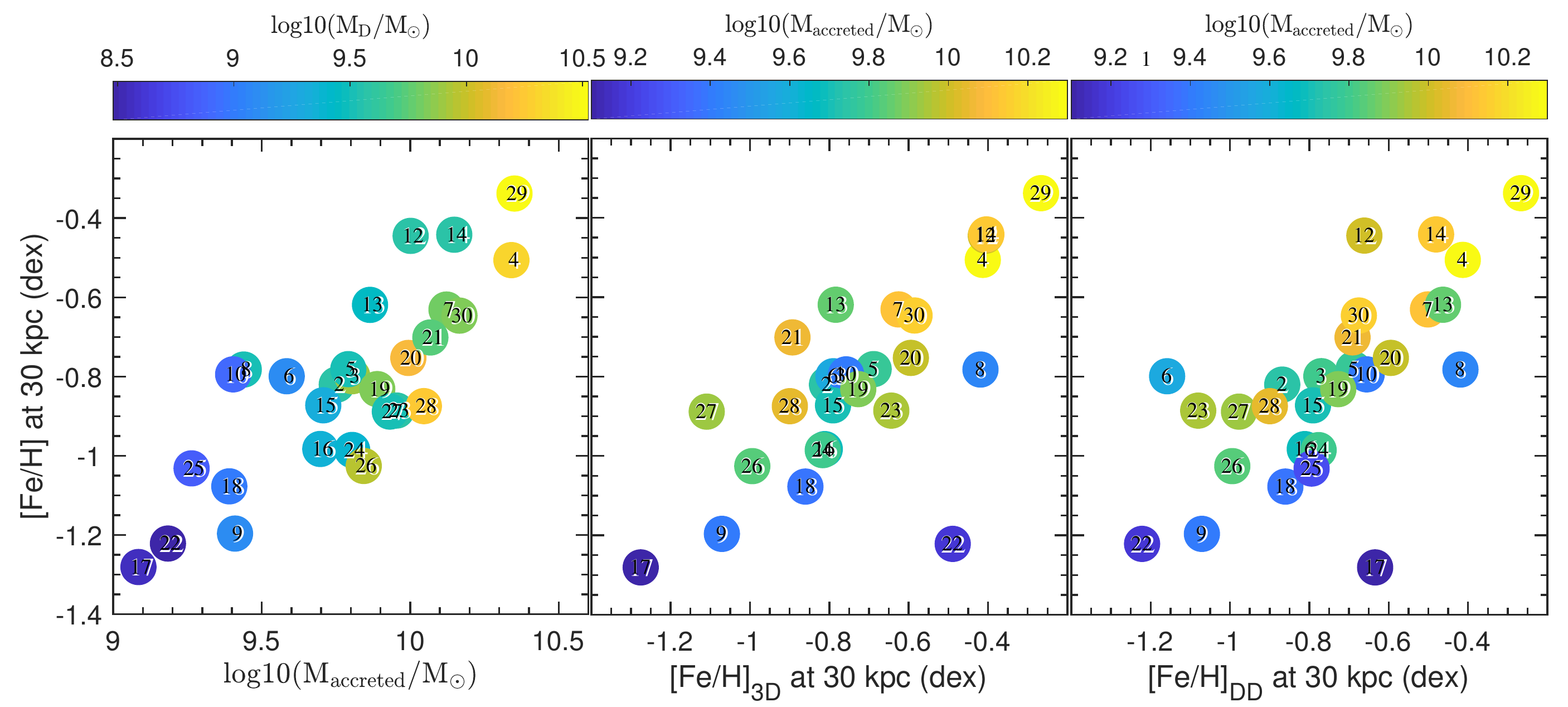}
	\caption{Left panel: stellar halo mass - [Fe/H] relation for the accreted stellar halo, as in the bottom-left panel of Figure~\ref{fig:scatterplot_acc_obs}. Circles are colored according to the stellar mass of the three most massive satellite debris contributors. 
	Middle (Right) panel: Median [Fe/H] of the overall halo at 30kpc as a function of the median [Fe/H] of the three most massive (single most dominant) contributing satellites. Circles are colored according to the total accreted stellar halo mass. Numbers inside each circle represent the galaxy label.
	The mass and [Fe/H] of the stellar halo reflect the mean properties of the more massive satellites that make it up. }
 \label{fig:3mostsign}
\end{figure*}

The left panel in Figure~\ref{fig:3mostsign} shows the stellar halo mass-[Fe/H] relation, with symbols color coded according to the median stellar halo mass of the three
most significant progenitors, i.e. the three most massive satellites accreted.  The middle panel shows the median stellar halo [Fe/H] at 30 kpc as a function of the median [Fe/H] at 30 kpc of the three
most significant progenitors while in the right panel we compare the same halo [Fe/H] with that of the single most significant (dominant) contributor. Both middle and right panels have their symbols color coded according to the total accreted halo mass of each galaxy.

The stellar mass of the three most significant progenitors, as well as that of the dominant contributor (not shown), of each galaxy correlates strongly
with the total stellar halo mass of the entire halo (see also Fig.~\ref{fig:mass_debris}). The median [Fe/H] of the three most significant progenitors also correlates strongly with the median [Fe/H] of the stellar halo. 
 
 However, the median [Fe/H] of the most significant (dominant) progenitor does not correlate that well with the median [Fe/H] of the stellar halo. 
 This is particularly the case for the lower mass haloes, where several progenitors
 contribute similarly to the total accreted stellar halo mass. These progenitors have similar masses, but their [Fe/H] may vary by up to 0.5 dex \citep{Kirby13}. 
 Thus, to relate the [Fe/H] of the stellar halo with that of the most dominant/significant progenitor is not always straighforward for lower mass stellar haloes such as that of the MW.
  
 We find with Auriga that the stellar halo mass and median [Fe/H] along the minor axis of the
 halo give us insight into the properties of the one to three most significant satellites that were accreted \citep[see also][]{Deason16, Bell17}.

\section{Discussion}
\label{sec:dis}

The stellar halo properties of the Auriga galaxies match quite well the properties of observed 
stellar haloes in individual nearby galaxies, i.e. M31 and those in the GHOSTS survey. 
The main discrepancy is in the total mass, where we find that simulated haloes are 
more massive than observed. This discrepancy is significantly reduced if only the accreted component of the Auriga haloes 
is considered, suggesting that their \emph{in-situ} component is much too massive. Nonetheless, 
even when only the accreted components are considered, haloes in the simulations are still slightly brighter and more massive 
than many of the observed haloes,
as shown in Figures~\ref{fig:surf_data} and \ref{fig:scatterplot_acc_obs}. 

 Interestingly, none of the simulated galaxies can reproduce the observed properties of the MW stellar halo. 
 Its mass and median metallicity are significantly lower than those estimated for nearby galaxies. 
 We note that the debris of the Sagittarius dwarf, currently disrupting, is generally disregarded when estimating the metallicity of the MW halo and that might partly explain the discrepancy in metallicity; however, the Sagittarius debris is included in the MW stellar halo mass estimate. This result suggests that  i) either the accreted satellites 
in the Auriga simulations are typically more massive (and therefore more metal rich) than those that primarily built the MW's halo, ii)  that the MW has a lower halo virial
  mass than that sampled by the Auriga galaxies, which is $1\times 10^{12} \rm{M_{\odot}}~ \rm{to}~ 2\times 10^{12} \rm{M_{\odot}}$, or iii) that the MW is an outlier in its halo properties and has had an unusually quiet history with little accretion. The virial halo mass of the MW is rather uncertain, with estimated values generally among those covered by Auriga simulations. However, there are several studies pointing 
  towards a slightly lower halo mass for the MW ($0.5\times10^{12} \rm{M_{\odot}}~ \rm{to}~ 1\times 10^{12} \rm{M_{\odot}}$, e.g., \citealt{ VeraCiro13, Diaz14, Cautun14}).  A lower virial halo mass would be in better agreement with the observed low stellar halo mass of the MW \citep[e.g.,][]{Pillepich14, Cooper13}.
  Regarding i), a recent study by \citet{Simpson18}  shows that most of the Auriga surviving satellites 
  have lost a substantial fraction of their peak mass ($>80\%$) by $z=0$ and have large stellar masses for their halo masses as compared to trends from abundance matching studies of central galaxies \citep{Guo10,Behroozi13}. 
   As previously discussed, this is partly due to dark matter stripping of the satellite galaxy as it enters the host virial radius, and the feedback model in Auriga, which does not model early stellar feedback. The Auriga haloes are also slightly younger than in other simulations (both overall and in their accreted component), which may be an indication  of longer star formation activity in the Auriga 
  satellites.  \citet{Simpson18} show that the feedback process implemented in Auriga allows higher-mass satellite galaxies ($> 10^8 \rm{M_{\odot}}$ in stellar mass) to be star formers until later times compared to observations and to other models implementing different feedback mechanisms. 

  Notwithstanding the disagreement with the data, we may attempt to shed some light onto the assembly history of the MW by combining the properties of the MW halo, together with the correlations shown in Figures~\ref{fig:mass_vs_prog}-\ref{fig:slopedens_vs_prog} and Figure~\ref{fig:3mostsign}. Considering its low stellar halo mass, its weak halo metallicity gradient and its 2D halo density power-law slope (see Figure~\ref{fig:scatterplot_acc_obs}), the correlations found in the Auriga simulations suggest that the number of significant progenitors contributing to the MW's  halo is $\gtrsim 6$; with stellar masses between $\sim 10^7-2\times10^8~\rm{M}_{\odot}$. Consequently the properties of the MW halo would be similar to those of Fornax-mass dwarf satellites (although see \citealt{Fiorentino17}) and rather different from the properties of very low mass satellites (such as Segue II, Bootes II, etc. see \citealt{McConnachie12}) or ultra-faint galaxies \citep[e.g.,][]{Belokurov06b,Bechtol15, Koposov15,Drlica-wagner16,Homma16}. This is in agreement with predictions from other models \citep[e.g.,][]{Robertson05, Font06a, Deason16, Amorisco17a}. We note again that this inference should be taken with some caution, given that the stellar mass and metallicity of the MW do not match individually any of those values from the Auriga simulations.

  Interestingly, M31 follows quite well the median SB profile of the Auriga simulations, and in particular the properties of the larger Auriga stellar haloes, which likely indicates that its aggregated stellar halo originates
 from the disruption of massive satellites. This is also reflected in the [Fe/H]/color profile, with median metallicity values typically higher than for the rest of the
 galaxies and with a strong negative [Fe/H] gradient with galactocentric distance. If we combine all these observed properties, our results suggest that most of the M31 stellar halo (90\% of its mass) was built from a few ($\lesssim 3$) satellites, and that its general properties are dominated by a single very massive, high [Fe/H] satellite (see also, e.g., \citealt{Gilbert12} for a more quantitative comparison of M31's halo properties with accreted stellar
 halo models and \citealt{Dsouza_bell18}).

\section{Summary and conclusions}
\label{sec:concl}

We analysed the stellar haloes of the Auriga
simulations, a large suite of high-resolution magnetohydrohynamical
cosmological simulations of
MW-mass galaxies with haloes in the mass range $1<M_{200}/10^{12}~\textrm{M}_{\odot}<2$.
Stellar halo particles are selected according to a commonly used
kinematical criterion to isolate and exclude rotationally supported disc
particles. In addition, for comparison with observational data, a 
spatial selection criterion was also applied to each galaxy. Star particles were classified
according to their origin as belonging to the accreted or \emph{in-situ} component of the stellar halo.

We characterise the stellar halo  using both spherically-averaged and projected quantities 
along one direction (the disc's minor axis). When required, the RGB component was extracted from the stellar population represented by each star particle and a similar star selection criteria and analysis as in the observations
were applied to the simulated galaxies. This allows a fair comparison to observations.

Our main results are as follows:

\begin{itemize}
    \item The Auriga stellar haloes display a great diversity in properties such as surface brightness profile slopes (between $-2$ and $-3.5$), median [Fe/H] at 30 kpc (between $-0.3$ dex and $-1.3$ dex), [Fe/H] gradients (from none to large negative gradients), median ages (from $\sim$6 to 12 Gyr), and accreted mass fractions. Some properties may differ quite significantly as a function of radius if constructed from spherical concentric shells or from line of sight projections. This is the case, in particular, for the metallicity and age profiles, and the fraction of accreted material. In addition, for each galaxy and at each radius, the spread in [Fe/H] and age is large.

    \item In general, the Auriga haloes are oblate in the inner 50 kpc and become prolate in the outer regions, at $\sim 100$ kpc. The typical values of $c/a$ across the entire distance range are $\sim 0.8$.

    \item Both the mass spectrum and number of accreted satellites that contribute 90\% of the total accreted stellar halo mass of each galaxy, $N_{\rm{sp}}$, also vary, with  $N_{\rm{sp}}$ ranging from 1 to 14 with a median of 6.5.  

    \item The values of the parameters that characterise the Auriga haloes, as well as their scatter, are generally in good agreement with the observed properties of nearby stellar haloes. 

    \item The most significant mismatch between our models and observations is the stellar halo masses, which are typically larger than the estimates for most nearby galaxies. The exception is M31, which has an estimated stellar halo mass that matches very well the values obtained with the Auriga galaxies. The discrepancy is significantly reduced, but not fully erased,  when only the accreted component is considered. This suggests that the \emph{in-situ} component of the simulated stellar haloes is too prominent and that satellites are possibly  too luminous. This is likely due to the feedback processes implemented in Auriga that may allow too much star formation at late times for the more massive satellites at given halo mass. Alternatively, the virial mass of the simulated haloes could be larger than those of the observed galaxies.

    \item We find correlations between accreted stellar halo mass, [Fe/H] gradients, and density slopes and $N_{\rm{sp}}$. The smaller $N_{\rm{sp}}$, the stronger the negative metallicity halo gradient, the steeper the halo density profile and the more massive the accreted stellar halo. These trends are not as marked when the overall halo (both accreted $+$ \emph{in-situ} components) is considered.
 
   \item Our results suggest that the MW stellar halo was built primarily from $N_{\rm{sp}} \gtrsim 6$ satellites with stellar masses $\lesssim 10^{8}~\rm{M}_{\odot}$. On the other hand, the M31 halo was likely formed from fewer satellites, $N_{\rm{sp}} \lesssim 3$, with one of them very massive and metal rich, dominating the properties of the overall stellar halo. 

   \item The stellar halo mass--[Fe/H] relation at 30 kpc (along the minor axis) found observationally in the GHOSTS survey is reproduced in the Auriga galaxies, both when only the accreted component and when the overall halo is considered. This relationship reflects the mass and metallicity of the most important satellites contributing to the stellar halo. We note nevertheless that the absolute metallicity values of the simulated haloes when both \emph{in-situ} and accreted components are considered are higher than those of the observed galaxies due to the higher stellar halo masses.
 \end{itemize}

It is important to highlight that the Auriga galaxies have not been specifically chosen to match the MW formation  and merger history
(as was done in e.g. \citealt{BJ05}) or the Local Group environment 
(as was done in the APOSTLE simulations of \citealt{Sawala16}). Thus, the diversity of accretion and merger histories
in Auriga is well suited to understand and interpret the diversity of stellar 
halo properties in the nearby Universe. The relatively large sample of thirty high-resolution simulations of late type galaxies 
allows us to study and characterise correlations between stellar halo properties and formation histories of individual galaxies. 
 
Observations of stellar haloes have increased in the past few years and will continue to increase over the next decades with current and
future programmes on HSC/Subaru, DESI, LSST, ELT, GMT, and WFIRST. These will provide panoramic views and information on the stellar
properties of individual stars in hundreds of galactic haloes, which will greatly improve the detailed information gained so far
from a handful of galaxies. The results and predictions from this work will help interpret those future observations.
 
The correlations uncovered in this work show that it is possible to learn about the accretion history of a galaxy
from the bulk properties of its stellar halo. In particular, it is possible to quantify the relative size and number of 
satellites that significantly contributed  to the accreted stellar material.  In a follow-up study, we will make use of these
correlations to infer the formation and assembly history of the GHOSTS stellar haloes (Monachesi et al. in prep.).

Finally, the \emph{in-situ} stellar halo population has not been fully addressed in this work; the details and channels describing
how this population formed and the role that it plays in defining the properties of stellar haloes will be addressed in a separate
paper. 

\section*{Acknowledgements}

We wish to thank  David Campbell and Adrian Jenkins for generating the initial conditions and selecting the sample of the Auriga galaxies. AM is grateful to Eric Bell for stimulating discussions and comments on an early draft version of this paper. We wish to thank the anonymous reviewer for a constructive report that helped us improve our manuscript. AM acknowledges partial support from CONICYT FONDECYT regular 1181797. FAG acknowledges partial support from CONICYT FONDECYT regular 1181264. RG and VS acknowledge support by the DFG Research Centre SFB-881 `The Milky Way System' through project A1. This work has also been supported by the European Research Council under ERC-StG grant EXAGAL-308037. CMS acknowledges support from the Klaus Tschira Foundation. SB acknowledges support from the International Max-Planck Research School for Astronomy 
and Cosmic Physics of Heidelberg (IMPRS-HD) and financial support from the Deutscher 
Akademischer Austauschdienst (DAAD) through the program Research Grants - Doctoral 
Programmes in Germany (57129429). This work used the DiRAC Data Centric system at Durham University,
operated by the Institute for Computational Cosmology
on behalf of the STFC DiRAC HPC Facility (www.dirac.ac.uk).
This equipment was funded by BIS National E-infrastructure capital
grant ST/K00042X/1, STFC capital grant ST/H008519/1, and
STFC DiRAC Operations grant ST/K003267/1 and Durham University.
DiRAC is part of the National E-Infrastructure. AM and FAG acknowledge funding from the Max Planck Society through a ''Partner Group'' grant. This work
was supported by the Science and Technology Facilities Council
ST/F001166/1 and the European Research Council under the European
Union's ERC Grant agreements 267291 ``Cosmiway".




\bibliographystyle{mnras}
\bibliography{stellhalos}








\label{lastpage}
\begin{landscape}
\begin{table}
\centering
\caption{Table of parameters of the simulated stellar haloes. The columns are 1) Model
  name; 2) Virial mass; 3) Virial radius; 4) Total stellar mass (enclosed within a virial radius); 5) \emph{In-situ} stellar halo mass; 6) Accreted stellar halo mass; 7) Total halo $\mu_{V}$ power-law fit slope; 8) Total halo fit normalization $\mu_V{}_0$; 9) Accreted halo $\mu_{V}$ power-law slope; 10) Accreted halo fit normalization $\mu_V{}_0^{\rm acc}$;  11) Total halo [Fe/H] gradient in units of $10^{-3}{\rm dex~ kpc^{-1}}$; 12) Total halo median [Fe/H] at 30 kpc in dex; 13) Accreted halo [Fe/H] gradient in units of $10^{-3}{\rm dex~ kpc^{-1}}$; 14) Accreted halo median [Fe/H] at 30 kpc in dex; 15) \# significant progenitors (contributing 90\% of the accreted stellar halo mass) and 16) Optical radius. \underline{Note:} Values in parentheses in columns 7-14 are projected quantities calculated along the disc's minor axis.}
\label{t1}
\begin{tabular}{c c c c c c c c c c c c c c c c}
\hline
Run & $\frac{M_{\rm vir}}{[\rm 10^{10} \rm{M_{\odot}}]}$ & $\frac{R_{\rm vir}}{[{\rm kpc}]}$ & $\frac{M_{*}}{[\rm 10^{10} \rm{M_{\odot}}]}$ & $\frac{M^{\rm SH}_{\rm in-situ}}{[\rm 10^{10} M_{\odot}]}$ & $\frac{M^{\rm SH}_{\rm acc}}{[\rm 10^{10} M_{\odot}]}$  & $\alpha_{\mu_V}^{\rm total}$ & ${\mu_V}^{\rm total}_{0}$ & $\alpha_{\mu_V}^{\rm acc}$ & ${\mu_V}^{\rm acc}_{0}$ &  $\Delta \rm [Fe/H]^{\rm total}$ &  $[\rm Fe/H]^{\rm total}_{30}$ &  $\Delta \rm [Fe/H]^{\rm acc}$ &  $[\rm Fe/H]^{\rm acc}_{30}$ & $N_{\rm sp}$ & $\frac{R_{\rm opt}}{[\rm kpc]}$\\
\hline

     Au2  &  191.46  &   261.75   &    9.41   &    1.05   &   0.56   &   $-2.7(-2.6)$   &    $19.5 (19.3)$  & $-2.0(-2.2)$ & $22.8 (21.3)$ &  $-3.1(-1.7)$& $-0.6(-0.8)$&  $-0.1(-1.1)$& $-0.8(-0.8)$  & 8  & 	37.0\\
     Au3  &  145.77  &   239.01   &    8.72   &   0.50   &    0.64    &   $-3.2(-3.1)$  &  $18.5 (17.9)$ & $-2.8(-2.9)$ & $20.6 (19.4)$  &   $-6.7(-8.9)$ & $-0.6(-0.8)$ &$-5.5(-8.3)$ & $-0.7(-0.8)$	&	3 &  31.0\\
     Au4  &  140.88  &   236.31   &    8.77   &    2.47   &    2.19  &  $-2.9(-2.7)$   & $17.1 (16.4)$  & $-2.4(-2.2)$ & $19.9 (19.1)$ &  $-4.1(-3.6)$& $-0.4(-0.4)$& $-3.0( -2.0)$& $-0.5( -0.5)$ 	& 6	 & 24.5 \\
     Au5  &  118.55  &   223.09   &    7.11  &    0.49   &    0.62    &   $-2.5(-2.1)$   &  $20.7 (20.6)$  & $-2.1(-1.9)$ & $22.3 (21.5)$  &   $ -3.8( -0.1)$& $-0.6( -0.8)$& $-2.7( 0.2)$& $-0.7( -0.8)$ &  8 & 	21.0\\
     Au6  &  104.38  &   213.82   &    5.41   &    0.64   &    0.38     &   $-2.8(-2.7)$ &  $19.7 (19.3)$ &   $-2.2(-2.4)$ & $22.3 (21.0)$ &$-5.1 (-1.6)$& $-0.7( -0.8)$& $-1.4( -1.2)$& $-0.8( -0.8)$  &	9  & 26.0\\
     Au7  &  112.04  &   218.93   &    5.87   &    0.93   &   1.32   &   $-2.9(-2.5)$ &  $18.0 (18.4)$ & $-2.5(-2.3)$ & $19.9 (19.4)$ & $-3.6( -0.7)$& $-0.4( -0.6)$& $-2.5( -0.3)$& $-0.5( -0.6)$ & 4  & 	25.0\\
     Au8  &  108.06  &   216.31   &    3.96   &    1.01   &    0.27   &   $-3.3(-2.8)$ & $17.2 (19.3)$ & $-2.5(-2.3)$ & $21.9 (21.6)$ & $-1.9( -6.4)$&$ -0.4( -0.8)$&$ -3.4( -5.1)$&$ -0.9( -0.8)$ & 4 & 	25.0\\
     Au9  &  104.97  &   214.22   &    6.29   &    0.58   &    0.26  &  $-3.4(-3.5)$ &  $18.1 (16.7)$ &  $-2.7(-2.9)$ & $21.5 (19.9)$  & $-8.8 (-7.1)$& $ -0.9( -1.0)$& $-3.8 (-2.6)$& $-1.1( -1.2)$	& 7  & 	19.0\\
     Au10 &  104.71  &   214.06   &    6.19   &    0.50   &    0.25   &   $-2.8(-2.9)$ & $19.6 (18.5)$ & $-2.0(-2.2)$ &  $23.3 (21.4)$ &    $-4.4( -1.1)$ &$-0.6 ( -0.8)$& $-0.9( 1.1)$& $ -0.7( -0.8)$	& 9 & 	16.0\\
     Au12 &  109.27  &   217.11   &    6.61   &   0.67   &    1.00  &  $-3.0(-2.6)$ & $18.2 (18.1) $ & $-2.6(-2.3)$ & $20.0 (19.3)$ &  $-4.6( -3.8)$& $ -0.4( -0.5)$& $-4.5 (-4.4)$& $-0.4( -0.5)$ & 4 & 	19.0\\
     Au13 &  118.90  &   223.32   &    6.72  &    0.83   &    0.73   &  $-2.8(-2.7)$ & $18.9 (17.9)$ & $-2.2(-2.2)$ &  $21.4 (20.1)$ &   $-3.2( -3.1)$& $-0.5( -0.6)$& $-3.2( -3.8)$ & $-0.5( -0.6)$ & 5  & 	15.5\\
     Au14 &  165.72  &   249.44   &   11.72   &    1.66   &    1.40   &   $-2.7(-2.3)$ & $18.7 (18.8)$ & $-2.2(-1.9)$ & $21.0 (20.8)$ &   $-2.3( -1.2)$& $-0.4( -0.5)$& $-1.1( -0.9)$& $-0.4 (-0.5)$& 6 & 	26.0\\
     Au15 &  122.24  &   225.40   &    4.33   &    0.69   &    0.51   &   $-3.0(-2.4)$ &   $18.6 (19.6)$ & $-2.5(-2.1)$ &   $21.1 (21.5)$ &  $-4.5( -0.2)$& $-0.7( -0.8)$& $-2.9( 1.7)$& $-0.8( -0.9)$ & 8 & 	23.0\\
     Au16 &  150.33  &   241.48   &    7.01   &    0.85   &    0.50  &   $-3.3(-2.6)$ &  $17.6 (19.3)$ & $-2.7(-2.4)$ & $20.3 (20.4)$ &    $-8.4( -3.6)$ & $ -0.7 (-0.9)$ & $-5.8( -2.1)$ & $-0.8( -1.0)$ & 8 & 	36.0\\
     Au17 &  102.83  &   212.76   &    7.91   &    0.62   &   0.12   &   $-3.5(-3.1)$ &  $17.4 (18.2)$&  $-2.2(-2.2)$& $23.6 (22.2)$ &$-12.5( -7.1)$ & $ -0.5( -1.1)$& $ -3.2( -1.6)$ & $ -1.1( -1.3)$ & 14 & 16.0 \\
     Au18 &  122.07  &   225.28   &    8.34   &    0.76   &    0.25  &   $-3.1(-3.3)$& $18.8 (17.2)$& $ -2.2(-2.5)$&  $22.8 (20.8)$  &  $-9.0( -5.7)$& $-0.7 (-1.0)$ & $ -4.4 (-3.4) $& $-0.9( -1.1)$ & 14  & 	21.0\\
     Au19 &  120.89  &   224.56   &    6.15   &    1.47   &    0.77  &  $-3.3(-3.3)$& $16.8 (16.6) $& $ -2.7(-2.9)$&$20.2 (18.9)$ & $-8.0 (-7.4)$ & $ -0.5( -0.7) $ & $-5.1( -6.1)$& $-0.7 (-0.8)$	& 5 & 	24.0\\
     Au20 &  124.92  &   227.02   &    5.56   &    0.89  &    0.98   &  $-3.1(-3.0)$& $17.6 (17.5)$&   $-2.9(-2.6)$& $19.4 (19.3)$ &  $ -5.4 (-6.3)$& $-0.4 (-0.7)$&$ -6.0( -4.5)$&$ -0.5( -0.7)$ & 3 &  	30.0\\
     Au21 &  145.09 &   238.64   &    8.65   &     1.03   &    1.17   &  $-2.5(-2.2)$& $19.4 (19.3 )$&  $-2.1(-1.9)$&$21.7 (20.7)$  &  $-2.4 (-1.0)$& $-0.5( -0.7)$& $-0.5( -0.9)$& $-0.6( -0.7)$ & 4 &  	24.0\\
     Au22 &   92.62  &   205.47   &    6.20   &    0.61   &    0.15   &  $-3.0(-3.2)$& $19.6 (17.9)$&$-2.0(-2.4)$&   $24.1 (21.4)$ &   $-7.6( -5.7)$ &$-0.8( -1.1)$& $2.8 (0.4)$& $-1.0( -1.2)$	& 13 & 	13.5\\
     Au23 &  157.53  &   245.27   &    9.80   &    0.79   &    0.90   &   $-2.6(-2.4)$& $19.7 (19.6)$& $-2.2(-2.3)$&$21.5 (20.3)$ & $-2.4( -3.5)$& $-0.6( -0.8)$ & $-0.8 (-3.2)$&$-0.6( -0.9)$ & 8 & 	25.0\\
     Au24 &  149.17  &   240.85   &    7.66   &    0.74   &    0.64   &  $-2.9(-2.5)$& $19.0 (19.4)$&   $-2.4(-2.3)$& $21.4 (20.9)$&  $-4.9( -2.6)$&$ -0.7( -0.9)$&$ -2.8(-0.9)$&$ -0.8(-1.0)$	& 8  & 30.0\\
     Au25 &  122.10  &   225.30  &    3.67   &    0.67   &    0.18   &   $-3.0(-2.6)$& $18.6 (20.7)$& $-1.2(-1.9)$&$27.2 (24.2)$ &  $-3.5( -6.4)$& $-0.3( -1.0)$& $7.2( -1.0)$& $-1.2( -1.0)$	& 7 & 	21.0\\
     Au26 &  156.38  &   244.68   &   11.36   &    0.73   &    0.70   &   $-3.3(-3.6)$& $17.4 (15.4)$&   $-2.6(-3.0)$ & $20.6 (18.2)$ &   $-6.5 (-8.4)$&$ -0.7( -0.9)$& $ -1.9( -5.8)$& $-0.9( -1.0)$  & 4 & 18.0\\
     Au27 &  174.54  &   253.80   &    10.27   &    1.02   &    0.85  & $-2.6(-2.4)$& $19.5 (19.3)$&  $-2.1(-1.9)$&$22.2 (21.5)$ &    $-4.5( -2.8)$&$ -0.7( -0.8)$&$ -0.5( -0.0)$& $-0.9( -0.9)$& 7 &	26.0\\
     Au28 &  160.53  &   246.83   &   10.99   &    1.08  &    1.11 &   $-3.0(-3.1)$& $18.0 (16.6)$ & $-2.5(-2.6)$&$20.5 (18.8)$&  $-6.3( -9.3)$&$ -0.6( -0.7)$& $-5.0( -7.6)$&$ -0.8( -0.9)$ & 4 & 	17.5 \\
     Au29 &  154.2   &   243.5    &    10.38    &   3.12    &   2.24  & $ -3.1(-3.0)$&$16.6 (15.1)$& $-2.8(-2.8)$&$18.6 (17.0)$ & $-6.5( -6.4)$& $-0.2 (-0.2)$& $-6.8( -5.7)$&$ -0.3( -0.3)$	& 1  & 	19\\
     Au30 &  160.53  &   246.83   &   5.40   &    1.33  &    1.46  &  $-2.5(-3.1)$& $18.8 (15.8)$&  $-1.7(-2.4)$&$22.3 (19.4)$ &   $-2.2 (-5.6)$& $-0.4 (-0.5)$& $0.3 (-3.2)$& $-0.5 (-0.6)$ & 4 & 17.5\\

\hline
\end{tabular}
\end{table}
\end{landscape}


\end{document}